\documentclass[twocolumn, nofootinbib, showpacs, superscriptaddress,floatfix]{revtex4} 
\usepackage{amsfonts}
\usepackage{amsmath}
\usepackage{ulem}
\usepackage{dsfont}
\usepackage[pdftex]{graphicx}
\usepackage{epstopdf}
\usepackage{textcomp} 
\usepackage{wasysym} 
\usepackage{graphicx}
\usepackage{bm}
\usepackage{cancel}
\usepackage{placeins}
\usepackage[varg]{txfonts} 
\usepackage{esint}
\usepackage{xcolor}
\usepackage{comment}
\usepackage{multirow}
\usepackage{hyperref}
\hypersetup{
    colorlinks=true,
    linkcolor=red,
    citecolor=blue,
} 

\specialcomment{supplement}{\begingroup\color{cyan}\footnotesize}{\endgroup}
\specialcomment{formula}{\begingroup\footnotesize}{\endgroup}
\excludecomment{details}

\newcommand{\Mpc}{{\,\mathrm{Mpc}/h}}

\newcommand{\del}[0]{\partial }
\newcommand{\sH}[0]{{\mathcal{H}}}
\newcommand{\hvz}{\hat{\v{z}}}
\newcommand{\vol}[2]{\hspace{-0.8mm}\mbox{$\text{d}^{\hspace{-0.0mm}#1}$}\hspace{-0.2mm}#2\hspace{0.8mm}\ }
\newcommand{\varvol}[2]{\hspace{-0.0mm}\mbox{$\text{d}^{\hspace{-0.0mm}#1}$}\hspace{-0.2mm}#2\hspace{0.8mm}\!}

\renewcommand{\v}[1]{\bm{#1} }

\newcommand{\lgM}{\mathrm{lgM} }

\makeatletter
\g@addto@macro\bfseries{\boldmath}
\makeatother

\begin{document}
\title{Choose to smooth: Gaussian streaming with the truncated Zel'dovich approximation}
\author{Michael Kopp} 
\email{kopp.michael@ucy.ac.cy}
\affiliation{Department of Physics, University of Cyprus, 1, Panepistimiou Street,
2109, Aglantzia, Cyprus} 
\author{Cora Uhlemann} 
\email{c.uhlemann@uu.nl}
\affiliation{Institute for Theoretical Physics \& Center for Extreme Matter and Emergent Phenomena, Utrecht University, Princetonplein 4, 3584CC Utrecht, The Netherlands}
\affiliation{Excellence Cluster Universe, Boltzmannstr. 2, 85748 Garching, Germany} 
\affiliation{Arnold Sommerfeld Center for Theoretical Physics, Ludwig-Maximilians-Universit\"at, Theresienstr. 37, 80333 Munich,
 Germany} 
\author{Ixandra Achitouv}
\email{iachitouv@swin.edu.au}
\affiliation{Centre for Astrophysics \& Supercomputing, Swinburne University of Technology, P.O. Box 218, Hawthorn, VIC 3122, Australia \\
ARC Centre of Excellence for All-sky Astrophysics (CAASTRO), 44 Rosehill St, Redfern, NSW 2016, Australia} 
\begin{widetext}

\end{widetext}
\begin{abstract}
We calculate the dark matter halo correlation function in redshift space using the Gaussian streaming model (GSM). To determine the scale dependent functions entering the streaming model we use local Lagrangian bias together with Convolution Lagrangian perturbation theory (CLPT) which constitutes an approximation to the Post-Zel'dovich approximation. 
On the basis of $N$-body simulations we demonstrate that a smoothing of the initial conditions with the Lagrangian radius improves the Zel'dovich approximation and its ability to predict the displacement field of proto-halos. 
Based on this observation we implement a ``truncated'' CLPT by smoothing the initial power spectrum and investigate the dependence of the streaming model ingredients on the smoothing scale. 
We find that the real space correlation functions of halos and their mean pairwise velocity are optimised if the coarse graining scale is chosen to be 1 $\Mpc$ at $z=0$, while the pairwise velocity dispersion is optimised if the smoothing scale is chosen to be the Lagrangian size of the halo. 
We compare theoretical results for the halo correlation function in redshift space to measurements within the Horizon Run 2 $N$-body simulation halo catalog. 
We find that this simple two-filter smoothing procedure in the spirit of the truncated Zel'dovich approximation significantly improves the GSM+CLPT prediction of the redshift space halo correlation function over the whole mass range from large galaxy to galaxy cluster-sized halos.

\end{abstract}
\maketitle
{\small 
\tableofcontents
}

\section{Introduction}

Current and upcoming surveys of the large scale structure  of the universe \cite{DES, Euclid} probe the dark matter (DM) distribution and dynamics in various ways, directly through gravitational lensing and indirectly through observation of luminous tracers like galaxies and clusters of galaxies.
 The aim is to employ the dark matter distribution and its dynamics to constrain parameters of and to search for physics beyond the cosmological standard model, $\Lambda$CDM, in which the main components of today's universe are a cosmological constant $\Lambda$ and cold dark matter (CDM). 
 In practice, simple analytical models are used to constrain model parameters, because computational and time expensive $N$-body simulations \cite{ KPetal09, GVetal14} although more accurate, cannot be run for each choice of parameters, see however \cite{KHetal13} for an approach for interpolating between $N$-body simulations.
 Another approach is the fast creation of mock catalogues from $N$-body initial conditions and approximate laws for the gravitational dynamics and identification of halos \cite{BM96, JW13, Chuetal14, TZE13, TEWZ15,FengChuSeljak2016}. 
 Those mock simulations are more ``brute-force'' and expensive than analytical models in the sense that they work at the level of realisations rather than statistical quantities, which makes the measurement of the latter a noisy extra step. 
 But mock simulations can be tested and compared  to $N$-body simulations much easier and more directly, and they require less approximations than analytical models. 
 Given the advance in computational power they might soon surpass analytical models in retrieving cosmological information from our data. 
 In this paper we will deal with Lagrangian perturbative analytical models whose physical ingredients are closely related to those used in realisation-based methods \cite{BM96, JW13, Chuetal14, TZE13, TEWZ15,FengChuSeljak2016}. 
 Therefore advancing analytical models will not only give us deeper insight into the large scale structure (LSS) formation but will continue to be an indispensable part of testing and constraining $\Lambda$CDM.

Simple models of LSS formation are successful because halos --  gravitationally bound objects hosting galaxies and clusters --  arise from halo progenitors, or proto-halos, whose initial distribution is tightly connected to the initial density perturbation \cite{BM96, LudlowPorciani2011} and which approximately behave as single-streaming collisionless ``particles'' following the large scale CDM flow \cite{BM96}.
While DM within a halo is multi-streaming and fully nonlinear and therefore can be described neither  as a pressureless perfect fluid nor using cosmological perturbation theory, the proto-halos themselves can be collectively treated as a pressureless dust fluid whose behaviour is  accessible through perturbation theory, with initial conditions linked to the initial statistical properties of the dark matter field \cite{BBKS86,BCEK91}, which is initially is linear and gaussian \cite{PlanckXX2015}. 
 
The combination of this with the empirical result that halos have universal density profiles \cite{NFW97} is known as the halo model, see \cite{CS02} for a review. 
The halo model allows to calculate correlation and cross-correlation functions between various probes of the dark matter field and its tracers like galaxies and clusters which usually reside in halos.
In this paper we focus on the clustering of halos, in particular the two-point statistics, such that the results can be applied to lensing or galaxy surveys after populating those halo ``particles'' with density profiles and galaxies.

A good starting point for the perturbative treatment of proto-halos is the Zel'dovich Approximation (ZA) \cite{Z70}, in which particles are displaced along straight trajectories parametrized by the linear growth function, because it is known to accurately describe gravitational dynamics over a surprisingly wide range of scales \cite{C93, Ta14}.
 Extending the ZA beyond linear order within Lagrangian Perturbation Theory (LPT) leads to the Post-Zel'dovich Approximation (PZA) which accordingly improves over the ZA.  
 Building upon the success of the (P)ZA, the truncated (Post-)Zel'dovich Approximation T(P)ZA has been proposed in \cite{C93,BMW94,M94,WGB95,Hamana1998} as phenomenological method to further improve the agreement between Zel'dovich and proper $N$-body simulations by artificially smoothing the initial power spectrum at the nonlinear scale of the time of interest. 
It might appear counterintuitive that smoothing the initial power spectrum and thereby decreasing the initial power on small scales, actually can increase the final power on those scales. 
 This is due to the fact that the smoothing reduces the velocity in high density regions and hence also the amount of shell-crossing events that subsequently would erase overdensities due the ballistic nature of LPT.

A phenomenological prescription of how to choose the filter size in a dark matter TZA simulation has been provided by \cite{C93}: the chosen filtering size is given by to the nonlinear scale of the initial power spectrum linearly extrapolated to the final time, which is around $1\Mpc$ for standard cosmological parameters at redshift $z=0$. 
On the other hand, for simulations of proto-halo displacements it appears that the Lagrangian size of the halo is the optimal filter size, which has been suggested for the ``peak-patch'' simulations to create mock halo catalogs  \cite{BM96} and which we confirm in Sec.\,\ref{sec:displace}.\\

 Since galaxy surveys observe in redshift space rather than real space we have to map the real space correlation function to redshift space taking peculiar velocities into account, which affect the observed clustering through a correlated Doppler shift along the line of sight \cite{K87}. 
 Many more accurate models for connecting the dark matter field to halos in redshift space have been developed and their performance tested against $N$-body simulations \cite{HeavensMatarreseVerde1998,M08,M11,NishimichiTaruya2011,KwanLewisLinder2012,delaTorreGuzzo2012,Gil-MarinWagnerVerdeEtal2012, CRW13, VSOD13}.
Here we focus on the Gaussian Streaming Model (GSM),  \cite{F94, S04, RW11},  according to which one can approximate the redshift space halo correlation function by a convolution of the real space correlation and an approximately Gaussian velocity distribution whose mean and variance are given by the scale-dependent mean and dispersion of the pairwise velocity. 
Different versions of streaming models and their ingredients have been successfully applied in cosmological parameter estimation \cite{RossEtal2007,GuzzoEtal2008,SamushiaEtal2014} justifying their further improvement given the increasing precision of measured redshift space correlation functions in the near future \cite{DES, Euclid}.
The ingredients of the GSM, in particular the pairwise velocity statistics of halos, also have other applications like the kinematic Sunyaev-Zel'dovich effect \cite{SoergelEtal2016}.
It has been demonstrated that the GSM is accurate to 1\% within statistical errors down to scales $s\approx 30\Mpc$ when the scale-dependent functions entering the model are determined from an $N$-body simulation  \cite{UKH15}, see also \cite{WRetal15}.
 A different version of the GSM, where the mean and variance are themselves Gaussian random variables has been proposed recently \cite{BCG15} and shown to be accurate down to even smaller scales \cite{Bianchi16}, albeit also containing some free parameters. The GSM employed here, Eq.\,\eqref{GSM}, does not have any free parameters.

To compute halo correlation functions in redshift space we follow closely \cite{RW11,WRW14,W14}, albeit our model is slightly different   \cite{UKH15}.
Also similar to \cite{RW11,WRW14,W14}, we apply the framework of Convolution Lagrangian Perturbation Theory (CLPT), an approximation to the PZA developed in \cite{CRW13}, in order to calculate perturbatively the ingredients of the GSM.
More precisely,  we use a particular coarse grained version called truncated CLPT (TCLPT), that has been studied before in \cite{UKH15}.
By using TCLPT we implement the smoothing procedure that is known to work on realisations and apply it to our analytical model  to  predict the halo correlation function and velocity statistics.
 CLPT recovers the ZA at lowest order while providing an approximation to PZA at higher order in perturbation theory. Since TPZA improves PZA it is therefore plausible that also the TCLPT will improve  CLPT.
 In this paper we will thus investigate how a smoothing of the input power spectrum involved in TCLPT improves the CLPT predictions by comparing to measurements done within the publicly available Horizon Run 2 halo catalog.

It turns out that a single filtering scale does not work for all relevant statistical quantities calculated using our analytical model based on local Lagrangian bias, such that we propose a smoothing procedure involving two filter scales: 
The best result for the real space correlation function and the mean pairwise velocity is achieved if a smoothing scale of around $1\Mpc$ is used similar to the smoothing scale of TZA simulations applied to DM \cite{C93}, which also optimises the agreement of the large scale vorticity with the $N$-body measurements \cite{UK14}. 
The pairwise velocity dispersion on the other hand is optimised with a smoothing scale given by the Lagrangian radius of the halo as suggested by peak-patch TZA simulations \cite{BM96} and our own realisation based studies. 
Our main result is that this easy to implement ``hybrid TCLPT,'' involving two smoothing scales, the nonlinear scale and the Lagrangian scale, significantly improves TCLPT and CLPT.
This is true both for the halo statistics in real space as well as in redshift space.  

\paragraph*{Structure}

This paper is organized as follows: In Sec.\,\ref{sec:displace}, we investigate the effect of the smoothing scale on the Zel'dovich Approximation (ZA) displacement field of proto-halos by comparing them to the actual displacements.
The proto-halos themselves and the actual displacements have been identified in the DEUS $N$-body simulation \cite{Raseraetal} by tracing back halo particles at the final time to the initial time.
 We recover the Lagrangian scale to be optimal \cite{BM96,AchitouvBlake2015}.
This gives us guidance for the choice of the appropriate smoothing scale for the theoretical evaluation of the halo correlation function in redshift space. In Sec.\,\ref{sec:GSM} we briefly introduce the Gaussian Streaming model (GSM) for redshift space distortions based on the pressureless fluid (dust) model and local Lagrangian bias. We then present the predictions of Convolution Lagrangian perturbation theory (CLPT) and its truncated version (TCLPT) for the halo correlation function in real space and the pairwise velocity statistics of halos for different smoothing scales.  We propose a new ``hybrid'' approach which introduces two smoothing scales, one to optimise the real space correlation function and the mean pairwise velocity and one to optimise the pairwise velocity dispersion. In Sec.\,\ref{sec:results}, equipped with the optimised ingredients for the streaming model, we finally determine the redshift space correlation function and compare the theoretical predictions to measurements of the halo correlation functions in the Horizon Run 2 (HR2) $N$-body simulation halo catalog. We find considerable improvement of hybrid TCLPT over the original CLPT results. 
We conclude in Sec.\,\ref{sec:concl} and describe possible further interesting lines of study in Sec.\,\ref{sec:outlook}. In App.\,\ref{sec:HR2} we describe how we extracted the correlation function from the publicly available HR2 catalog \cite{KPetal09, KPetal11} and discuss the halo bias model \cite{CA2}. 

\section{Zel'dovich simulations}
\label{sec:displace}
In this section we focus on the displacement of proto-halos using the truncated Zel'dovich approximation \cite{Z70,C93} for different smoothing scales. We will find that the Lagrangian size of the halo stands out giving optimal results compared to $N$-body simulations. This has motivated the consideration of different smoothing procedures in perturbation theory involving the Lagrangian scale in \cite{UKH15} and motivates our choice of the smoothing scale in \ref{sec:velstat}.

\subsection{Zel'dovich approximation}
The displacement field
 \begin{equation}
\bm{\varPsi}(\bm{q},z) = \bm{x}(\bm{q},z) - \bm{q} \label{displacEL} 
\end{equation}
 connects the final Eulerian positions $\bm{x}$  of dark matter tracers at their current redshift $z$ to their initial Lagrangian positions $\bm{q}$.
In the Zel'dovich approximation, the displacement field $\bm{\varPsi}$ is determined by the gradient of the linearly evolved gravitational potential related to the linearly extrapolated density field $\delta_{\rm lin}(\bm q, z) := D(z)/D(z_{\rm i})\, \delta_{\rm lin}(\bm q, z_\mathrm{i})$ via
\begin{equation}
\bm{\nabla}_{\bm{q}}\bm{\cdot} \bm{\varPsi}_{\rm Z}(\bm{q},z) = - \delta_{\rm lin}(\bm q, z_\mathrm{i})\, D(z)/D(z_{\rm i})\label{Poisson} \,,
\end{equation}
where $D(z)$ is the linear growth factor and $z_\mathrm{i}$ the redshift after recombination when all relevant scales are still linear and the subscript `lin', stands for 1st order or linear perturbation theory. A particularly nice feature of the Zel'dovich approximation is that particles are displaced along straight lines parametrized by $D(z)$ and all statistical properties of $\bm{\varPsi}_{\rm Z}$ are inherited from the linear density field $\delta_{\rm lin}(\bm q, z_\mathrm{i})$, which we assume to be a gaussian random field.
In order to describe an object corresponding to a certain spatial scale we expect that some smoothing or averaging of the initial conditions should be applied. Therefore, we replace the linear density field $\delta_{\rm lin}(\bm q, z_\mathrm{i})$ by its at scale $R$ smoothed version
\begin{subequations}
\begin{equation}
\label{deltaR}
\delta_R(\bm q, z) := D(z)/D(z_{\rm i})\, \int d^3\!q' \,W(|\bm q- \bm q'|,R) \delta_{\rm lin}(\bm q', z_\mathrm{i})\,,
\end{equation}
where $W(|\bm q- \bm q'|,R)$ is a window or filter function that implements the smoothing at scale $R$ and where for later convenience the linear density has been extrapolated by the linear growth function $D$.\footnote{The linear extrapolation is physically meaningless once the physical density contrast has become nonlinear. Nevertheless it is a useful quantity because it approximately absorbs the time dependence of the barrier $\delta_c$ of spherical collapse and the time dependence of ZA displacements \eqref{Poisson} and truncated ZA displacements \eqref{Poissonsmooth}.}
For notational simplicity we dropped the subscript `lin' from the smoothed and linearly extrapolated density field $\delta_R$.
This determines, in analogy to Eq.\,\eqref{Poisson}, the at scale $R$ smoothed displacement field
\begin{equation}
\bm{\nabla}_{\bm{q}}\bm{\cdot} \bm{\varPsi}_{\rm Z}(\bm{q},z,R) \equiv - \delta_R(\bm q, z)\,.\label{Poissonsmooth}
\end{equation}
\end{subequations}

If the matter density field $\delta_{\rm lin}(\bm q, z_\mathrm{i})$ has Gaussian initial conditions, the same is true for the at scale $R$ smoothed displacement field $\bm{\varPsi}_{\rm Z}(\bm{q},z,R)$. We will measure the displacement field of Eq.\,\eqref{displacEL} from $N$-body simulations and compare them to the prediction of the truncated Zel'dovich approximation \eqref{Poissonsmooth}. 
Using Eq.\,\eqref{Poissonsmooth} and therefore $\bm{\varPsi}_{\rm Z}(\bm{q},z,R) = D(z)/D(z_{\rm i})\, \bm{\varPsi}_{\rm Z}(\bm{q},z_{\rm i},R)$ as well as $\v{v} = a \dot{\v{\varPsi}}_{\rm Z} = a\, \partial_t D\, \partial_D\v{\varPsi}_{\rm Z}$, the displacement can be rewritten as function of the initial smoothed peculiar velocity $\bm{v}_{\rm i}(\bm{q},R)$
\begin{equation}
\bm{\varPsi}_{\rm Z}(\bm{q},z,R) =\frac{\bm{v}_\mathrm{i}(\bm{q},R)}{a_\mathrm{i}  H(a_\mathrm{i})f(a_\mathrm{i})}\frac{D(z)}{D(z_\mathrm{i})} \,, \label{Zelvelo}
\end{equation}
where $a_i$ is the initial scale factor and
$f= d\ln D/d\ln a$ the linear growth rate. This law applied to each particle is the truncated Zel'dovich approximation. 
For convenience and since there is no possibility for confusion within this section, we label the displacement field with a Z, although it refers to TZA.\\

In what follows, we will not apply the displacement Eq.\,\eqref{Zelvelo} to every particle of the initial conditions of the $N$-body simulation but only to the special points corresponding to the centers of proto-halos $\tilde{\bm{q}}$. 
This is somewhat different from the truncated Zel'dovich approximation since picking those special points of the random field will affect the statistics of $\bm{\varPsi}_{\rm Z}$. 
We will generate those Zel'dovich simulations, (Z-simulations), by displacing proto-halo positions using Eq.\,\eqref{Zelvelo} for different smoothing scales $R$. 
The comparison of the final halo positions from the $N$-body simulation to the Z-simulations will allow us to find the optimal smoothing scale for displacing proto-halos. 

\subsection{Displacement of proto-halos}
\label{sec:DEUSS}
For this comparison we use the simulation data from the DEUS consortium, described in \cite{Raseraetal,Courtinetal,Alimietal}. The simulation boxes are $648 \Mpc$ and $2592 \Mpc$ on a side with $1024^3$ particles, realized using the RAMSES code \cite{RAMSES} for a $\Lambda$CDM model calibrated to WMAP 5yr cosmology, see App.\,\ref{sec:HR2}. Halos are identified with the Friend of Friend (FoF) algorithm with a linking length parameter $b=0.2$. For each halo identified at $z=0$ we compute the center-of-mass which is considered as final coordinate $\v{x}$ of our halo. 
Then we label each particle that belongs to the $n$-th halo to compute the center-of-mass $\v{\tilde q}$ in the initial conditions -- the position of the proto-halo. Starting from this center-of-mass we re-compute the center of mass $\v{q}$ by considering all particles within a sphere of radius $R_{\rm L}$ around $\v{\tilde q}$, where $R_{\rm L}$ is the Lagrangian size of the halo. The difference between the two initial center of mass $\v{q}$ and $\v{\tilde q}$ is negligible as we would expect when dealing with spherical proto-halos. Hence, we obtain a direct measure of the displacement vector $\v{\varPsi}=\v{x}-\v{q}$ for each halo which allows us to measure the probability distribution function (PDF) of $\v{\varPsi}$ for a given halo mass. 

Within the Z-simulation we expect smaller displacements for large mass halos for two reasons.
{\it i}) There is less power on large scales and (proto-)halo displacements are not expected to be sensitive to power on scales smaller than the size of the object and {\it ii}), large halos correspond to rare high variance peaks such that they are more likely to be already aligned with the skeleton of the final cosmic web and move less then a randomly selected point of the smoothed density field. 
That means that $\v{\varPsi}$ evaluated at those points will have a different ``peak'' statistic \cite{BBKS86,DS10}. 
If we neglect point {\it ii}), that halos form on average at initial peaks in the matter density field, then it is very simple to predict the PDF of the displacement for any patch of matter with radius $R$. 
We introduce the absolute value of the smoothed displacement field
\begin{equation}
\varPsi_{\rm Z}\equiv|\bm{\varPsi}_{\rm Z}|=\sqrt{(x_1-q_1)^2+(x_2-q_2)^2+(x_3-q_3)^2}\label{Deltar} \,,
\end{equation}
for which we want the compute the probability distribution. 
With this assumption and due to Gaussian initial conditions, the distribution of displacements lenghts is a Maxwell-Boltzmann distribution
\begin{subequations}
\label{PDFeqs}
\begin{equation}
\Pi\left(\varPsi_{\rm Z},\sigma_{\varPsi_{\rm Z}}(R)\right)=\sqrt{\frac{2}{\pi}}\left(\frac{\sqrt{3}}{\sigma_{\varPsi_{\rm Z}}}\right)^3 \varPsi_{\rm Z}^2 \exp\left({-\frac{3\varPsi_{\rm Z}^2}{2\sigma_{\varPsi_{\rm Z}}^2}}\right)\label{PDFeq} \,,
\end{equation}
whose variance and mean are given by
\begin{align}
\sigma_{\varPsi_{\rm Z}}^2(R) &=\frac{1}{2\pi^2}\int dk\, \tilde{W}^2(k,R) P_{\rm lin}(k,z) \ , \label{sigma} \\
\langle \varPsi_{\rm Z}\rangle &=\frac{2\sqrt{2}}{\sqrt{3\pi}}\,\sigma_{\varPsi_{\rm Z}}\label{meanPsi}\,,
\end{align}
\end{subequations}
where $P_{\rm lin}(k,z)$ is the linear matter power spectrum at redshift $z$ and $\tilde{W}(k,R)$ is the Fourier transform of the filter function $W(r,R)$. 
This is a universal prediction of the truncated Zel'dovich approximation applied to random positions. In addition to caveat {\it ii}) we expect {\it iii}) that this linear approximation breaks down at $z=0$ due to the presence of non-linearities in the halo field itself and effectively non-deterministic effects arising on much smaller scales. 
Taking into account {\it ii}) simply requires replacing $\sigma_{\varPsi_{\rm Z}}$ in Eq.\,\ref{PDFeq} by $ \sigma_{\varPsi_{\rm Z}} \sqrt{1- \gamma^2_v}$, where 
\begin{equation}
\gamma_v(R) = \frac{\sigma^2}{\sigma_{\varPsi_{\rm Z}} \sigma_{1}}\,, \label{Peakgamma}
\end{equation}
where $\sigma^2$ and $\sigma^2_{1}$ are obtained by appending to the integrand in \eqref{sigma} a factor $k^{2}$ and a factor $k^{4}$, respectively \cite{BBKS86}.

 \begin{figure*}[t]
\centering
\includegraphics[scale=0.303]{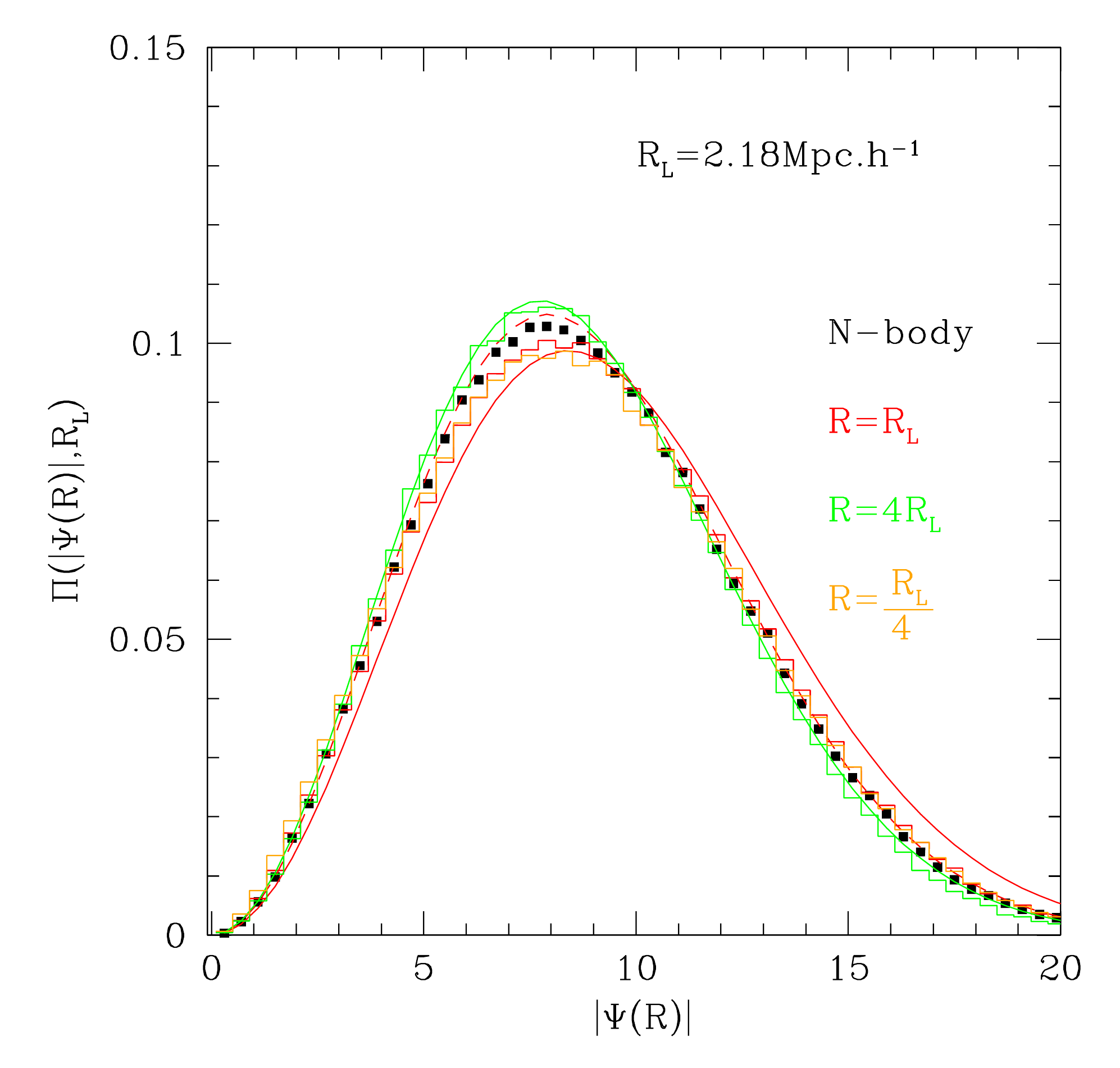}\includegraphics[scale=0.3]{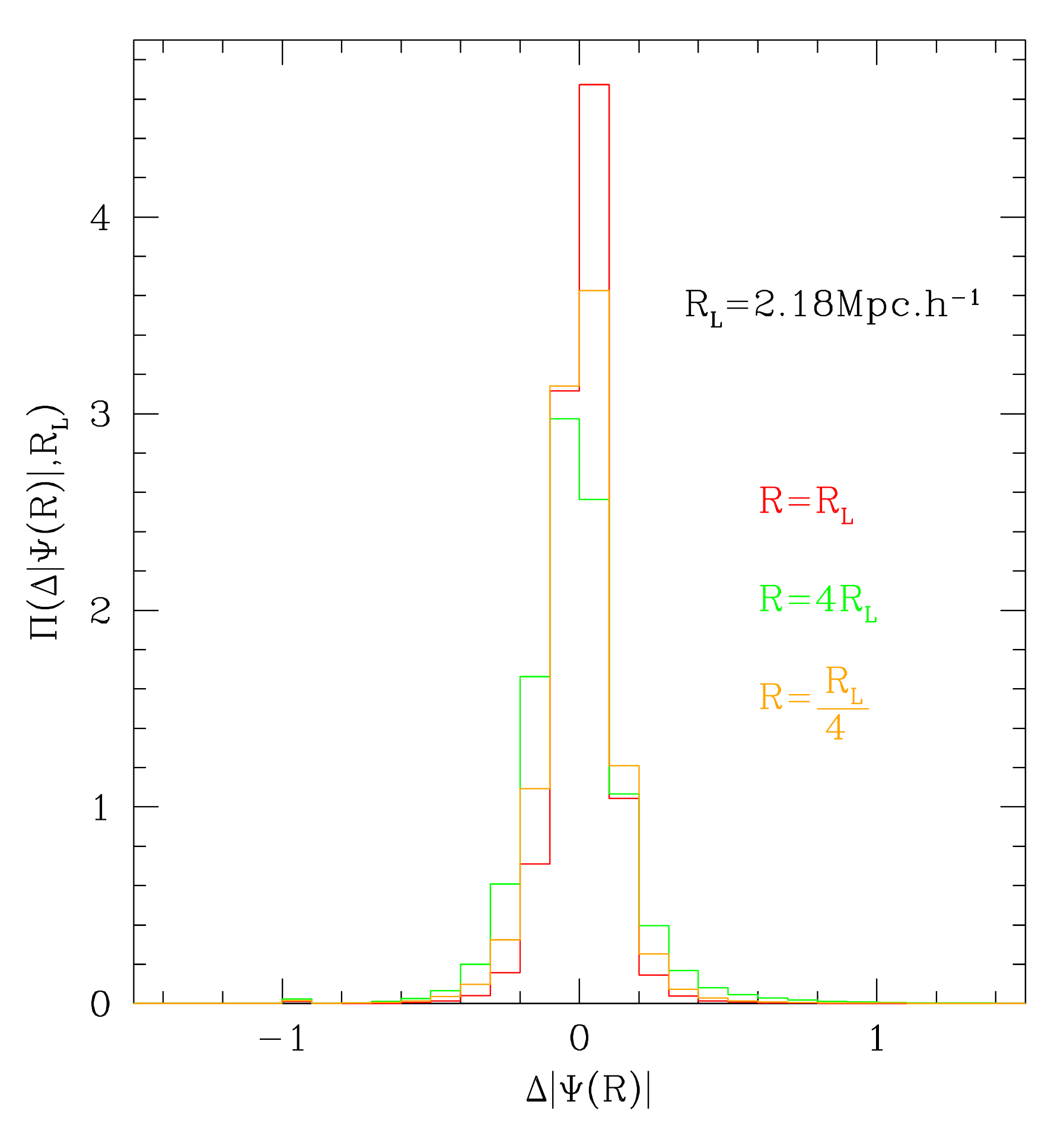}\includegraphics[scale=0.3]{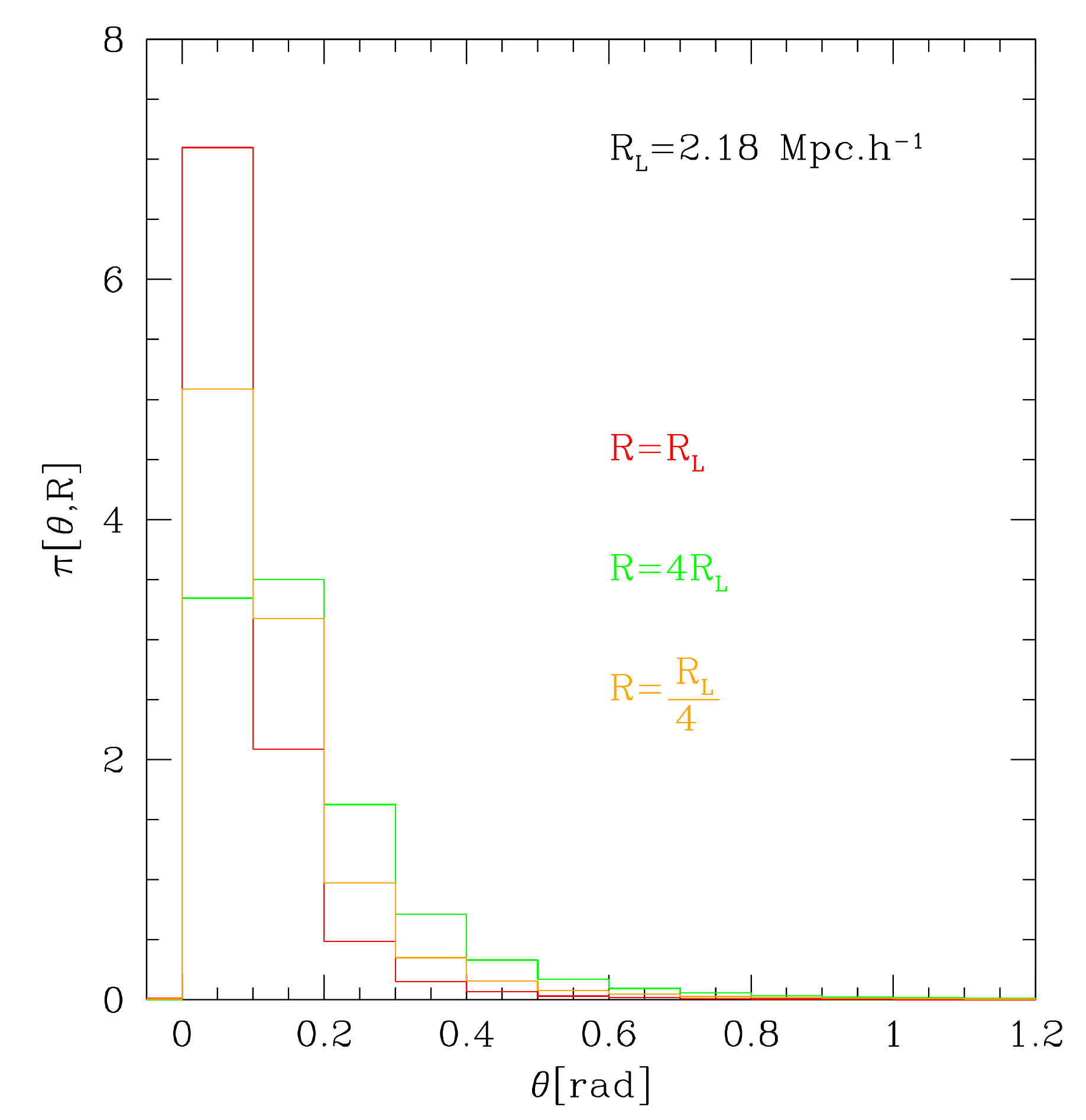}\\
\includegraphics[scale=0.303]{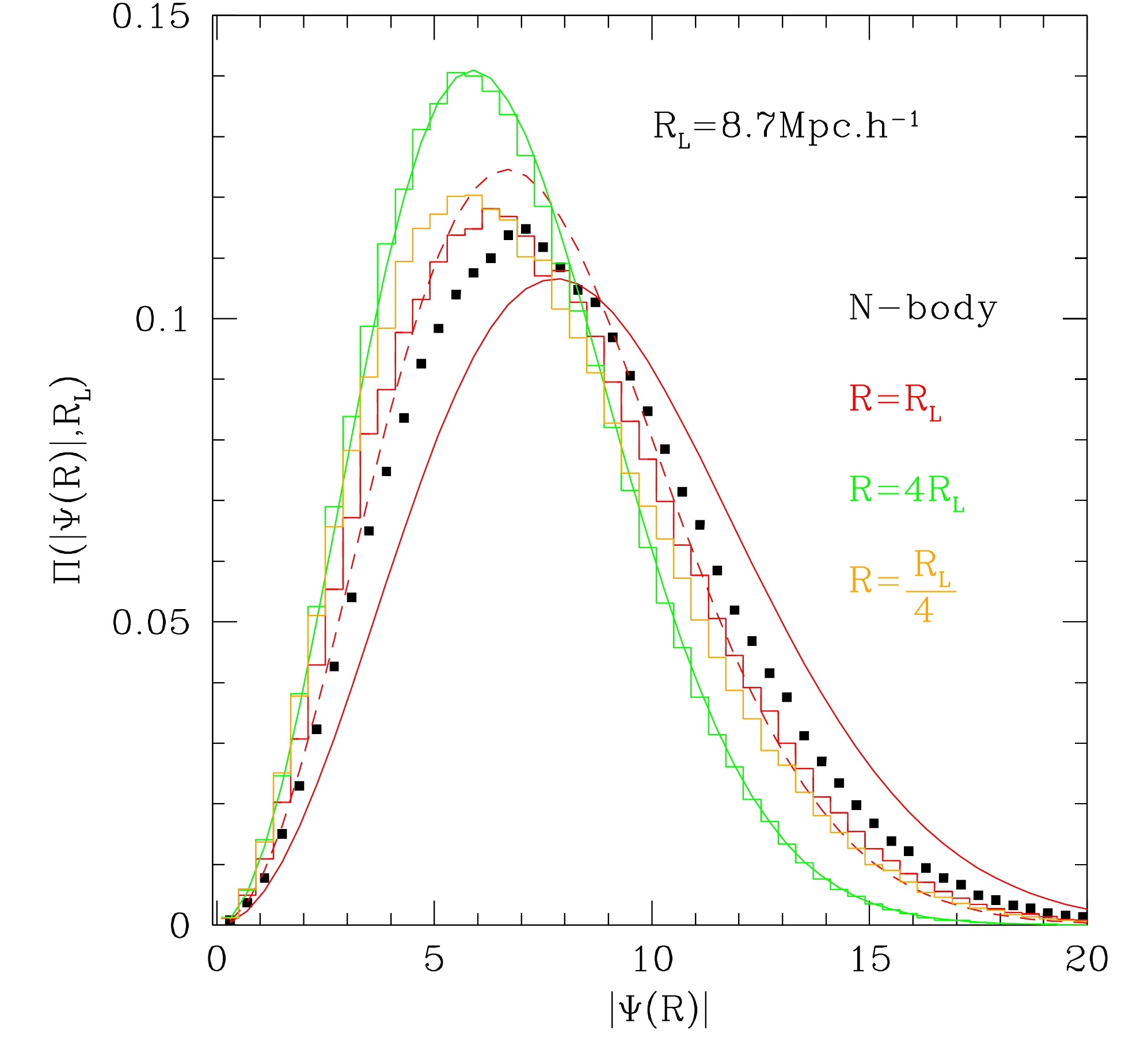}\includegraphics[scale=0.3]{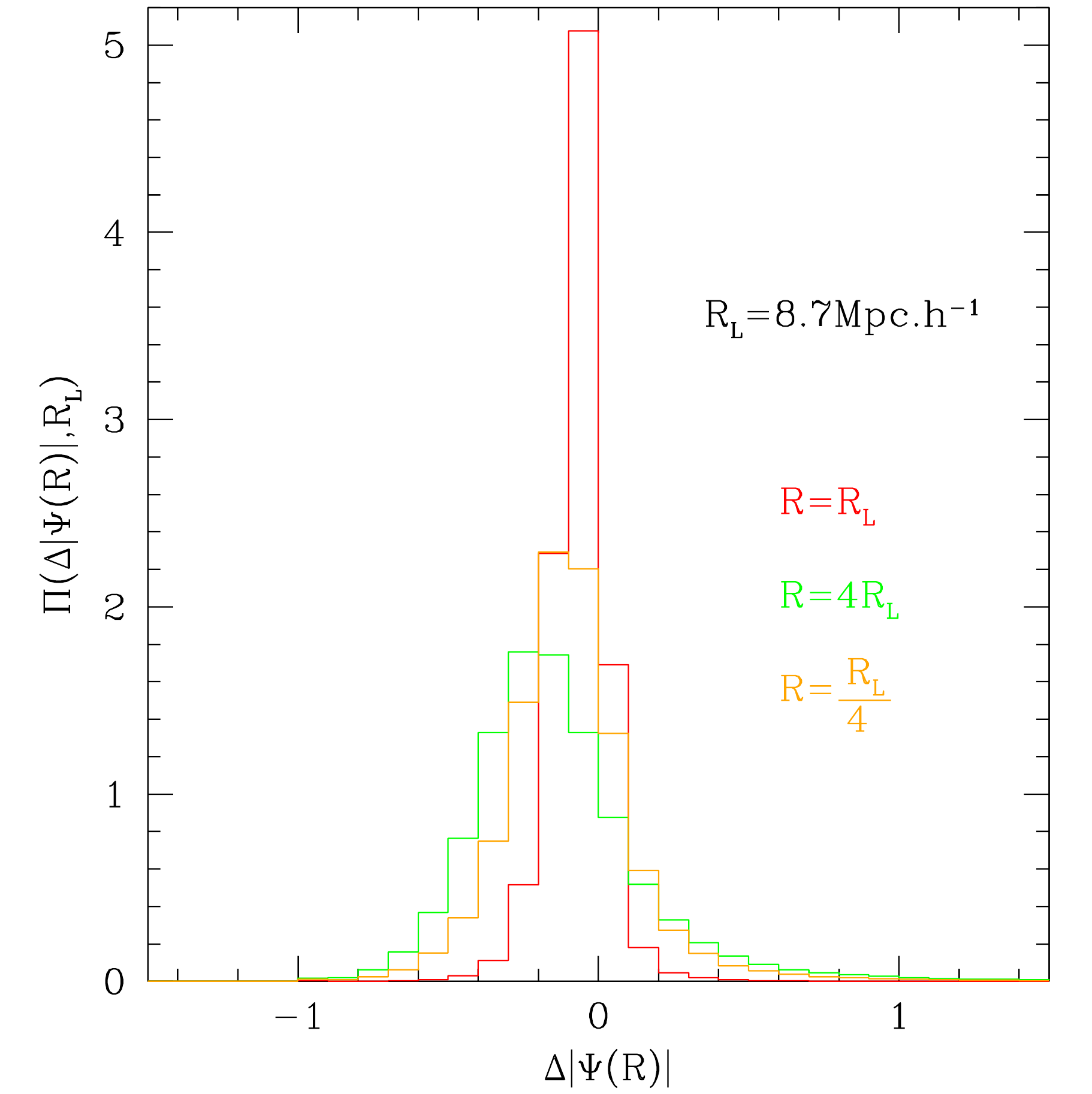}\includegraphics[scale=0.3]{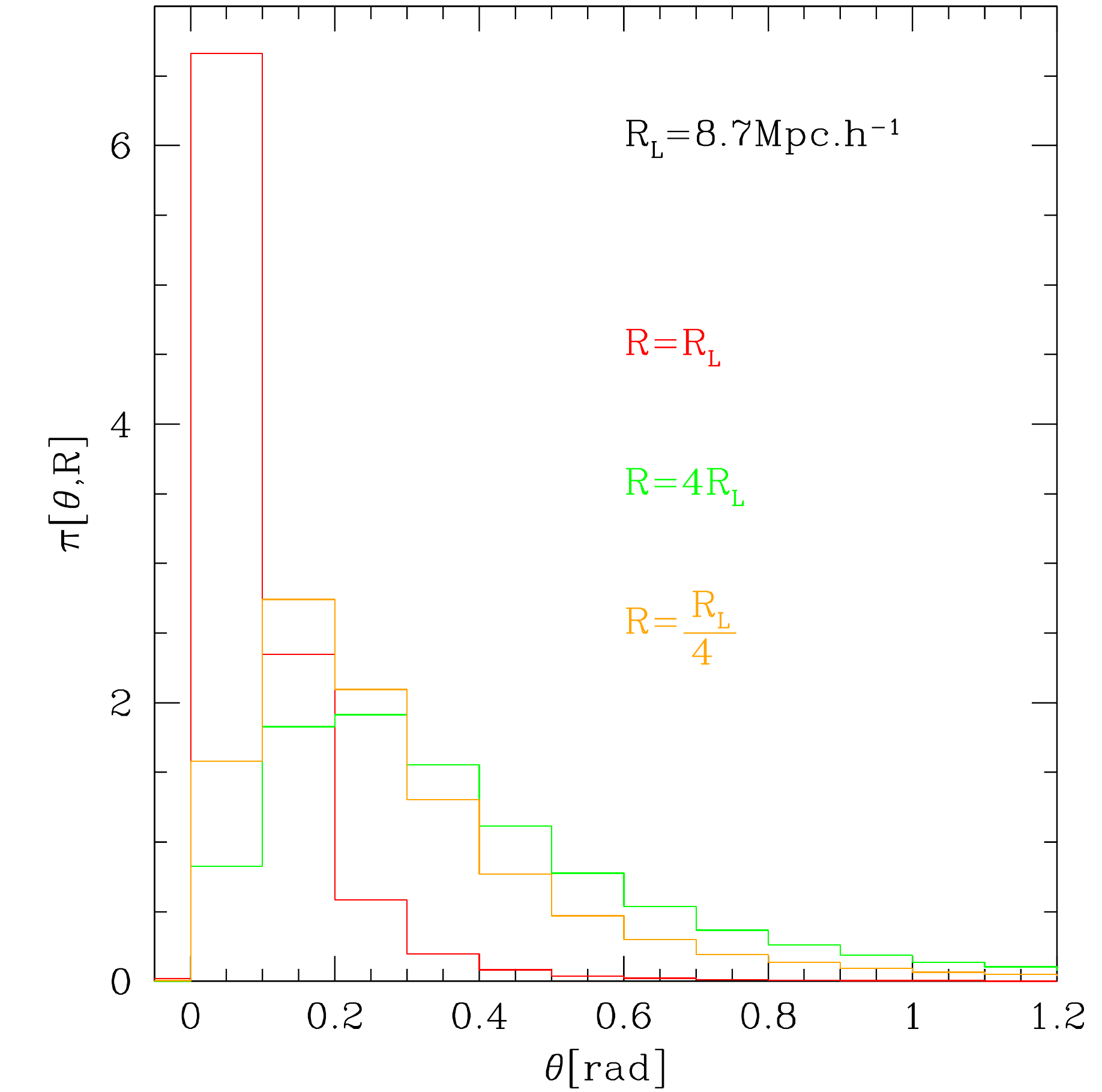}
\caption{Comparison of the displacement field $\v{\varPsi}$ obtained from the Z-simulations  ({\it colored histograms}) using a smoothing length of size $R=R_{\rm L}$ ({\it red}), $R=4R_{\rm L}$ ({\it green}) and $R=R_{\rm L}/4$ ({\it yellow}) to the exact result from the DEUS $N$-body for proto-halos with Lagrangian radii $R_{\rm L}=2.18 \Mpc$ ({\it upper panels}) and $R_{\rm L}=8.7 \Mpc$ ({\it lower panels}). 
{\it left panels:} Distribution of the absolute value of the displacement $|\v{\varPsi}|$ measured in the $N$-body simulation ({\it black squares}) and in the Z-simulations for different smoothing lengths ({\it colored histograms}) together with the theoretical prediction of Eqs.\,\eqref{PDFeqs} at the smoothing lengths $R=R_{\rm L}$ and $R=4R_{\rm L}$ ({\it colored solid lines}).
For $R=R_{\rm L}$ we show show also Eqs.\,\eqref{PDFeqs} with $\sigma_{\varPsi_{\rm Z}}$ replaced by $ \sigma_{\varPsi_{\rm Z}} \sqrt{1- \gamma^2_v}$ in {\it red dashed}.
{\it middle panels:} Distribution of the mismatch amplitude $\Delta|\bm{\varPsi}|$ \eqref{deltaPsi} for different smoothing lengths in the Z-simulations ({\it colored histograms}).
{\it right panels:} Distribution of the misalignment angle $\theta$ \eqref{thetaPsi} for different smoothing lengths in the Z-simulations ({\it colored histograms}) }\label{figDEUS}
\end{figure*}

A natural scale for a smoothing is the Lagrangian size of a proto-halo
\begin{equation}
\label{LagrangeR}
R_{\rm L}(M) \equiv R_{L}\equiv \left( \frac{3M}{4\pi\rho_0} \right)^{1/3}
\end{equation}
 which is defined as the radius in which the final mass $M$ of the halo is enclosed in a homogeneous universe with comoving matter density $\rho_0$ and therefore corresponds to size of a spherically symmetric proto-halo in the initial conditions.
This implements the intuitive picture presented in point {\it i)}.

Since we neglect that proto-halos are identified with special points and that their displacements are affected by non-linearities, neglecting points {\it ii)} and {\it iii)}, we do not expect Eqs.\,\eqref{PDFeqs} to exactly hold when applied to proto-halos.
However in the limit $R \gg R_{\rm L}(M)$ it should work because then both non-linearity and specialness of proto-halos are erased. 
This property will serve as a test of our method.

 In order to investigate the validity of Eqs.\,\eqref{PDFeqs} and the robustness of the Z-simulations, we measure the PDF $\Pi(\varPsi_Z)$ of the absolute value of the displacement field Eq.\,\eqref{Deltar} for different halos masses and different smoothing lengths in Eq.\,\eqref{Zelvelo}.

 In Fig.\,\ref{figDEUS} we compare the result of the Z-simulations  ({\it colored histograms}) using a smoothing length of size $R=R_{\rm L}$ ({\it red}), $R=4R_{\rm L}$ ({\it green}) and $R=R_{\rm L}/4$ ({\it yellow}) to the exact result from the DEUS $N$-body simulation using two halo mass bins corresponding to small proto-halos with Lagrangian radius $R_{\rm L}=2.18 \Mpc$ ({\it upper panels}) and larger proto-halos with Lagrangian size $R_{\rm L}=8.7 \Mpc$ ({\it lower panels}), respectively. 

The left panel of Fig.\,\ref{figDEUS} shows the absolute value of the displacement field determined from the $N$-body simulation ({\it black squares}), the Z-simulations ({\it colored histograms}) and the theoretical prediction Eq.\,\eqref{PDFeqs} evaluated at the smoothing lengths $R_{\rm L}$ and $4R_{\rm L}$ ({\it red and green solid lines}). 
For larger proto-halos, the amplitude of the distribution is higher and consequently the variance of the displacement is  smaller compared to the smaller mass halos.
This is a consequence of having filtered out more power for the large halos in $\sigma_{\varPsi_{\rm Z}}^2(R)$, Eq.\,\eqref{PDFeqs}. 
While for the lower halo mass the histograms with the different smoothing scales are similarly close to the $N$-body result, it becomes apparent for the larger halo mass that $R=R_{\rm L}$ performs significantly better than $R=4R_{\rm L}$ and improves over $R=R_{\rm L}/4$.
 We observe that the theoretical prediction from Eq.\,\eqref{PDFeqs} is in good agreement with the Z-simulation for $R=4R_{\rm L}$ and as mentioned before serves as a consistency check.
 For $R=R_{\rm L}$ we observe a small mismatch in agreement with the expectation from {\it ii)} that a proto-halo does not behave like an average particle since the peak-property is not taken into account in \eqref{sigma}. 
To illustrate this we also show the theoretical prediction ({\it red dashed line}) from the Zel'dovich approximation taking into account the peak correction $\sigma_{\varPsi_{\rm Z}}  \rightarrow  \sigma_{\varPsi_{\rm Z}} \sqrt{1- \gamma^2_v}$ which reduces the velocity dispersion of peaks compared to a randomly selected point, see Eqs.\,(4.24-4.26) of \cite{BBKS86} and Eq.\,(22) in \cite{D08}.

 Furthermore we assess the accuracy of the displacement $\v{\varPsi}_{\rm Z}$ determined from our Z-simulations by determining the relative difference $\Delta|\bm{\varPsi}|$ and the relative angle $\theta$ with respect to the $N$-body simulation, defined as
\begin{subequations}
 \label{diffZ}
\begin{align}
\Delta|\bm{\varPsi}| &\equiv \frac{ |\v{\varPsi}_{\rm Z}| } {|\v{\varPsi}|} -1 \,, \label{deltaPsi}\\
\theta \equiv \angle (\v{\varPsi},\v{\varPsi}_{\rm Z})&=\rm{Arccos}\left[\frac{\vert \bm{\varPsi}_{\rm Z} \cdot\bm{\varPsi}\vert}{|\bm{\varPsi}_{\rm Z}|\cdot |\bm{\varPsi}|}\right] \label{thetaPsi} \,.
\end{align}
\end{subequations}
A perfect agreement between the Zel'dovich prediction and the $N$-body simulation data would yield a $\delta_{\rm D}$ distributions with a zero mean for both the mismatch amplitude $\Delta|\bm{\varPsi}|$ and the misalignment angle $\theta$.

In the middle and right panel of Fig.\,\ref{figDEUS} we illustrate the difference between the Z-simulation and the $N$-body result by depicting the distributions of the mismatch amplitude $\Delta|\v{\varPsi}|$ \eqref{deltaPsi} and the misalignment angle $\theta$ \eqref{thetaPsi}. 
It shows again that the amplitudes determined from the Z-simulations are in overall good agreement with the exact $N$-body result for $R<4R_{\rm L}$. Choosing $R=4R_{\rm L}$ leads to a significant underestimate of the amplitude for the displacement field while choosing $R=R_{\rm L}/4$ is less biased but has a significantly larger variance than $R=R_{\rm L}$.
For the misalignment angle we observe once again that the Z-simulation with $R=R_{\rm L}$ provides the best and $R=4R_{\rm L}$ the worst match with the $N$-body result. 
Those trends are more pronounced for the large mass halos.
The distributions for $\theta$ show a systematic bias between the displacement of $N$-body and Z-simulation. 
This is simply because $\theta$ as defined in \eqref{thetaPsi} is constrained to be non-negative.  
We observe that for $R=R_{\rm L}$ the peak of the distribution is closer to zero compared the other cases. 
In summary for all the quantities we studied, the distributions of $\varPsi$, $\Delta|\bm{\varPsi}|$ and $\theta$ show that best agreement between Z-simulations and $N$-body simulations is achieved for $R=R_{\rm L}$, see also \cite{AchitouvBlake2015}.

Having established the Lagrangian radius of halos as optimal smoothing scale to predict proto-halo displacements using the Zel'dovich approximation, we will apply this know\-ledge to predict the redshift space halo correlation function in Section \ref{sec:GSM}. Therein, the Gaussian Streaming Model (GSM) for redshift space distortions is combined with the Post-Zel'dovich approximation through the Convolution Lagrangian Perturbation Theory (CLPT) and a smoothing of the initial power spectrum leading to the `truncated' CLPT (TCLPT).
We be implement the proto-halo identification through a local Lagrangian bias model based on an accurate halo mass function described in detail in App. \ref{app:halobias}.
We will benchmark the combined GSM+TCLPT predictions for the halo correlation functions in real and redshift space (see Sec.\,\ref{sec:results}) using measurements from the Horizon Run 2 halo catalog described in App.\,\ref{sec:HR2}.


\section{Gaussian Streaming model and its ingredients}
\label{sec:GSM}
In order to infer predictions for the halo correlation function in redshift space we use the Gaussian streaming model, originally derived in \cite{F94} and studied in \cite{RW11,WRW14} for linear and a more complicated Lagrangian bias models \cite{M08, Mat12}.
 We present a concise but self-contained derivation of the Gaussian streaming model (GSM) starting from a phase-space distribution function for single-streaming dark matter tracers. For details and a more general derivation of the streaming model, including possible multi-streaming, we refer to our accompanying paper \cite{UKH15}. We then shortly summarize how the ingredients of the streaming models can be inferred from CLPT and its truncated version TCLPT, which has been described in more detail in \cite{UKH15}.
 Our local Lagrangian model is implemented as in \cite{M08, WRW14} with the exception of the conditional mass function used. Instead of the Sheth-Tormen \cite{ST99} mass function we choose a more accurate model that we calibrate to the mass function and describe in App.\,\ref{app:halobias}.

\subsection{Gaussian Streaming Model}
In the following we will consider single-streaming tracers of dark matter, like proto-halos, whose phase-space distribution is given by $f_X(\v{r},\v{u},t)=\left(1+\delta_X(\v{r},t)\right)\delta_{\rm D}(\v{u}-a \v{v}_X(\v{r},t))$. We assume that the tracer density in real space is statistically homogeneous and isotropic such that the two-point correlation function depends only on the relative distance $r= |\v{r}_2 - \v{r}_1|$
\begin{equation}
\label{2ptcorrfctreal}
1+\xi_X(r,t) =  \Big\langle (1+\delta_X(\v{r}_1,t)) (1+\delta_X(\v{r}_2,t)) \Big\rangle \,.
\end{equation}

To map from real to redshift space we use the distant observer approximation, where the line of sight is assumed to be a fixed direction $\hvz$, without loss of generality chosen as the direction of the $z-$axis. Hence, the observed comoving distance in redshift space $\v{s} = \v{r}+ \sH^{-1} (\v{v}\cdot \hvz)\ \hvz $ is affected by the peculiar velocity $\v{v}\cdot \hvz =v_z$ of the tracer along the line of sight via the Doppler effect,
where $\mathcal{H} = a H = \dot{a}$ is the conformal Hubble constant. The observed position of the tracer perpendicular to the line of sight remains unaffected if we neglect gravitational lensing such that $\v{s}_\perp  =\v{r}_\perp$. In contrast, the redshift space coordinate parallel to the line of sight $\hvz$  depends on the peculiar velocity $v_z$ via $s_{||}=\v{s}\cdot \hvz =r_{||} + \sH^{-1} v_z$\,.
Since objects cannot disappear going from real space to redshift space (assuming that all objects remain observable) we have the following relation between the densities in real and redshift space
\begin{align}
(1+\delta_X(\v{s},t))\, \varvol{3}{s} &= (1+\delta_X(\v{r},t))\, \varvol{3}{r} \,.\label{realtoredshift}
\end{align}
In the distant observer approximation, the density fluctuation for single-streaming tracers in redshift space \eqref{realtoredshift} is given by
\begin{align}
1+\delta_X(\v{s},t) = &\int \vol{3}{r}\! (1+\delta_X(\v{r},t))\, \delta_{\rm D}\left(\v{s} -\v{r} -  \frac{v_z(\v{r},t)}{\sH} \hvz\right) \label{deltasr} \,.
 \end{align} 
 
Obviously, this mapping will introduce anisotropies in the redshift space two-point correlation function
\begin{equation}
\label{2ptcorrfctgeneral}
1+\xi_X(\v{s},t) =  \Big\langle (1+\delta_X(\v{s}_1)) (1+\delta_X(\v{s}_2)) \Big\rangle \,,
\end{equation}
where $\v{s}=\v{s}_2-\v{s}_1$. For the calculation we use cylindrical coordinates $\v{s} =  s_{\perp} [\cos (\phi)\hat{\v{x}} + \sin (\phi)\hat{\v{y}} ] + s_{||} \hvz $ because $\xi_X(\v{s},t)$ does not depend on the angle $\phi$.  By inserting \eqref{deltasr} in \eqref{2ptcorrfctgeneral} and re-expressing the delta functions in Fourier space and integrating over $\v{R}=\v{r}_1+\v{r}_2$ and one momentum variable the correlation function can be brought into the following form
 \begin{subequations} \label{xizspacegeneral}
\begin{align}
\label{xizspaceexprgeneral}
1+\xi_X(\v{s},t) &=\int \vol{3}{r}\!\!\! \int \frac{\vol{3}{k}}{(2\pi)^3} e^{i \v{k}\cdot (\v{r} - \v{s}) } Z\Big(\v{r}, \v{J} = k_z\, \hvz,t\Big)\,,\\
\notag Z(\v{r}, \v{J},t) &= \Bigg\langle [1+\delta_X(\v{r}_1,t)][1+\delta_X(\v{r}_2,t)]\\
&\quad \times \exp\left[i \frac{[\v{v}_X(\v{r}_2,t) - \v{v}_X(\v{r}_1,t)] \cdot \v{J}}{\sH}\right] \Bigg\rangle\label{defZss}\,.
\end{align}
\end{subequations}
When we Taylor expand $\ln Z(\v{J})$ around $\v{J}=0$ and keep only the terms up to second order we obtain the Gaussian streaming model (GSM)
\begin{align} 
1+\xi_X(s_{||},s_\perp,t)&= \int^{\infty}_{-\infty} \frac{ \vol{}{r_{||}}}{\sqrt{2 \pi} \sigma_{12}(r,r_{||},t)}  (1+\xi_X(r,t)) \notag\\ 
&\times\exp\left[-\frac{\left(s_{||}- r_{||} - v_{12}(r,t) r_{||}/r\right)^2}{2 \sigma_{12}^2(r,r_{||},t)}\right] \label{GSM} \,,
\end{align}
with the expansion coefficients that have been projected onto the line of sight,
\begin{subequations}
\label{GSMparam}
\begin{align}
1+ \xi_X(r,t) &:= Z\, |_{J=0} \,,\label{xirel}\\
\v{v}_{12}(\v{r},t) \cdot \hvz& := \frac{\frac{\partial Z}{(\partial i \v{J})}\big|_{J=0}\cdot \hvz}{(1+ \xi_X(r,t))}=: v_{12}(r,t) \frac{r_{||}}{r}\,,\label{v12rel}\\
\hvz^{T} \v{\sigma}^2_{12}(\v{r},t)  \hvz  & :=  \frac{\hvz^{T}\frac{\partial^2 Z}{(i\partial \v{J})^2}\big|_{J=0} \hvz}{(1+ \xi_X(r,t))} -v_{12}(r,t)^2\left(\frac{r_{||}}{r}\right)^2 \label{sigma12rel}\\
 &= \sigma_{||}^2(r,t) \left(\frac{r_{||}}{r} \right)^2 +\sigma_\perp^2(r,t)\left[1-\left(\frac{r_{||}}{r} \right)^2\right] \label{sigperpparsplit} \\
 &=: \sigma_{12}^2(r,r_{||})\,. \notag
\end{align}
\end{subequations}
In the GSM, all redshift space distortion induced clustering is encoded in the scale dependent mean and variance given by the pairwise velocity $v_{12}$ and its dispersion $\sigma_{12}^2$. 

\subsection{Streaming model ingredients from CLPT}
\label{sec:corrfuncandpairw}
The quantities $1+\xi_X(r)$, $v_{12}(r)$ and $ \sigma_{||/\perp}^2(r)$ \eqref{GSMparam} entering the Gaussian streaming model \eqref{GSM} will be calculated within CLPT using local Lagrangian bias \eqref{deltarq} similarly to \cite{WRW14}. We will compare predictions for the streaming model ingredients from ordinary Convolution Lagrangian perturbation theory (CLPT) with its truncated version (TCLPT) employing a smoothed input power spectrum which is straightforward to implement. This approach is motivated by \cite{Mat12, Chuetal14} and the observation that a smoothing improves the prediction of proto-halo displacements as demonstrated in Sec.\,\ref{sec:displace}. We will employ two different smoothing scales {\it (i)} the Lagrangian radius $R=R_{\rm L}(M)$ as suggested by our investiation in Sec.\,\ref{sec:displace} and which has been investigated in the context of perturbation theory in \cite{UKH15} and {\it (ii)} the phenomenological smoothing scale $R=1\Mpc$ which is similar to the smoothing scale of dark matter TZA simulations \cite{C93} and optimises the agreement of the large scale vorticity with the $N$-body measurements, see \cite{UK14}.

We will benchmark the CLPT predictions for the GSM ingredients, the real space correlation function  $\xi_X$ and the pairwise velocity statistics $v_{12}$, $\sigma_{\perp}^2$ and $\sigma_{||}^2$, to measurements within the publicly available Horizon Run 2 (HR2) simulation halo catalog \cite{KPetal09, KPetal11} described in App.\,\ref{sec:HR2}. 
Those results are then used for the comparison of the combined predictions of the GSM + CLPT to the HR2 measurements in Sec.\,\ref{sec:results}.
In the next subsections we introduce and calibrate the bias model, developed in \cite{CA2} and extended in \cite{ABPW}. 

\subsection{Halo bias model}
The local Lagrangian bias model is implemented by replacing in Eqs.\,\eqref{xizspacegeneral} the Eulerian tracer density $\delta_X(\v{r},t)$ by an expression in terms the initial (Lagrangian) density field $F[\delta_R(\v{q},t)]$ and the proto-halo displacement field $\v{\varPsi}_X(\v{q},t)$. These two building blocks will be reviewed in the following paragraphs. 
\subsubsection{Lagrangian halo density field and bias}
The first building block is the Lagrangian, or initial, halo density field 
\begin{align}
1+\delta_X(\v{q},t \rightarrow 0 |t_c) =  F[\delta_R(\v{q},t_c),...]\ \label{LagrangianDensityField} 
\end{align}
which is a functional of the linear density field extrapolated to the time $t_c$, at which the halos with a specific mass related to the smoothing scale $R$ will form. $F$ depends also on other time dependent functions like the critical threshold for spherical collapse and the variance of the linearly extrapolated density perturbation as well as other parameters related to the halo mass function. We will present more details of the proto-halo density field and its connection to the halo mass function in  App.\,\ref{app:halobias}.
The bias parameters $b_n$ arise from a local expansion of the Lagrangian halo density field in terms of 
\begin{align}
F[\delta_R(\v{q},t_c)]=1+\sum_{n=1}^\infty \frac{1}{n!} b_n(M,t_c) \delta_R^n(\v{q},t_c)\,. \label{biasexp}
\end{align}
In our case, since we specify $F$ through a conditional mass function \eqref{condf}, the $b_n$ are known and given by Eqs.\,\eqref{IxBias}. Therefore the bias coefficients that apply for halos within a finite range of masses \eqref{averbias} can be predicted without any free parameters.
They are however off by a few percent such that even on largest scales, the theoretical halo correlation function would not fit the measured one.
Therefore we will not use \eqref{averbias} and 
instead calibrate the bias model following the procedure of \cite{CRW13, WRW14} by treating the mass $M$ appearing in \eqref{IxBias} as a free parameter and find the optimal mass $M_{\rm opt}$ by fitting the TCLPT correlation function $\xi_X(r,M)$ with unspecified $M$ to the HR2 real space correlation function. 
Once this optimal mass $M_{\rm opt}$ is determined for each mass bin, the statistical averages $\langle \partial^n_{\delta_R}F\rangle$ that also enter TCLPT expressions for $v_{12}(r)$, $\sigma^2_{\perp}(r)$ and $\sigma^2_{||}(r)$ will be identified with $b_n(M_{\rm opt})$ as suggested in \cite{M08, CRW13, WRW14}. 
Thus  $v_{12}(r)$, $\sigma^2_{\perp}(r)$ and $\sigma^2_{||}(r)$ as well as the GSM redshift space correlation function $\xi_X(\v{s})$ will then be determined without any further fits. 
Although somewhat unsatisfying,  the model is still highly predictive since there is only one parameter fitted per mass bin.

\subsubsection{Eulerian halo density field}
\label{SmoothingscaleAsPeakscale}
The other building block of the bias model is the halo displacement field $\v{\varPsi}_X$
that is the integral of the halo velocity field $\v{v}_X = a \dot{\v{\varPsi}}_X$. It connects the initial proto-halo center $\v{q}$ to its final position $\v{r}= \v{q}+ \v{\varPsi}_X $.
The proto-halo fluctuation field $F[\delta_R(\v{q},t_c)]$ encodes only the initial clustering for halos of size $R$ that collapse at time $t_c$, while the subsequent gravitational evolution $\delta_X(\v{r},t |t_c) $ for $t<t_c$ is described by the proto-halo displacement field $\v{\varPsi}_X(\v{q},t)$.

For $t < t_c$ the interpretation of $\delta_X(\v{r},t|t_c)$ is the conserved but non-linearly evolving proto-halo density field of halos with mass $R$ that form halos at time $t_c$. For $t=t_c$ we use the shorthand notation $\delta_X(\v{r},t) = \delta_X(\v{r},t|t)$ for the halo field at time $t$, such that 
\begin{align}
1+\delta_X(\v{r},t)= \int \vol{3}{q} F[\delta_R(\v{q},t)]\, \delta_{\rm D}\left(\v{r} -\v{q} - \v{\varPsi}_X(\v{q},t)\right)\,.\label{deltarq} 
\end{align}

The interpretation of this formula is the same as the connection between real and redshift space halo density fields, and we can write analogously to \eqref{realtoredshift} 
\begin{align}
(1+\delta_X(\v{r},t|t_c))\, \varvol{3}{r} &= F[\delta_R(\v{q},t_c),t_c]\,\varvol{3}{q} \label{locLagBias}
\end{align}
making manifest the proto-halo conservation for $t<t_c$.

Following \cite{CLMP98} we can establish the connection between Eulerian and Lagrangian bias by assuming zero velocity bias $\v{v}_X=\v{v}$, such that $\v{\varPsi}_X=\v{\varPsi}$, where $\v{\varPsi}$ can be calculated using the LPT kernels of a dust fluid, see for instance \cite{RB12}.

Using this in Eq.\,\eqref{deltarq} and setting  $t= t_c$ gives
\begin{align}
1+\delta_X(\v{r},t)&= \int \vol{3}{q} F[\delta_R(\v{q},t)]\, \delta_{\rm D}\left(\v{r} -\v{q} - \v{\varPsi}(\v{q},t)\right) \,.\label{deltarq2} \\
&=F[\delta_R(\v{q},t),t]
\big(1+\delta(\v{r},t)\big)\label{nonlinnonlocbias}
\end{align}
 with initial conditions $\delta(\v{r},t\!\!\rightarrow\!\!0)=0$ and $\v{r}(\v{q},t\!\!\rightarrow\!\!0)=\v{q}$ and Eq.\,\eqref{LagrangianDensityField}. 
 The second equality, Eq.\,\eqref{nonlinnonlocbias}, follows from the continuity equation for CDM \cite{CLMP98}. 

\subsubsection{The effect of the smoothing scale}
If $R \simeq R_{\rm L}$ is kept consistently within the whole perturbation theory calculation and is furthermore applied also to $\v{\varPsi}$, as suggested by our findings in Sec.\,\ref{sec:DEUSS}, then from Eq.\,\eqref{biasexp} and \eqref{nonlinnonlocbias} the halo density perturbation $\delta_X$ assumes its standard linear SPT form
\begin{align}
\delta^{\rm lin}_{X}(\v{r})& = [1+ b_1(M)]\, \delta_{R}( \v{r})\,.
\end{align}
In Fourier space $\delta_{R}(\v{k}) = \tilde W(k R )\delta_{\rm lin}(\v{k})$ 
such that the linear halo power spectrum becomes
\begin{align}
P^{\rm lin}_{X}(\v{k}) &= [1+b_1(M)]^2 \tilde W(k R )^2 P_{\rm lin}(\v{k})\,. \label{Plin}
\end{align}

It is clear that the shape of $\xi^{\rm lin}_{X}(r)$, the Fourier transform of $P^{\rm lin}_{X}$,  appears smeared out on scales below $R$ compared to the unsmoothed linear correlation function  in which $\tilde W =1$.  
 But maybe somewhat surprisingly $\xi^{\rm lin}_{X}(r)$ is affected by $W$ even for $r\gg R$. In particular the BAO peak is smeared out, as can be clearly seen comparing $\xi^{\rm lin}_{X}(r)$ for $R=10\Mpc$ (corresponding to $R_{\rm L} (\lgM \!\!= \!\!14.5)$), the thick dashed line in Fig.\,\ref{fig:z0realAllmass}  to $R=0$, the thin black line.
 In fact the at scale $R=10\Mpc$ smoothed linear theory fits the measured halo correlation function (the thin colored lines) significantly worse than the unsmoothed linear one which explains why the smoothing scale is usually dropped from the random field $\delta_R \rightarrow \delta_{\rm lin}$ and only kept as implicitly as mass-dependence of the bias coefficients $b_n(M)$, although this  mass dependence arises precisely through the same window function that is commonly dropped from the random field!
 The dotted lines shows again $\xi^{\rm lin}_{X}(r)$ but smoothed at the nonlinear scale 
   \begin{equation} \label{RNL}
  R_{\rm NL}(z) \equiv \sqrt {\langle \v{\varPsi}^2_{\rm Z}\rangle/3}\,,
 \end{equation} 
  which provides the best fit using linear theory \eqref{Plin} on large scales.
  The choice of the smoothing scale for the real space correlation function obtained from CLPT will be discussed in the following.

\subsection{Real space correlation function  $\xi_X(r)$}
\label{sec:realcorrfct}
\subsubsection{Choosing the smoothing scale}

\begin{figure*}[t!]
\raisebox{0.2cm}{\includegraphics[width=0.18\textwidth]{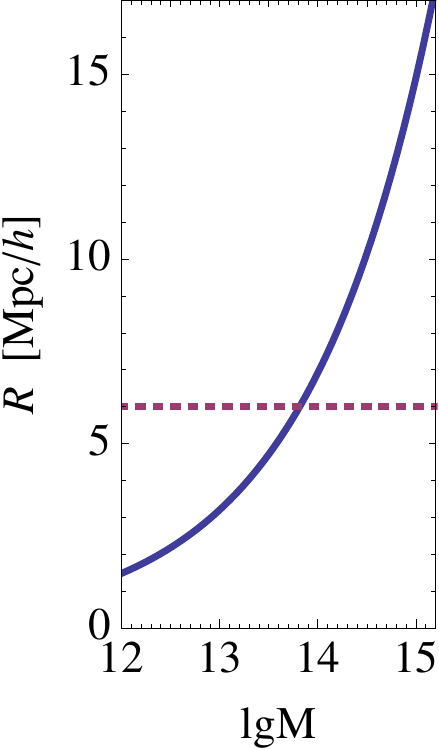}}\qquad \qquad
\includegraphics[width=0.74\textwidth]{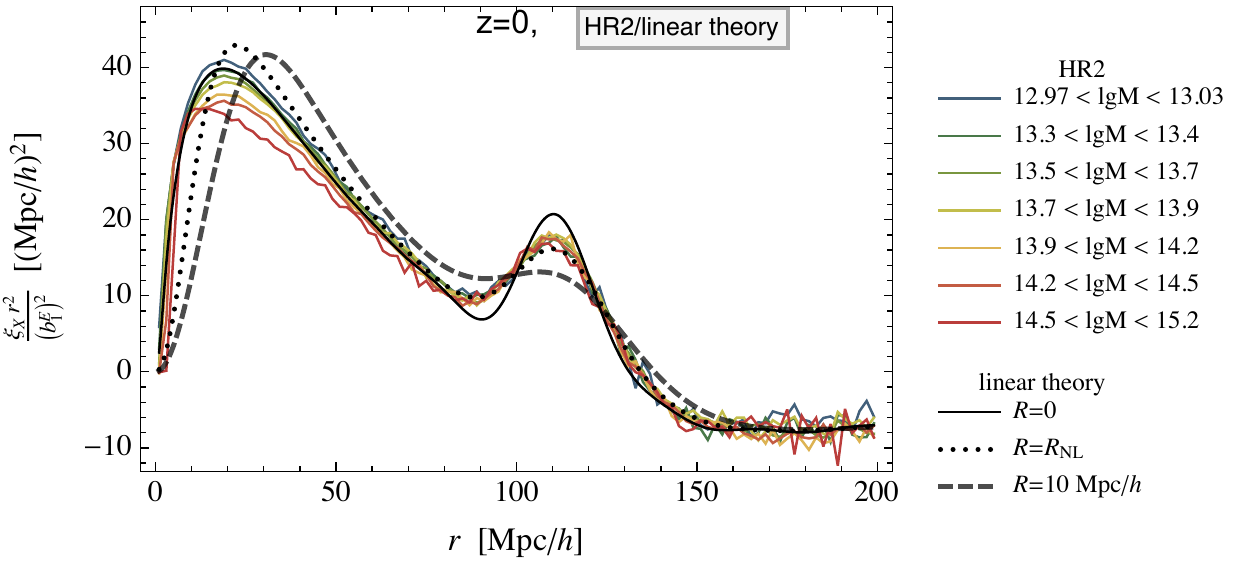}
\caption{{\it left}: Lagrange radius ({\it full}) as a function of lgM and $R_{\rm NL}=6\,\mathrm{Mpc}/h$ ({\it dashed}). {\it right} Comparison between real space halo correlation function times $r^2$ for 7 mass bins measured in HR2 ({\it thin lines}). All curves are rescaled by the best fitting $b_1(M_{\rm opt})$ for the nonlinear theory shown in Fig.\,\ref{fig:allreal}. Also shown are the linear correlation function ({\it black, thick}), the linear correlation function convolved with $W(k\, 6\,\mathrm{Mpc}/h)^2$ ({\it black, thick dashed}) and $W(k\, 10\,\mathrm{Mpc}/h)^2$ ({\it black, thin}). }
\label{fig:z0realAllmass}
\end{figure*}

Although, as we have seen in Sec.\,\ref{sec:displace}, a smoothing at the peak scale $R_{\rm L}(M)$ is physically meaningful \cite{BM96}, the CLPT correlation function $\xi_X(r)$ shows significant deformations if the Lagrangian smoothing scale $R_{\rm L}(M)$ is used in the window function, similar to the linear correlation function $\xi^{\rm lin}_{X}(r)$ shown in Fig.\,\ref{fig:z0realAllmass}. 
For masses  $\lgM>13.7$, the Lagranigan scale $R_{\rm L}(M)$ will be larger than the nonlinear scale
$R_{\rm NL}$
related to the average displacement of a particle. 

The problem this causes becomes manifest in an approximation to CLPT, called integrated perturbation theory (iPT) \cite{M08},  where it turns out that a Gaussian window function with width $R_{\rm NL}$ multiplies the whole expression for the nonlinear halo correlation function:
\begin{equation}\label{iptpower}
P_{\mathrm{iPT},X} = W(k R_{\rm NL})^2 \left\{(1+b_1(M))^2W(k R)^2P_{\rm lin}(k)+ ...\right\}\,,
\end{equation}
with $R=R_{\rm L}$ and where we omitted other nonlinear terms and restored the window function in front of $P_{\rm lin}$ \cite{M08}.  At $z=0$ the  average displacement   $R_{\rm NL}\simeq 6\,\mathrm{Mpc}/h$  is responsible for the nonlinear smoothing of the BAO peak in the Zel'dovich approximation. One can see from Eq.\,\eqref{iptpower} that any mass-related smoothing $R_{\rm L}(M)$ adds to the already existing $R_{\rm NL}$-smoothing, leading to an effective smoothing $\sqrt{R_{\rm{NL}}^2+R_{\rm L}(M)^2}$ in case of Gaussian filters.\footnote{Note that also $R_{\rm NL}$ slightly depends on the scale $R$ and therefore on mass if $R=R_{\rm L}(M)$. But this mass dependence cannot compensate the second term.} 
In contrast to our expectations, no extra mass-dependent smoothing is observed in the HR2 simulation; all mass bins show a similarly smoothed-out BAO peak after rescaling with the best fitting linear bias $b^{\rm E}_1 = 1+b_1(M_{\rm opt})$, see Fig.\,\ref{fig:z0realAllmass}.\footnote{Note that $M_{\rm opt}$ was obtained by fitting the TCLPT model $\xi_X$ to the real space correlation function $\hat \xi_X$ of the simulation, see the next paragraph, and not with the aim to overlap all BAO peaks for the different mass bins.} 

\begin{figure}[t!]
\includegraphics[width=0.48\textwidth]{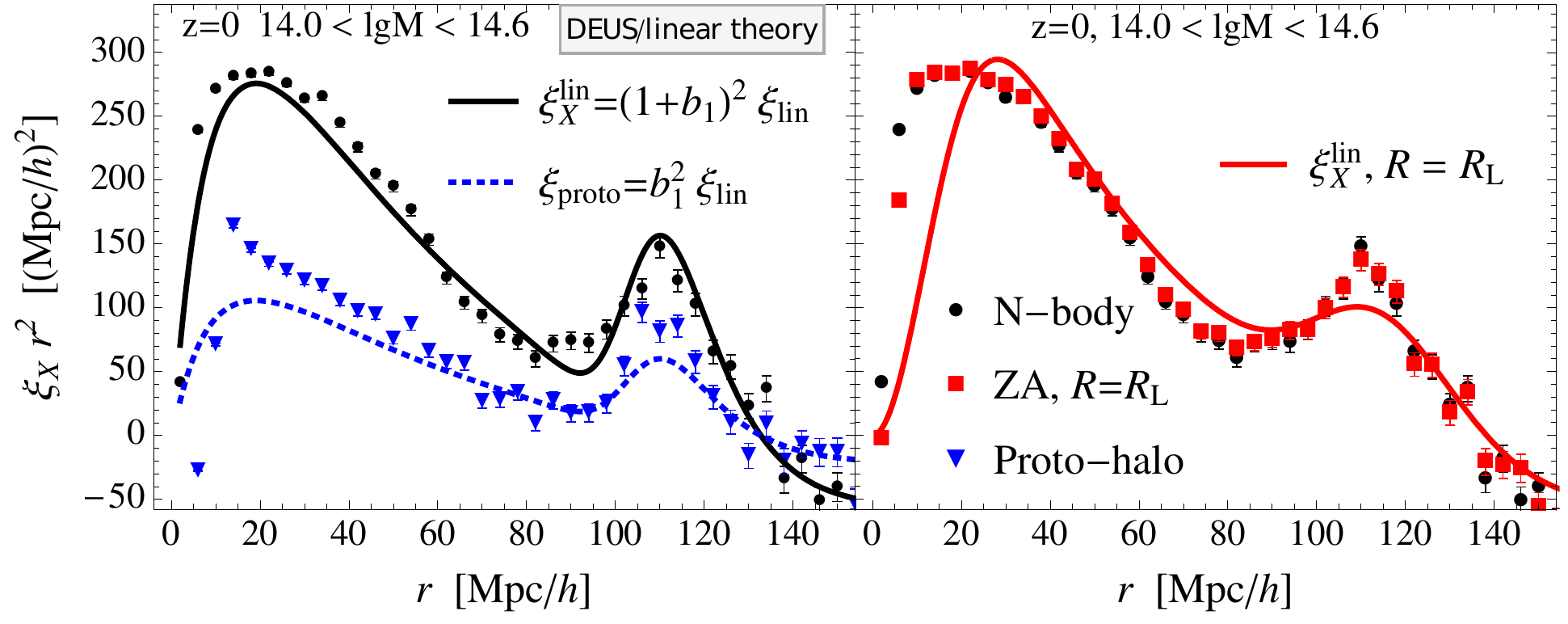}
\caption{Real space halo correlation function times $r^2$ from the DEUS simulation for the mass bin denoted in Fig.\,\ref{figDEUS} $R_{\rm L}=8.7\Mpc$.  Data points show the correlation function measured from DEUS simulation data with Poissonian estimate for the error bars.
{\it left panel} Comparison of correlation function for halos ({\it black dots}) and proto-halos ({\it blue triangles}) within the DEUS simulation together with the prediction from linear theory for halos ({\it black line}) and proto-halos ({\it blue dotted line}).
{\it right panel} 
Comparison between the DEUS $N$-body simulation ({\it black dots}), Zel'dovich simulations ({\it colored symbols}) and linear theory ({\it colored lines}) for $R=R_{\rm L}$.
}
\label{fig:corrDEUS}
\end{figure}

In Fig.\,\ref{fig:corrDEUS} we show the correlation functions of the DEUS simulation studied in Sec.\,\ref{sec:DEUSS}. In the right panel we observe that for the Zel'dovich approximation applied to actual proto-halos (colored data points) nearly no damping of the BAO peak occurs for a smoothing at $R=R_{\rm L}$ (red squares) despite the large halo mass.
This is in stark contrast to the linear theory prediction, where a smoothing at $R=R_{\rm L}$ has a much stronger impact on the BAO peak, see the full red curve.
 Remember that due to Eq.\,\eqref{iptpower} the corresponding curve in the analytical Lagrangian formalisms like TZA and TCLPT would look even worse due to the presence of the extra smoothing scale $R_{\rm NL}$

So the situation is follows:  theoretical considerations of the proto-halo density field $F$ and displacement field $\v{\varPsi}_X$ as well as numerical experiments, see  Sec.\,\ref{sec:DEUSS} and \cite{AchitouvBlake2015},  suggest that a smoothing at the peak scale should improve a modelling based on the ZA, while on the other hand a direct comparison of the theoretical correlation functions with those measured in $N$-body simulations show that a smoothing at the peak scale is worse than not smoothing at all.

A pragmatic way to avoid this problem at the BAO scale is the widespread negligence of the window function $W(k R_{\rm L})$ on the random field. 
For instance, in \cite{M08} the problem is mentioned but it is argued that the second window function in Eq.\,\eqref{iptpower} can be set to 1 (or $R_{\rm L}=0$), at the end of the calculation if one is interested only in scales much larger than $R_{\rm L}$. 
Although this argument sounds reasonable, in reality the result for $\xi_X$ is changed dramatically if the window function is kept, even on scales seemingly large compared to $R_{\rm L}$; the BAO peak is smeared out much stronger than observed in simulations. 
This happens equally in linear perturbation theory, iPT and CLPT once the window function with the halo scale coming from the local Lagrangian bias is not discarded.\footnote{We compared the two cases $R=0$ and $R=R_{\rm L}(M)$ within TCLPT previously, see  Figs.\,4 and 5 of \cite{UKH15}.}
This suggests that it is the bias model that causes the problem.

The discrepancy between analytical and numerical implementations of ZA suggests that the effect of considering peaks instead of random particles, referred to as peak bias, has to be taken into account for the analytical models based on the ZA.
Our analytical bias model does not take into account that the displacements of proto-halos $\v{\varPsi}_X$ is not the same as the displacement $\v{\varPsi}$ of average points, even though our bias model based on a conditional mass function and excursion sets takes into account some aspects of specialness of proto-halo positions within $F$, but not all of them.
In the left panel of Fig.\,\ref{fig:corrDEUS} we show the initial correlation function of proto-halos and the prediction our linearised local bias model evaluated at the mean mass of the  mass bin without any fitting. 
It is apparent that our local $F$ predicts a proto-halo correlation function $\xi_{\rm proto}$ that seems less sharp, even with $R=0$,  than the one reconstructed from the DEUS initial conditions (blue triangles).
This extra sharpness of $\xi_{\rm proto}$ arises in the peak bias model \cite{D08} despite having a smoothing $R\simeq R_{\rm L}$.
 So there are two parts missing from our bias model: one in the proto-halo distribution $F$, leading to the enhanced sharpness of the BAO peak for the proto-halo correlation function, and a modification of displacement field  $\v{\varPsi}_X \neq \v{\varPsi}$  (velocity bias) \cite{BDS14}. 

 We are thus convinced that missing the peak nature of proto-halos is responsible for some of our problems. In particular we hope that one can maintain the sharpness of the BAO peak and a smoothing at the physical proto-halo scale $R\simeq R_{\rm L}$ for analytical calculations based on the ZA  within peak bias approach  \cite{D08}.

Include the peak property of proto-halos can be done as described in \cite{D08,BDS14} or by a more agnostic effective theory approach \cite{MirbabayiSchmidtZaldarriaga2015}. 
In practice this means that one needs to include considerably more terms in the bias, such that $F$ and $\v{\varPsi}_X$ depend now on $\delta_R,\partial_i\delta_R,\partial_i\partial_j\delta_R$ and the tidal gravitational field $\partial_i\partial_j \Phi$ which then demands computing a whole new set of correlators within CLPT. 
This goes beyond the scope of this paper and we leave it for future investigations.

We will in the following adopt a pragmatic approach that relies on the observation that peak bias typically leads to a re-sharpening of the BAO feature that is reversing the smoothing caused by the physical peak scale \cite{D08,BDS14}. 
Hence, the sharpeness of the BAO peak that arises when using a peak-bias model  together with  Lagrangian scale smoothing a can be mimicked by using instead a local Lagrangian bias and a smaller smoothing (somwhat smaller than $R_{\rm NL}$).
 We find that the best correspondence between the nonlinear halo density correlation $\xi_X$ and simulations, see Fig.\,\ref{fig:allreal}, is obtained by choosing the smoothing scale to be $R\simeq 1\Mpc$ for all masses which slightly improves over $R=0$ which is the most common choice in perturbation theory.
  For $R=0,1,2,3 \Mpc$ differences in the correlation function are visible for small $r$, see Fig.\,\ref{fig:smoothingDensitysmall}, while the smearing of the BAO peak is not strongly affected by those values of $R$. 
  For the HR2 cosmological parameter, see App.\,\ref{sec:HR2}, and at $z=0$ we find that $R\simeq1 \Mpc$ gives largest $\xi_X$ for small $r$ and the best agreement with the HR2 simulation.
This sharpening effect on the small scales of the density field through smoothing of the input power spectrum was originally observed in the ``truncated Zel'dovich approximation'' (TZA) \cite{C93,M94} and investigated in detail in \cite{Hamana1998}.

\subsubsection{Fitting the bias model using $\xi_X$ and $\hat \xi_X$}

  \begin{figure*}
\includegraphics[width=0.32\textwidth]{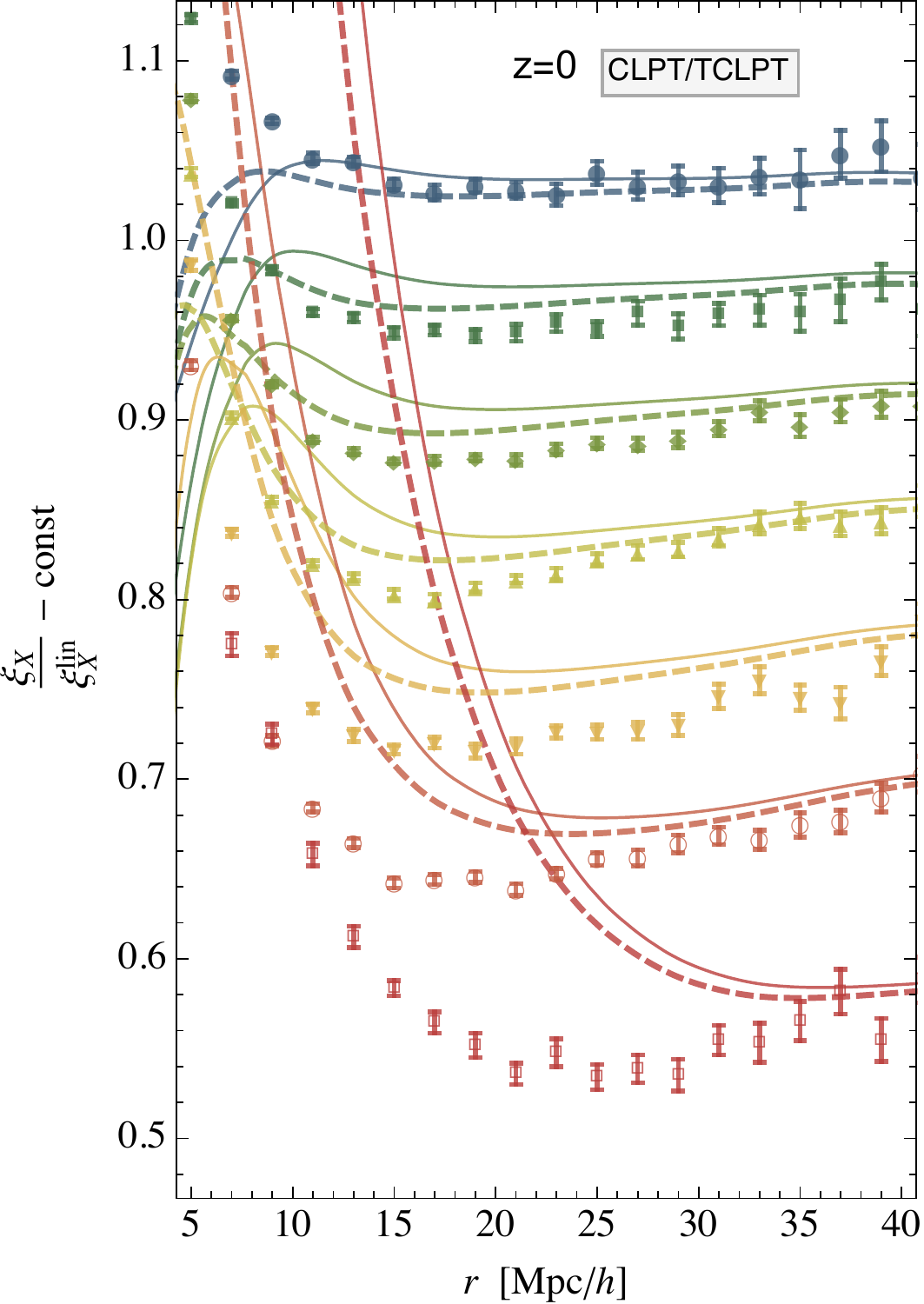}
\hspace{-0.1cm}
\includegraphics[width=0.32\textwidth]{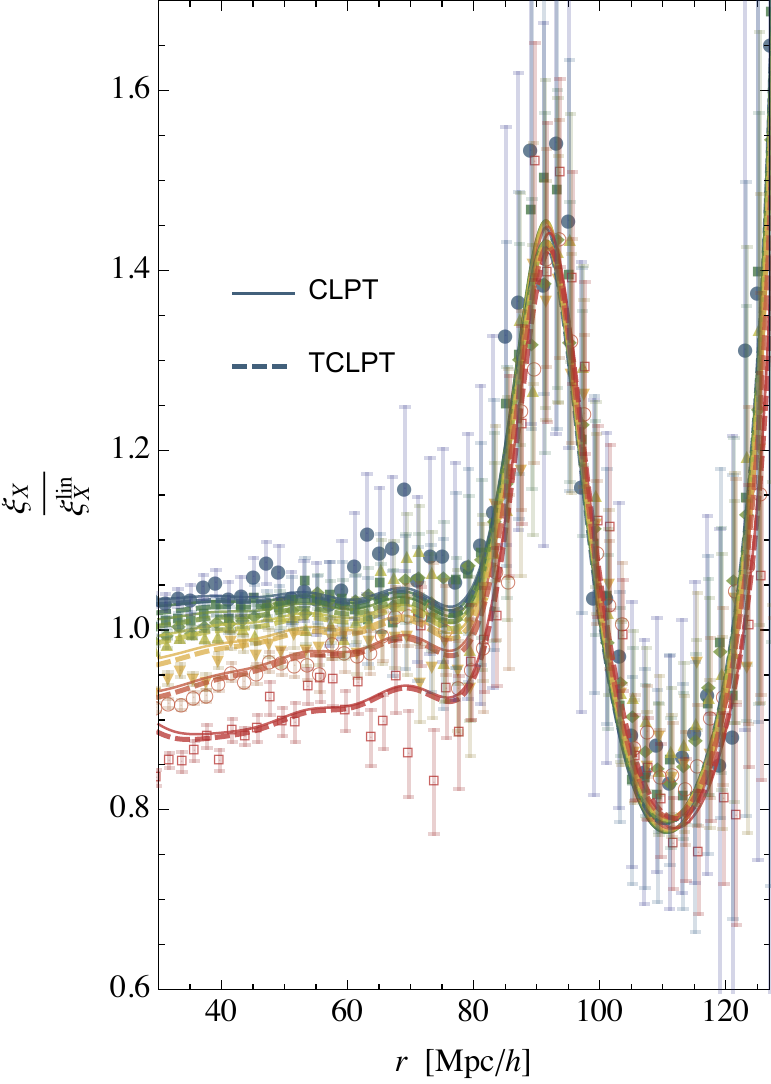}
\hspace{0.3cm}
\includegraphics[width=0.32\textwidth]{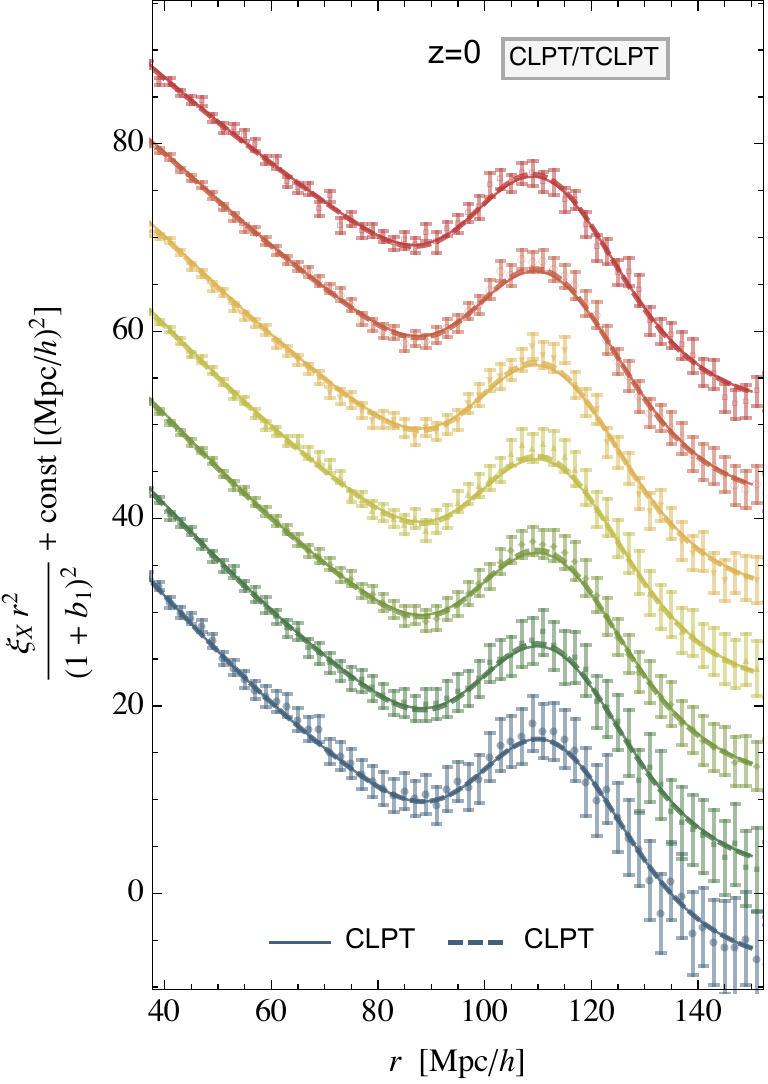}
\caption{Comparison between real-space halo correlation function $\xi_X$ obtained from TCLPT with smoothing scale $R=1\Mpc$ ({\it thick dashed} lines), CLPT ({\it thin} lines) and as measured in HR2 (data points).  For each curve, $M_{\rm opt}$ was fitted by minimizing the $\chi^2$ \eqref{chisq} between $r=40\Mpc$ to $r=200\Mpc$. Theory curves and data points have been divided by the linear halo correlation function $\xi^{\rm lin}_{X} $ in the left and middle panel, see Table \ref{biasfittab}. In the right panel we show $\xi_X r^2$ normalized by the respective linear Eulerian bias and shifted by mass bin dependent constant $10 i (\Mpc)^2$, $i=0,...,6$ for better visibility. In the left we also shifted the result by mass-bin dependent constant $0.05 i$ for better visibility. In both cases $i=0$ corresponds to the lowest mass, and $i=6$ to the highest.}
\label{fig:allreal}
\end{figure*}


In this paragraph we fix our TCLPT model for the real space correlation $\xi(r)$ to $R=1\,\Mpc$ and fit $b_1(M)$ and $b_2(M)$ for each mass bin to the HR2 correlation function $\hat \xi_X$ using only scales $40 \Mpc < r < 200 \Mpc$. The best fitting values $M_{\rm opt}$ will then determine
\begin{equation}
\langle F'\rangle \equiv b_1(M_{\rm opt})\quad\mathrm{and}\quad
\langle F''\rangle \equiv b_2(M_{\rm opt})
\end{equation}
appearing as parameters in TCLPT \cite{CRW13}. As CLPT and TCLPT give virtually identical results on those large scales, the same bias parameter apply to CLPT.
The results for the best-fitting masses as well as the corresponding bias factors are summarized in Table\,\ref{biasfittab} together with the average masses of the bins and the corresponding average bias.
We repeat the fit for ZA and TZA, which prefer slightly different values for the $M_{\rm opt}$ and give a slightly worse fit compared to (T)CLPT, as can be seen when comparing the $\chi^2$ in Table\,\ref{biasfittab}.
We refer the reader to App.\,\ref{app:fitbias} for details on the fitting procedure.
The bias parameters are kept fixed for the prediction of the velocity statistics in Section \ref{sec:velstat} and therefore also for the evaluation of the redshift-space halo correlation function in Sec.\,\ref{sec:results}.

\begin{figure}[t!]
\includegraphics[width=0.47\textwidth]{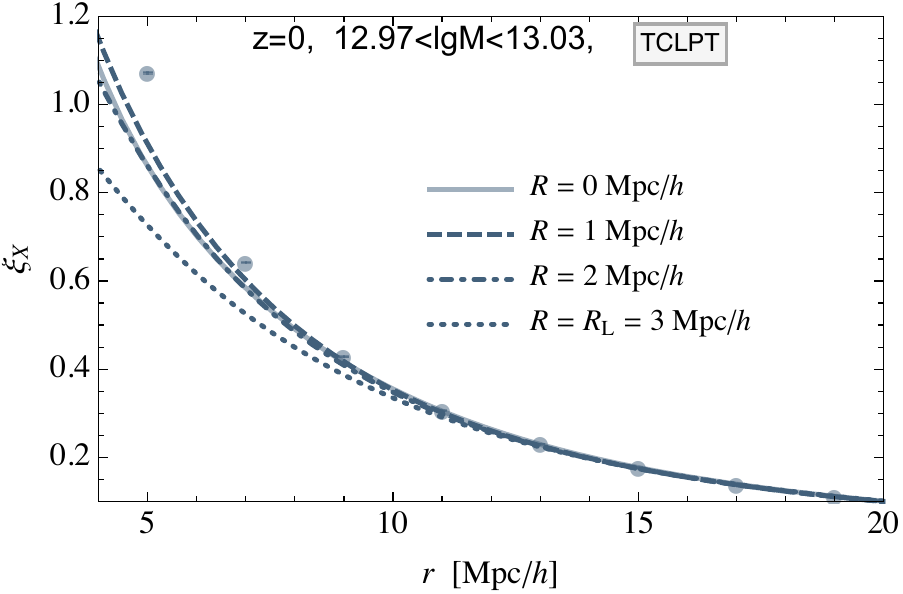}
\caption{The halo correlation for $\overline{\lgM} = 12.97$ at small scales. It is clearly visible that that a smoothing of the initial conditions as done in TCLPT with $R= 1\Mpc$ {\it (dashed)} increases the power on small scales and improves the agreement to HR2 (data points) compared to CLPT {\it (solid)} and TCLPT with $R= 2\Mpc$ {\it (dot-dashed} and $R= 3 \Mpc \simeq R_{\rm L}(M)$ {\it (dotted)}. }
\label{fig:smoothingDensitysmall}
\end{figure}

  \begin{table}[t!]
\begin{tabular}{| l | l | l | l | l | l | l | l |}
\hline
$\overline \lgM$               & 13.00 & 13.35 & 13.59 & 13.79 & 13.99 &14.25  & 14.67  \\\hline
$R_{\rm L}(M)\,[\Mpc]$ & 3.21 & 4.20 & 5.05 & 5.89 & 6.86 & 8.38 & 11.57 \\\hline
$\bar b_1$           & 0.02 & 0.25 & 0.47 & 0.69 &  0.98 & 1.49 & 2.71   \\
$\bar b_2$     & -0.76 & -0.83 & -0.80 & -0.68 &  -0.36 & 0.64 & 5.37    \\\hline
\hline
\multicolumn{8}{|c|}{CLPT/TCLPT} \\ \hline
\hline
\hline
$\lgM_{\rm opt}$  & 12.93 & 13.35 & 13.61 & 13.85 &  14.07 & 14.34 & 14.79    \\ \hline
$R_{\rm L}(M_{\rm opt})\,[\Mpc]$ & 3.04 & 4.19 & 5.15 & 6.14 & 7.28 & 9.01 & 12.73\\\hline
$b_1(\lgM_{\rm opt})$            & -0.01 & 0.26 & 0.51 & 0.77 &  1.11 & 1.68 & 3.17   \\
$b_2(\lgM_{\rm opt})$     & -0.74 & -0.83 & -0.79 & -0.61 &  -0.18 & 1.06 & 7.44    \\
\hline
$\chi^2$  & 27.4 & 9.5 & 19.7 & 24.9 &  13.9 & 19.7 & 23.3    \\\hline
\hline
\multicolumn{8}{|c|}{ZA/TZA} \\ \hline
\hline
\hline
$\lgM_{\rm opt}$  & 12.92& 13.34& 13.6& 13.83& 14.05& 14.32& 14.77    \\ \hline
$R_{\rm L}(M_{\rm opt})\,[\Mpc]$ & 3.01& 4.16& 5.1& 6.07& 7.21& 8.87& 12.54\\\hline
$b_1(\lgM_{\rm opt})$            & -0.01& 0.25& 0.49& 0.75& 1.08& 1.64& 3.09   \\
$b_2(\lgM_{\rm opt})$     & -0.74& -0.83& -0.8& -0.63& -0.22& 0.96& 6.94    \\
\hline
$\chi^2$  & 30.7& 10.5& 27.4& 39.9& 25.8& 38.8& 28.1    \\
\hline
\end{tabular}
\caption{Best fit values for the mass $\lgM_{\rm opt}$ of the given mass bin for $z=0$, denoted here by the average mass $\overline \lgM$ \eqref{meanmass} and the corresponding Lagrangian radius $R_{\rm L}(M)$. Also given are the bias parameters \eqref{IxBias} evaluated at $\lgM_{\rm opt}$ as well as the average bias \eqref{averbias}. There is good agreement between $\overline \lgM$ and $\lgM_{\rm opt}$ as well as the resulting best fitting $b_{1,2}(\lgM_{\rm opt})$ and model predictions $\bar b_{1,2}$ for small to intermediate masses which gets worse for the largest masses.}
\label{biasfittab}
\end{table}

In Fig.\,\ref{fig:allreal} we show the measured HR2 correlation functions as well as the model with best fitting $M=M_{\rm opt}$. 
The left and middle panel show the halo correlation $\xi_X$ normalized by the linear theory $\xi^{\rm lin}_{X} = \xi_{\rm lin} (1+b_1(M_{\rm opt}))^2$ \eqref{Plin} for all mass bins (see the legend on the right of Fig.\,\ref{fig:z0realAllmass}). The data points show the  $\hat \xi_X$ estimated using the HR2 halo catalogue using \eqref{correst} with error bars \eqref{correrror}. 
The full lines show the model predictions with the best-fitting $M_{\rm opt}$. The middle panel nicely shows that both CLPT and TCLPT fit $\hat \xi_X$ on large scales and are virtually identical. It also becomes apparent that all scales smaller than $90\Mpc$ are affected by the nonlinear halo bias and its mass dependence is clearly visible. 
In more detail we observe at scales $40\Mpc< r<90\Mpc$ that the resulting Eulerian bias is non-local as it suppresses power even on very large scales, with highest masses having the smallest power and the biggest deviation with respect to linear theory (see the middle panel of Fig.\,\ref{fig:allreal}).
 In the right panel we normalize all correlation functions by the linear local Eulerian bias $b^{\rm E}_1 = 1+ b_1(M_{\rm opt})$, multiply by $r^2$ and shift the result  for the various masses for better visibility.
Otherwise all masses would overlap at scales larger than $100 \Mpc$, see Fig.\,\ref{fig:z0realAllmass}  and the middle panel of Fig.\,\ref{fig:allreal}. Regarding the BAO peak we observe that  ({\it i}) BAO peak height and shape are not affected by non-local and non-linear biasing, and ({\it ii}) the nonlinear theory matches the shape of the BAO peak for all masses.  
Turning to the smallest scales, it becomes apparent from the left panel of Fig.\,\ref{fig:allreal} that the nonlinear model has problems on small scales $r<40\Mpc$ in particular for the largest mass bins, corresponding to galaxy clusters (shown in red and orange). This problem persists regardless of the chosen model, TCLPT with $R=1\Mpc$ or CLPT, and regardless of the chosen conditional mass function. 
In  Appendix \ref{app:fitbias} we discuss the possibility to fit for both $b_1$ and $b_2$ independently which improves the agreement on small scales, see Fig.\,\ref{fig:fitfail}.
It is not obvious that giving up the connection between $b_1$ and $b_2$ is physically reasonable, so we will not make use of this fit in the following and either limit our analysis to smaller halos or only consider large enough scales for the largest two mass bins.
We also point out that the bias model is well behaved for small $r$ if the Lagrangian scale $R=R_{\rm L}(M)$ is used, see Fig.\,\ref{fig:velstatcomp} for the velocity statistics and Fig.\,5 of \cite{UKH15} for $\xi_X$. 
We therefore expect this problem to disappear when using peak bias, where we are allowed to consistently use $R=R_{\rm L}(M)$.

Our final observation regarding the small $r$ behaviour of $\xi_X$ concerns the performance of CLPT versus TCLPT. 
While on large scales TCLPT with $R=1\Mpc$ and CLPT (TCLPT with $R=0$) are virtually identical, on small scales there are differences. 
Generically TCLPT with $R=1\Mpc$ gives the highest power on the smallest scales and  a better fit for $r< 40 \Mpc$ compared to CLPT, see also Fig.\,\ref{fig:smoothingDensitysmall} where more smoothing scales are compared for the single mass bin $12.97 < \lgM < 13.03$. 
$R=0$ and $R=2\Mpc$ have approximately the same shape, while $R=3\Mpc \simeq R_{\rm L}(M)$ is clearly less power and is farthest away from the simulation.

\subsection{ Real space velocity statistics: $v_{12}(r)$, $\sigma^2_\perp(r)$ and $\sigma^2_{||}(r)$}
\label{sec:velstat}
An important ingredient of the Gaussian streaming model \eqref{GSM} is the statistics of the mass weighted pairwise velocity $\v{v}_2(\v{r}_2) - \v{v}_{1}(\v{r}_1)$. Its first two cumulants are given by the mean pairwise velocity  $\v{v}_{12}(\v{r})  = v_{12}(r) \hat{\v{r}}$, where $\hat{\v{r}}$ is the normalised halo separation vector $\v{r}=\v{r}_2 -\v{r}_1$  and the pairwise velocity dispersion $\v{\sigma}_{12}^2(\v{r}) = \sigma_{||}^2(r) \hat{\v{r}} \hat{\v{r}} +  \sigma_{\perp}^2(r) (
\v{1} -\hat{\v{r}} \hat{\v{r}})$ which can be calculated from theory according to Eq.\,\eqref{sigperpparsplit} with the help of CLPT \cite{WRW14} and measured in a halo catalogue according to Eq.\,\eqref{HR2v12}.

\subsubsection{Comparison of CLPT and TCLPT}

\begin{figure}[t!]
\includegraphics[width=0.5\textwidth]{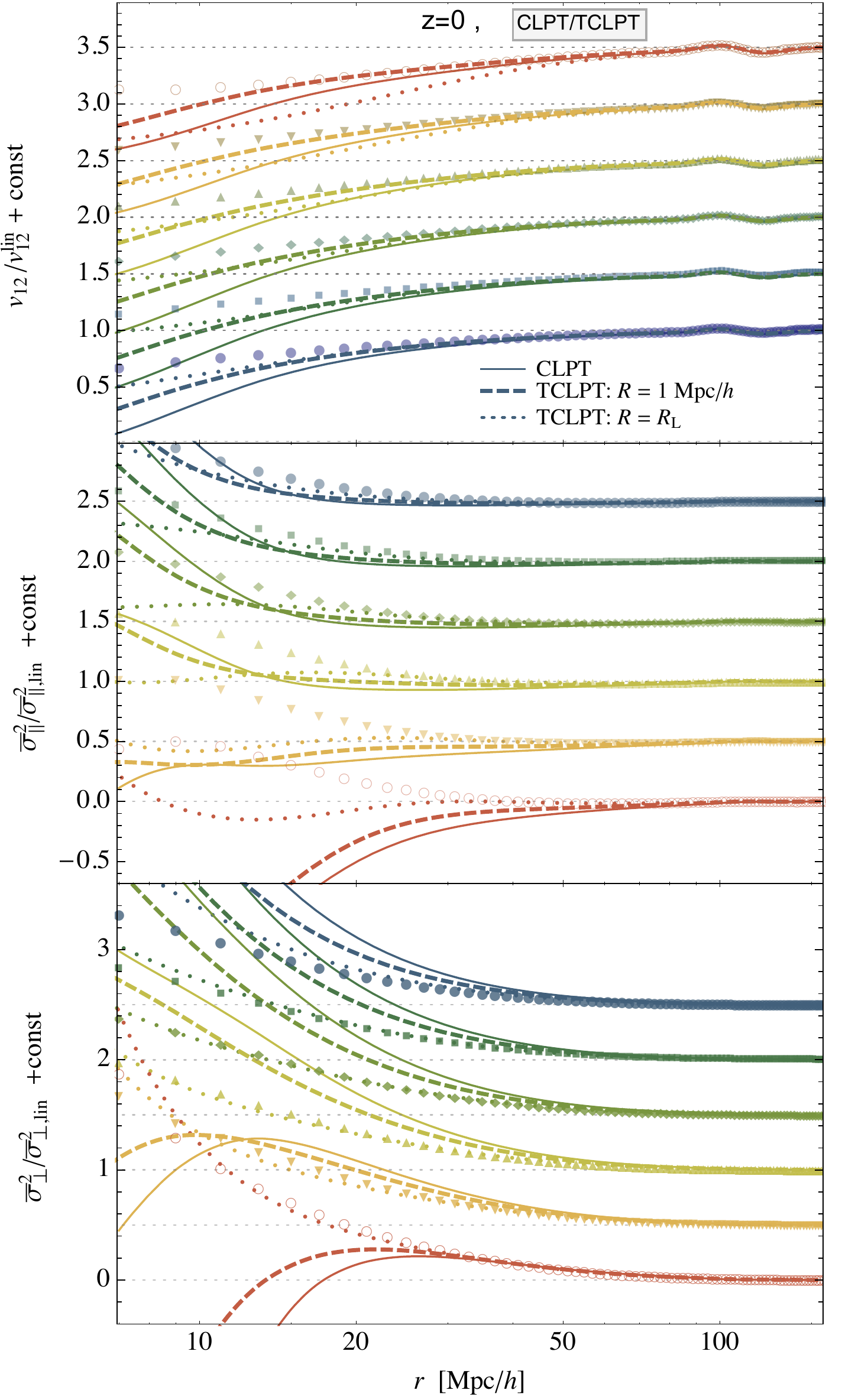}
\caption{Velocity statistics predicted from perturbation theory together with the HR2 measurements ({\it data points}) comparing CLPT ({\it thin solid}), TCLPT with $R=1\Mpc$ ({\it thick dashed}) and the Lagrangian scale $R=R_{\rm L}(M)$ ({\it thick dotted}) for 3 selected halo masses (low, intermediate, high). {\it upper panel} The mean pairwise velocity $v_{12}$ normalized by linear theory $v_{12}^{\text{lin}}$. {\it lower panel} The parallel part of the pairwise velocity dispersion  $\sigma_{\|}^2$.}
\label{fig:velstatcomp}
\end{figure}

\begin{figure}[t!]
\includegraphics[width=0.46\textwidth]{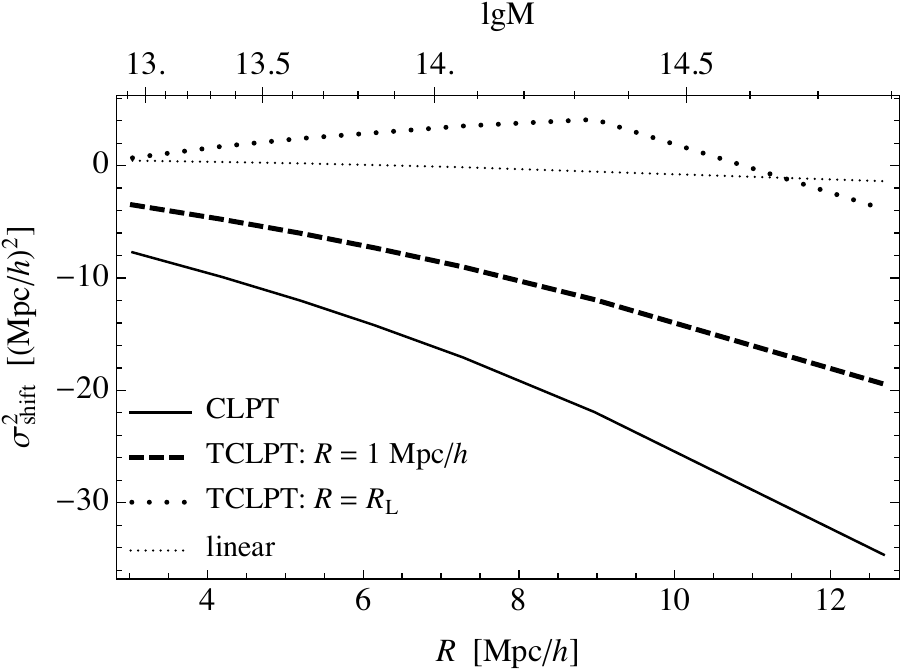}
\caption{The residual or `shift' between the HR2 $\hat \sigma^2_{\perp}$ and several theoretical predictions for $\sigma_{\perp}^2$. The curves for $ \sigma^2_{\rm shift, ||}$ and $ \sigma^2_{\rm shift, \perp}$ cannot be distinguished by eye.}
\label{fig:sigmashift}
\end{figure}

The pairwise velocity mean $v_{12}$ and dispersion $\v{\sigma}_{12}^2$ defined in Eq.\,\eqref{GSMparam} are computed in (T)CLPT and shown in Fig.\,\ref{fig:velstatcomp} for CLPT as well as TCLPT with a smoothing at scales $R=1\Mpc$ and $R_{\rm L}(M)$ together with the HR2 data points. 
The upper panel shows the mean pairwise velocity $v_{12}$ predicted from CLPT and TCLPT compared to linear theory, see \cite{RW11},
\begin{align}
v^{\rm lin}_{12} &= -2\sH f b_1^{\rm E}\ \frac{1}{2\pi^2} \int_0^\infty d k\, P_{\rm lin}(k) j_1(k r) \,. \label{linv12}
\end{align}
TCLPT with $R = 1\Mpc$ consistently improves the agreement with the data on small scales $r \lesssim 30\Mpc$ compared to CLPT for all masses, whereas TCLPT with a smoothing at the Lagrangian scale $R_{\rm L}(M)$ shows an improvement over CLPT for the smaller halo masses but a growing disagreement for larger halo masses. 
The lower two panels show the components of the pairwise velocity dispersion $\v{\sigma}_{12}^2$.
We do not directly display the theoretical predictions but instead show 
\begin{equation}
\bar \sigma^2_\perp \equiv \sigma^2_\perp - \sigma^2_{\rm shift, \perp}~~\mathrm{and}~~\bar \sigma^2_{||} \equiv \sigma^2_{||} - \sigma^2_{\rm shift, ||}\,,
\end{equation}
 where 
 \begin{equation}
 \sigma^2_{\rm shift, \perp} \equiv  \hat \sigma^2_{\perp}(149\Mpc) -  \sigma^2_{\perp}(149\Mpc)
  \end{equation}
  and similarly for $\sigma^2_{\rm shift, ||}$.
Therefore $\bar \sigma^2_\perp$ and $\bar \sigma^2_{||}$ are constructed to agree with the HR2 at large $r$.
This way we can compare separately how well the shape and how well the amplitude of $\v{\sigma}_{12}^2$ fits the data $\hat{\v{\sigma}}_{12}^2$.
The reason why this separate comparison is useful, is that that shape of $\v{\sigma}_{12}^2$, and therefore $\bar \sigma^2_\perp$ and $\bar \sigma^2_{||}$ are mostly responsible for the redshift space correlation function, while the amplitude of $\v{\sigma}_{12}^2$ and therefore how accurately $\sigma^2_{\rm shift}$ vanishes is less important, at least if $\sigma^2_{\rm shift}$ is not too large compared to $\v{\sigma}_{12}^2$.
The reason for this is that  $\sigma^2_{\rm shift}$  is completely irrelevant in the Kaiser limit once the GSM is linearized, see \cite{RW11}.
  In Fig.\,\ref{fig:sigmashift} only $\sigma^2_{\rm shift, \perp}$ is shown, which is virtually the same as $\sigma^2_{\rm shift, ||}$ for all models and masses.
 It is interesting that linear theory gives the best prediction in terms of having the smallest $\sigma^2_{\rm shift}$, while CLPT is worst having also a strong mass dependence. 
 On the other hand TCLPT with Lagrangian smoothing $R=R_{\rm L}(M)$ is not too far away from $\sigma^2_{\rm shift}=0$.
 
The reason for the wrong prediction of the absolute value of $\v{\sigma}_{12}^2$ and its strong dependence on $R$ is its sensitivity on UV modes. Some of the CLPT kernels $X^{(11)}(r)$, $X^{(22)}(r)$, $X^{(13)}(r)$ and $X^{(12)}_{10}(r)$ defined in Eqs.\,B41-B43 of \cite{CRW13} survive the limit $r \rightarrow \infty$, and it is only for $\v{\sigma}_{12}^2$ that they appear unsurpressed \cite{WRW14}. While $X^{(11)}(r \!\! \rightarrow \!\! \infty)$ is proportional to $\int_0^\infty P_{\rm lin}(k,R) dk$, producing the constant pieces in Eq.\,\eqref{sig12lin}, the other expressions are schematically proportional to $\int_0^\infty P^2_{\rm lin}(k,R)dk$, see App.\,B3 of \cite{CRW13} for details. This explains why even on the largest scales $\v{\sigma}_{12}^2$ is highly sensitive to UV modes and therefore also to the smoothing scale $R$, in contrast to $v_{12}$ and $\xi_X$.

\begin{figure}[h!]
\includegraphics[width=0.47\textwidth]{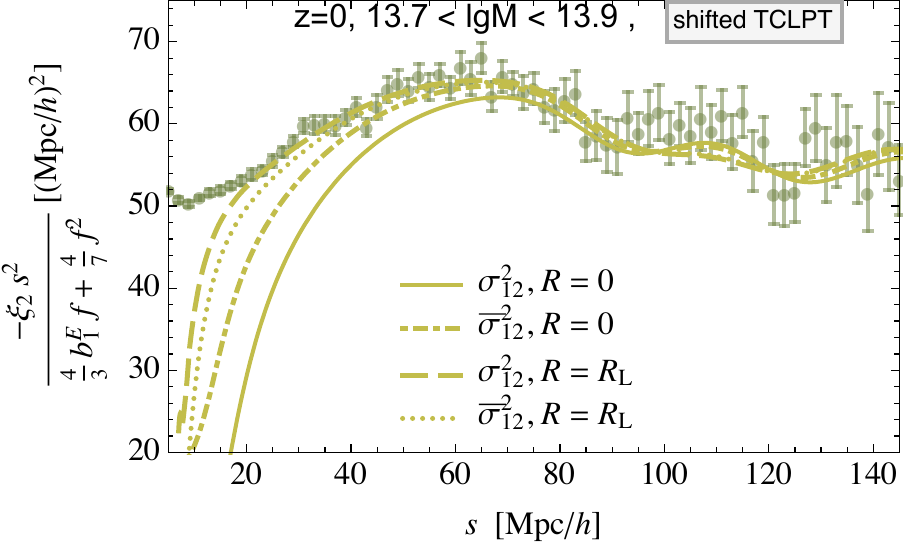}
\caption{The effect of shifting CLPT predictions for $\v{\sigma}_{12}^2$ to agree with HR2 at $149\Mpc$ on the quadrupole of the redshift space halo correlation function predicted by the GSM \eqref{GSM}. All models use the same function $\xi_X$ and $v_{12}$ calculated with TCLPT at $R=1 \Mpc$, while $\v{\sigma}_{12}^2$ is either calculated with CLPT ($R=0$) or TCLPT with $R=R_{\rm L}(M)$ and either used directly (`$\v{\sigma}_{12}^2$') or shifted (`$\bar{\v{\sigma}}_{12}^2$') to match HR2 at larger scales.}
\label{fig:z0quadrupoleShift}
\end{figure}

In Fig.\,\ref{fig:velstatcomp}  the results for $\bar \sigma^2_\perp$ and $\bar \sigma^2_{||}$ and the data points are normalised by the mass-independent result of linear theory, see \cite{RW11},
\begin{align}
\label{sig12lin}
\notag  \sigma^2_{||,\rm lin}&=  2\sH^2 f^2 \left(R_{\rm NL}^2 -\frac{1}{2\pi^2} \int_0^\infty d k\, P_{\rm lin}(k) \left( j_0(k r)- \frac{2 j_1(k r)}{k r}\right) \right)\\
 \sigma^2_{\perp,\rm lin}& =  2\sH^2 f^2 \left(R_{\rm NL}^2 - \frac{1}{2\pi^2} \int_0^\infty d k\, P_{\rm lin}(k) \frac{ j_1(k r)}{k r}\right) \\
\notag R_{\rm NL}^2&= \frac{1}{6 \pi^2} \int_0^\infty dk\, P_{\rm lin}(k) \,.
\end{align}
 One can clearly see that the shape of $\v{\sigma}_{12}^2$, namely $\bar \sigma^2_\perp$ and $\bar \sigma^2_{||}$, matches best the $N$-body data for TCLPT with a Lagrangian smoothing scale $R_{\rm L}(M)$. 
 Thus we observe that within TCLPT both the shape and amplitude of $\v{\sigma}_{12}^2$ are optimised for $R=R_{\rm L}(M)$ compared to TCLPT with $R=1\Mpc$ or CLPT. 
 Hence, we conclude that the optimal smoothing scale is $R=1\Mpc$ for the mean pairwise velocity (just as for the real space correlation function) but $R_{\rm L}(M)$ for the pairwise velocity dispersion. 

Since the $\bar{\v{\sigma}}_{12}^2$ of CLPT is better than linear perturbation theory, at least for $r>30 \Mpc$, but $\sigma_{\rm shift}^2$ is very large, see Fig.\,\ref{fig:sigmashift}, one might be tempted to simply use $\bar{\v{\sigma}}^2_{12}$ in the GSM, as has been done in \cite{RW11,WRW14}. Although at linear order, any constant, isotropic velocity dispersion does not alter the redshift space correlation function as explained in \cite{RW11} it does affect the GSM.
 In Fig.\,\ref{fig:z0quadrupoleShift} we investigate the impact of using different versions of the pairwise velocity dispersion, while keeping all other ingredients of the GSM fixed, in particular $\xi_X$ and $v_{12}$ have been calculated with TCLPT and $R=1\Mpc$. 
We observe that for $R=R_{\rm L}(M)$ switching between the  shifted $\bar{\v{\sigma}}_{12}^2$ and the original $\v{\sigma}_{12}^2$ does have a smaller impact than switching between shifted and original prediction of CLPT ($R=0$) as is expected from the fact that $\bar{\v{\sigma}}_{12}^2$ and $\v{\sigma}_{12}^2$ are closer to each other in the former case and very different in the latter.
In to addition to not being well justified, using  the quantity $\bar{\v{\sigma}}_{12}^2$ can be dangerous since it can become easily negative for relevant scales and therefore requires some further ad-hoc fudging when used in the GSM.
In the rest of this paper and in particular for all results presented in Sec.\,\ref{sec:results} we therefore use the un-shifted $\v{\sigma}_{12}^2$.

\subsubsection{Comparison of TCLPT and TZA}

In \cite{W14} the `Zel'dovich Streaming Model' was proposed, which is identical to combining the GSM and CLPT from \cite{WRW14}, with the exception that only first order (or ZA) displacements $\v{\Psi}_{\rm Z}$ are considered, instead of $\v{\Psi}_{\rm Z,2,3}$ as done in CLPT. 
In \cite{W14} it was found that for the redshift space quadrupole the ZA exhibits a better agreement with the $N$-body data than CLPT when galaxy-sized halos are considered. 
Note however, that in \cite{W14} the linear input power spectrum has been additionally smoothed on a scale $R=1\Mpc$ compared to \cite{WRW14}, such that in our nomenclature that model considered in \cite{W14} is TZA with $R=1\Mpc$.

We therefore investigate in more detail the relative performance of TZA and TCLPT.
From the $\chi^2$ values in Table \ref{biasfittab} it follows that TZA/ZA is slightly worse than CLPT/TCLPT for the real space halo correlation function on large scales. 
Now we demonstrate that TZA is also worse than TCLPT for the correlation function on small scales and the mean pairwise velocity.

\begin{figure}[t!]
\includegraphics[width=0.48\textwidth]{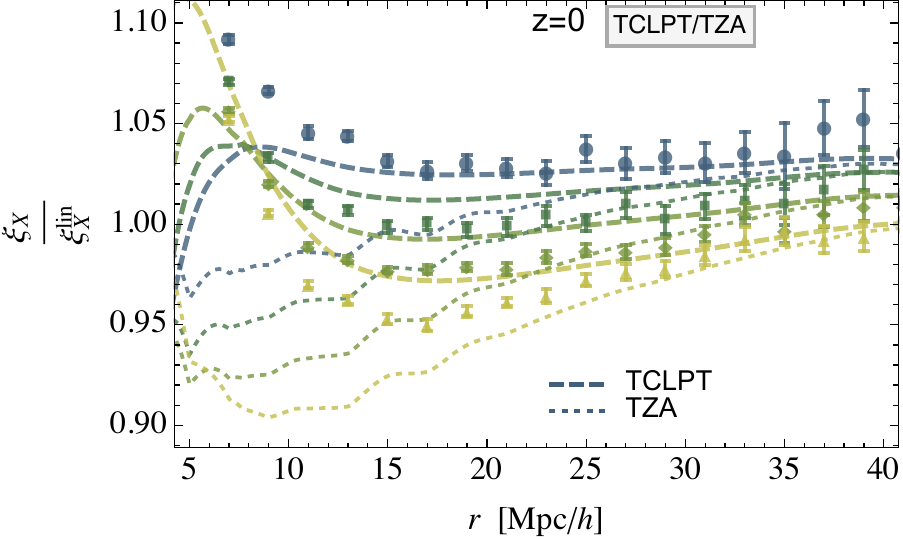}
\caption{Comparison of the real space correlation function between TCLPT ({\it thick dashed}) and TZA ({\it thin dotted}) for small scales and  the four lowest halo masses, both with $R=1 \Mpc$. The halo correlation function $\xi_X$ is normalized by linear theory $\xi_{X}^{\text{lin}}$.}
\label{fig:xiRealZeldo}
\end{figure}

\begin{figure}[h!]
\includegraphics[width=0.45\textwidth]{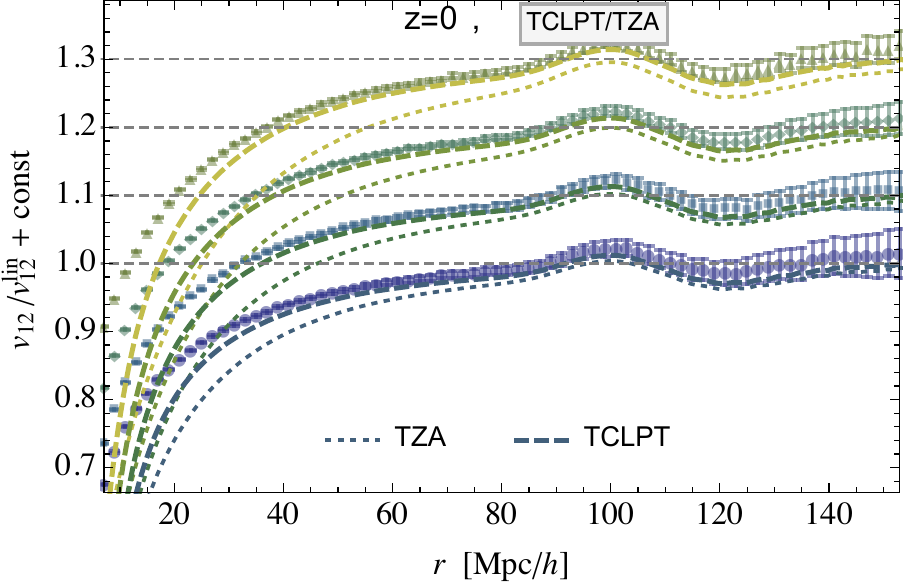}\\[6pt]
\caption{Comparison of the pairwise velocity statistic predicted by TCLPT ({\it thick dashed}) and TZA ({\it thin dotted}), both with $R=1 \Mpc$, together with measurements from HR2 ({\it data points}) for the four lowest halo masses. The mean pairwise velocity $v_{12}$ is normalized by linear theory $v_{12}^{\text{lin}}$ and shifted by mass bin dependent constant for better visibility.}
\label{fig:velstatZeldo}
\end{figure}

In Figs.\,\ref{fig:xiRealZeldo} and \ref{fig:velstatZeldo} we compare the prediction computed from TCLPT with that from TZA, both with $R=1\Mpc$ for $\xi_X$ and $v_{12}$ respectively.\footnote{Note that $v_{12}^{\rm lin}$ does not coincide with the $v_{12}$ of ZA because the mean pairwise velocity from \eqref{GSMparam} is weighted with the density and not linearised in ZA. The same is true for $\v{\sigma}_{12}^2$.} 
We do not show the pairwise velocity dispersion $\v{\sigma}_{12}^2$ because it resembles the result from linear theory on large scales and hence gives a reasonable amplitude but a less accurate scale dependence than TCLPT.
We also studied the redshift space correlation function obtained when feeding those ingredients into the GSM. As expected TCLPT always performs better than TZA.
Thus the Zel'dovich approximation is clearly improved by the inclusion of higher order kernels appearing in TCLPT, at least for the halo masses and redshift ($z=0$) considered by us. 

\section{Results}
\label{sec:results}
\subsection{Redshift space correlation function} 
\label{zspacecomp}
In the following we will employ the TCLPT results from the previous section, in particular what we learned about the smoothing scale $R$, to determine the redshift space correlation function $\xi(s,s_{\|})$ using the GSM \eqref{GSM}.
These GSM ingredients are the real space correlation $\xi_X(r)$, the mean pairwise velocity $v_{12}(r)$ and the pairwise velocity dispersion $\sigma_{12}^2(r,r_{\|})$ given according to \eqref{GSMparam} and evaluated in TCLPT, see \cite{UKH15}, as shown in Figs.\,\ref{fig:allreal} and \ref{fig:velstatcomp}.
 The streaming model ingredients obtained from TCLPT are then employed to compute the correlation function in redshift space $\xi_X(s,s_{\|})$ via the GSM.

In order to evaluate and compare  $\xi_X(s_{||},s_\perp,t)$ from Eq.\,\eqref{GSM} to $N$-body simulations we will expand the redshift-space halo correlation function in Legendre polynomials to obtain its multipoles, see Eq.\,\eqref{ximultipoles}. The magnitude of $\xi_n$ rapidly decreases with $n$ as can be seen from Fig.\,\ref{fig:allzspacedirect} in App.\,\ref{sec:HR2}. In linear perturbation theory, the only non-zero moments are the monopole $\xi_0$, quadrupole $\xi_2$ and hexadecapole $\xi_4$ which are according to \cite{RW11} given by
\begin{align}
\notag \xi^{\rm lin}_0(s) &= 
\left( (b_1^E)^2 + \frac{2}{3} b_1^E f + \frac{1}{5} f^2 \right)  \frac{1}{2\pi} \int d k\, k^2 P_{\rm lin}(k) j_0(ks)  \\
 \xi^{\rm lin}_2(s)& = 
 - \left(  \frac{4}{3} b_1^E f + \frac{4}{7} f^2 \right)\frac{1}{2\pi}\int d k\, k^2 P_{\rm lin}(k) j_2(ks)  \label{linearZspace} \\
\notag  \xi^{\rm lin}_4(s)& = 
  \frac{8}{35} f^2 \frac{1}{2\pi}\int d k\, k^2 P_{\rm lin}(k) j_4(ks)     \,,
   \end{align}
where $f$ is the linear growth rate and $j_n(x)$ are the spherical Bessel functions. 
We use  prefactors of the $ \xi^{\rm lin}_n(s) $ as a normalization when plotting multipoles. Those are entirely determined from the fit involving only the the real space correlation function, see Tab.\,\ref{biasfittab}. 

\begin{figure}[h]
\includegraphics[width=0.48\textwidth]{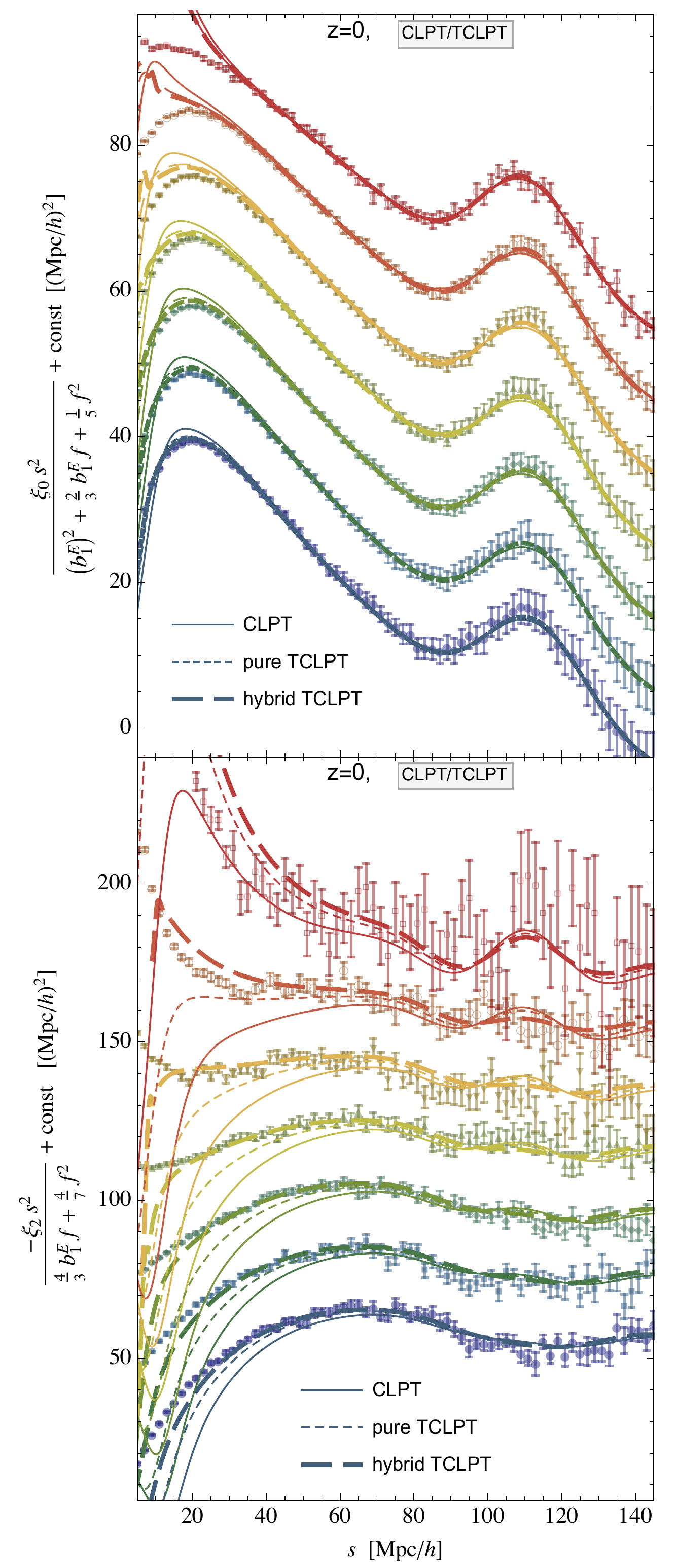}
\caption{The monopole $\xi_0$ and quadrupole $\xi_2$ times $s^2$ predicted by the GSM \eqref{GSM} for all masses rescaled by linear bias comparing predictions from CLPT {\it (thin solid)}, TCLPT with smoothing on $R=1\Mpc$ {\it (thin dashed)} and TCLPT with hybrid smoothing on $R=1\Mpc$ for $\xi_X$ and $v_{12}$ and  $R=R_{\rm L}(M)$ for $\v{\sigma}_{12}^2$ {\it (thick dashed)}. We shifted all values for $\xi_0$ by $10(i-1)(\Mpc)^2$ and those for $\xi_2$ by $20(i-1)(\Mpc)^2$ according to the $i$-th mass bin for better visibility.}
\label{fig:z0xi02}
\end{figure}

In Fig.\,\ref{fig:z0xi02} we show the monopole $\xi_0(s)$ and quadrupole $\xi_2(s)$ times $s^2$ for all masses.  We rescaled the HR2 points and the theory as just explained. 
Like it was the case for the real space correlation Fig.\,\ref{fig:allreal} the rescaling reveals that for $\xi_0$, the bias is linear and local bias factor for scale  $s>90\Mpc$.
Below that scale a non-local mass dependence appears.
Let us emphasise that the theory lines are predictions based on the fitting of $\lgM_{\rm opt}$ in real space, no further fit was performed. The quadrupole $\xi_2(s)$ and the hexadecapole $\xi_4(s)$ are similarly shown in Fig.\,\ref{fig:z0xi24} for three different mass bins. 
It can be clearly seen that for all masses CLPT performs less satisfying than TCLPT on small scales. This can be most easily seen when focusing on $r=20 \Mpc$ for $\xi_0$  in Fig.\,\ref{fig:z0xi02}  and  is obvious for $\xi_2$ and $\xi_4$ in Figs.\,\ref{fig:z0xi02} and \ref{fig:z0xi24}.
We denote by `pure TCLPT' the model where for all ingredients the same smoothing scale $R=1 \Mpc$ has been used, and `hybrid TCLPT', where we used a combination suggested by the results of  Secs.\,\ref{sec:realcorrfct} and \ref{sec:velstat}, namely $R=1 \Mpc$ for $\xi_X$ and $v_{12}$ as well as $R=R_{\rm}(M)$ for $\sigma_{\perp}^2$ and $\sigma_{||}^2$. 

\subsection{Status of the truncated (Post-)Zel'dovich approximation}
\label{truncZeld}

It was recently observed in \cite{W14} that a smoothing of the initial power spectrum $P_{\rm lin}$ does not change or improve the resulting redshift space quadrupole $\xi_2$ within TZA evaluated in redshift space. 
In contrast both in real space in redshift space via the GSM, we find a mild improvement between of TZA over ZA, and a significant one for TCLPT. 
This is expected since for the truncated Post-Zel'dovich approximation, which TCLPT approximates, it is well known that once the linear power spectrum is smoothed with a Gaussian filter on an optimal scale (depending on redshift and cosmology, we find $1\Mpc$) the result  improves significantly \cite{C93,M94,BMW94,WGB95,Hamana1998}.

The reason for the improvement is that within Lagrangian perturbation theory the displacements never stop and therefore, after the occurrence of shell crossings, structures which should be held in place by gravity do instead disperse indefinitely.
We saw this improvement already in Fig.\,\ref{fig:smoothingDensitysmall}, where the power on small scales increased for TCLPT with a smoothing scale of $R=1\Mpc$ compared to CLPT and TCLPT with a smoothing scale of $R=2\Mpc$. 
Therefore contrary \cite{W14}, we find that the truncation does improve analytical computations. 

Although we are not the first to observe the better performance of T(P)ZA in analytical calculations for the real space \cite{MHP93, P97}, we want to advocate that the truncation also significantly improves the redshift space correlation functions and is at the same time very easy to implement in existing Lagrangian codes, since it involves only the multipilication of the linear power spectrum $P_{\rm lin}$, that is read in by the code with a gaussian window function $W(R k)^2 = \exp(- k^2 R^2)$.

\begin{figure}[h!]
\includegraphics[width=0.47\textwidth]{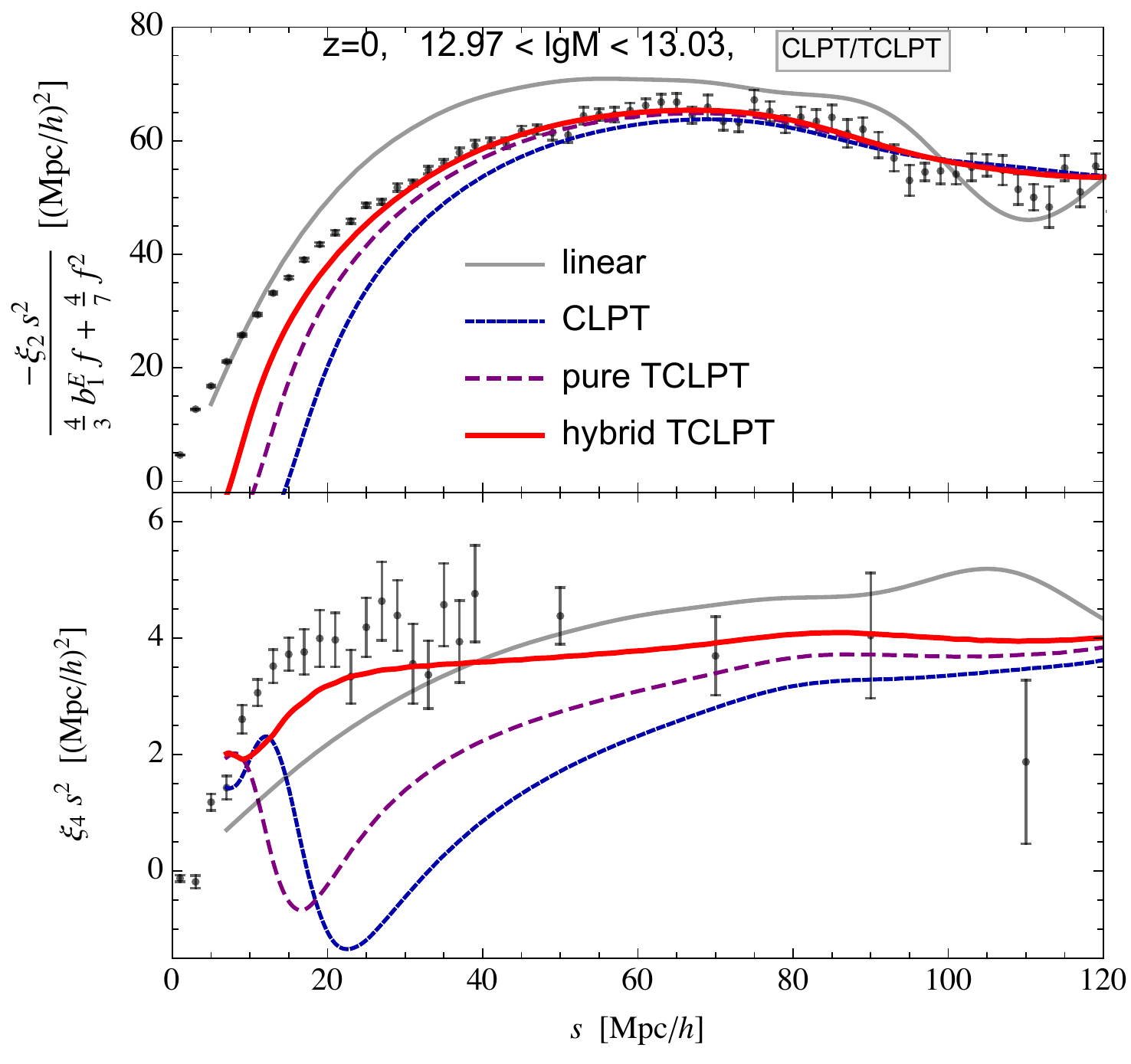}\\
\vspace{-0.2cm}
\includegraphics[width=0.47\textwidth]{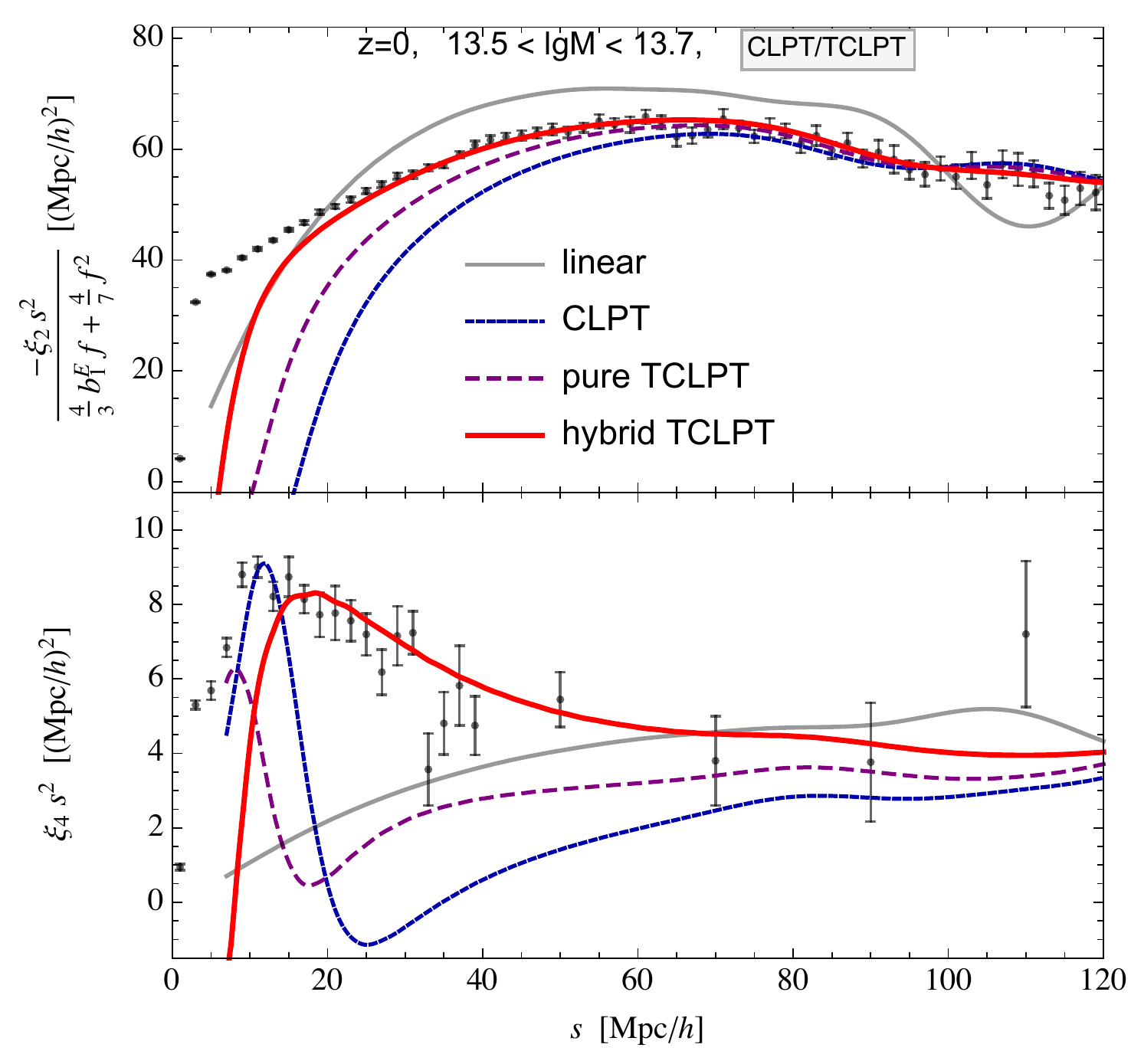}\\
\vspace{-0.2cm}
\includegraphics[width=0.47\textwidth]{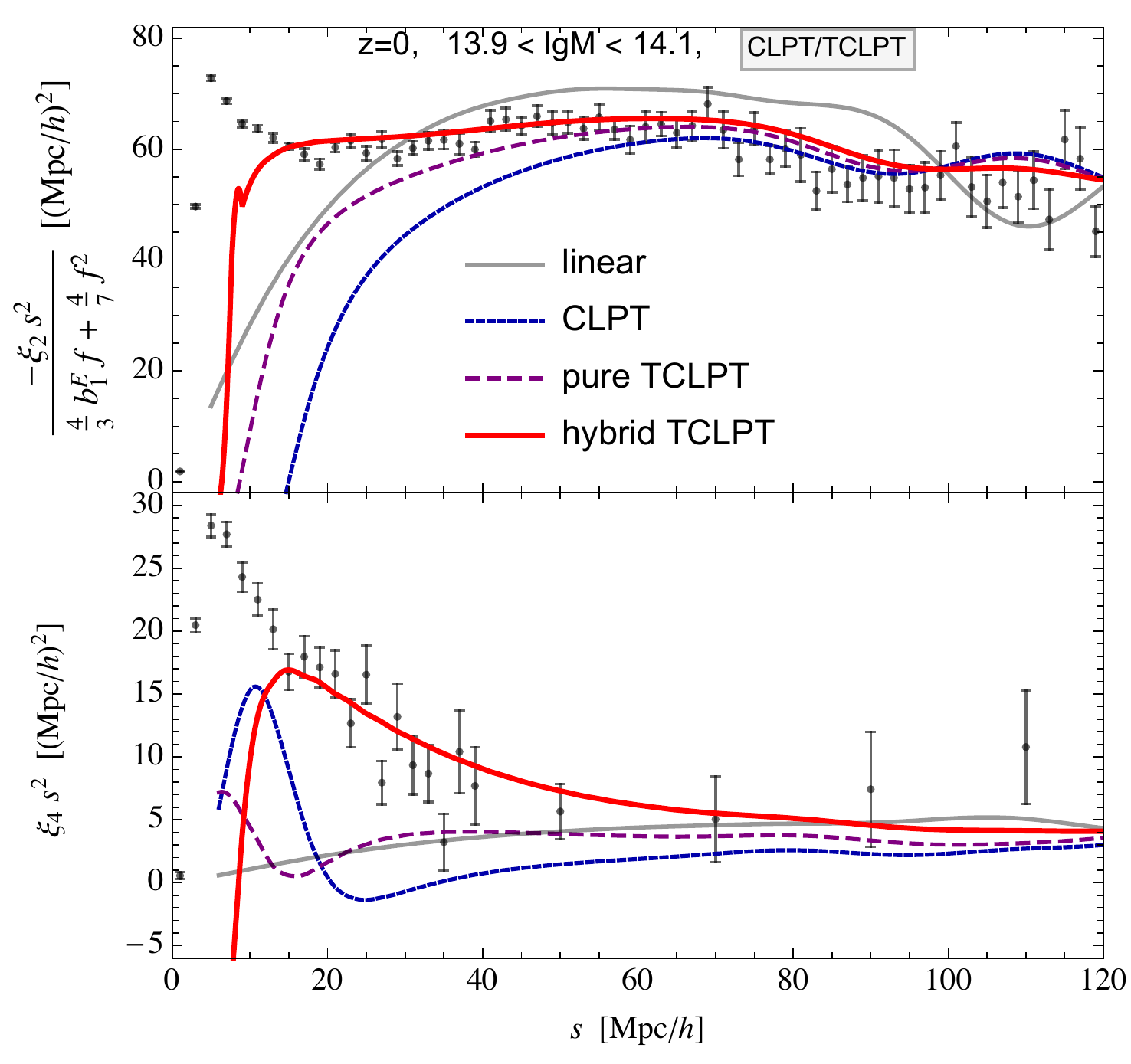}
\caption{Comparison of the quadrupole $\xi_2$ and hexadecapole $\xi_4$ times $s^2$ rescaled by linear bias as measured in HR2 {\it (black data points)} and predicted by the GSM \eqref{GSM} using linear theory {\it (grey solid)}, CLPT {\it (blue dashed)}, pure TCLPT with smoothing on $R=1\Mpc$ {\it (magenta long dashed)} and hybrid TCLPT with smoothing on $R=1\Mpc$ for $\xi_X$ and $v_{12}$ and  $R=R_{\rm L}(M)$ for $\v{\sigma}_{12}^2$ {\it (red solid)}.}
\label{fig:z0xi24}
\end{figure}

 We show the significant improvement obtained for the quadrupole and hexadecapole if the linear input power spectrum $P_{\rm lin}$ (CLPT) is smoothed $P_{\rm lin} W(kR)^2$ (`pure TCLPT' and `hybrid TCLPT') and fed into the TCLPT+GSM model, see Fig.\,\ref{fig:z0xi24}.
The black thin line shows  linear theory Eq.\,\eqref{linearZspace} as reference. The blue lines come from  CLPT, the dashed purple lines show the case where $R=1\Mpc$ is the same for calculating $\xi_X(r)$, $v_{12}(r)$, $\sigma^2_{\perp}(r)$ and $\sigma^2_{\|}(r)$ within TCLPT, which already improves over CLPT.  Finally, the thick red line shows the hybrid TCLPT approach in which we smooth  at $1\Mpc$ to obtain $\xi(r)$ and $v_{12}(r)$, but at $R_{\rm L}(M)$ to obtain $\sigma^2_{\perp}(r)$ and $\sigma^2_{\|}(r)$.
This again significantly improves the prediction of the redshift space correlation function, in particular  $\xi_4(s)$.

\section{Conclusion}
\label{sec:concl}

On the basis of the DEUS $N$-body simulations\footnote{http://www.deus-consortium.org/data/} we determined the Zel'dovich displacement \cite{Z70} of proto-halos using different smoothing scales and used them to displace the proto-halo centers. Comparing the result of this Zel'dovich simulation with $N$-body simulations, we found that the Lagrangian size of the halo stands out as smoothing scale optimizing the prediction of the Zel'dovich approximation \cite{AchitouvBlake2015}. This seconds the heuristic argument from \cite{UKH15} that the Lagrangian halo size should be accounted for in the fluid description of halos and guide the choice of the physical coarse-graining scale.

This observation inspired us to predict the redshift space halo correlation function using the Gaussian Streaming Model, known to be accurate on scales $s\gtrsim 30\,\text{Mpc}/h$ \cite{UKH15} combined with the Post-Zel'dovich approximation. Therefore we used Convolution Lagrangian Perturbation Theory (CLPT) where the coarse-graining scale can be straightforwardly implemented by a smoothing of the initial power spectrum leading to the `truncated' CLPT (TCLPT). We benchmarked the CLPT and TCLPT predictions for the streaming model ingredients against results from the Horizon Run 2 halo catalog and optimised the smoothing scale accordingly. 

We found that for the halo correlation function in redshift space $\xi_X(s,s_{\|})$, a hybrid approach delivers the best results, where the calculation is performed partially with $R \simeq 1\Mpc$ to obtain the real space halo correlation $\xi_X(r)$ and the mean pairwise velocity $v_{12}(r)$, and partially with $R= R_{\rm L}(M)$, where $R_{\rm L}(M)$ is the Lagrangian radius of halos with mass $M$, to obtain the pairwise velocity dispersion $\sigma_{12}^2(r,r_{||})$ of halos. 
The improvement achieved by the smoothing of the linear power spectrum is well known in the context of the Zel'dovich approximation as truncated Zel'dovich approximation.
 We showed that our hybrid approach involving two different smoothing scales, $R=1\Mpc$ and the Lagrangian radius $R=R_{\rm L}(M)$, outperforms the truncated Zel'dovich approximation. 
 This GSM+TCLPT model with a hybrid smoothing on the level of the input power spectraum is an easy-to-implement modification of the original GSM+CLPT model \cite{WRW14} and hence of interest for improving cosmological constraints obtained based on this formalism \cite{Alam15}.
 We expect that the somewhat strange necessity of two filters is caused by using a local Lagrangian bias model detailed in App.\,\eqref{LocLagrBiasModel} based on \cite{BCEK91,M11}, that does not explicitly take into account that proto-halos are predominantly peaks of the initial density field.
 Furthermore we expect that a single smoothing scale close to the Lagrangian scale, the natural scale associated with a proto-halo, will suffice once the peak bias model, \cite{BBKS86,ParanjapeShethDesjacques2013}, is implemented in TCLPT which we leave for the future.
\FloatBarrier
\section{Outlook}
\label{sec:outlook}
More work is necessary to better model all the ingredients of the Gaussian streaming model, in particular the small scales, see Figs.\,\ref{fig:velstatcomp} and \ref{fig:allreal}. 
We believe that many of the observed discrepancies between the simulation results and the theoretical model that is based on local Lagrangian halo bias, can be overcome if  peak bias, as advocated in \cite{D08}, is included in CLPT. 
We expect that upon considering density peak correlations, similar as done for linear theory in \cite{DS10}, the Lagrangian scale, the only physically meaningful scale associated with a dark matter proto-halo, will turn out to be only relevant smoothing scale entering this peak-CLPT.

In \cite{AchitouvBlake2015} a refined study similar to the one presented in Sec.\,\ref{sec:displace} has shown that part of the late time non-linear contributions to the halo displacement cannot be  described by a deterministic  approximation (Zel'dovich or higher order LPT). Those stochastic contributions can however be reduced by considering only under-dense environments (e.g.\,large underdense regions) and therefore it would be interesting to test the limitations of the (T)CLPT in different environments.


So far we focused on two-point statistics, the two halo term, but in principle the (T)CLPT formalism allows to calculate higher correlation functions which are necessary to predict higher order statistics like the bispectrum \cite{RW12} and covariances \cite{LHT14}. They are particularly relevant to upcoming redshift surveys which will contain enough statistics to measure these higher order correlation functions. In addition, when two probes of the LSS like cluster counts $n(M)$ and the cluster correlation function $\xi(r)$ are combined to infer cosmological parameters, as done for instance in \cite{Manaetal13}, it is desirable to have analytic predictions not only for the observables but also for covariance matrices like $C_{\xi} (r,r')$ \eqref{covmatrix} and cross covariances between $n(M)$ and $\xi(r)$, including correlations for different mass bins. 
These covariance matrices are required for correct  error estimation of the inferred cosmological parameters. Analytic estimates of those covariance matrices require the knowledge of the connected three $\zeta(r_1,r_2,r_3)$ and four-point  $\eta(r_1,r_2,...,r_6)$ halo correlation functions, where the $r_i$ are the edge lengths of the observed triangle or tetrahedron, respectively \cite{SM14}. It would be interesting to check whether (T)CLPT evaluation of these quantities is feasible and whether a Gaussian streaming model can be developed to convert the real space results into the redshift space quantities. The results for halo correlation function could then be implemented into the halo model \cite{MF00,CS02} to additionally obtain correlation functions for the dark matter field which is required for predicting statistics involving weak gravitational lensing \cite{TB07}.

\section*{Acknowledgement}
The work of MK \& CU was supported by the DFG cluster of excellence ``Origin and Structure of the Universe''. Part of this research was conducted by the Australian Research Council
Centre of Excellence for All-sky Astrophysics (CAASTRO), through
project number CE110001020. IA also acknowledges
support from the Trans-Regional Collaborative Research
Center TRR 33 ``The Dark Universe'' of the
Deutsche Forschungsgemeinschaft (DFG).

We would like to thank Tobias Baldauf, Julien Bel, Davide Bianchi, Vincent Desjacques, Luigi Guzzo, Eiichiro Komatsu, Roman Scoccimarro, Uros Seljak and Ravi Sheth for their  interesting discussions.  
We also like to thank Lile Wang for making publicly available the codes \url{https://github.com/wll745881210/CLPT_GSRSD} on which our TCLPT and GSM codes are based.


\newcommand{\apjl}{Astrophys. J. Letters}
\newcommand{\apjs}{Astrophys. J. Suppl. Ser.}
\newcommand{\mnras}{Mon. Not. R. Astron. Soc.}
\newcommand{\pasj}{Publ. Astron. Soc. Japan}
\newcommand{\apss}{Astrophys. Space Sci.}
\newcommand{\aap}{Astron. Astrophys.}
\newcommand{\physrep}{Phys. Rep.}
\newcommand{\mpla}{Mod. Phys. Lett. A}
\newcommand{\jcap}{J. Cosmol. Astropart. Phys.}

\appendix
%
\section{$N$-body halo catalog statistics and bias model fit}
\label{sec:HR2}
The HR2 has an enormous size of 7200\,Mpc/$h$ and consists of $6000^3$ particles of mass $\lgM = 11.097$, where we introduced the notation $\lgM \equiv \log_{10}(M h/M_\odot)$. We measured halo correlation functions and velocity statistics from large galaxy-sized halos $\lgM = 13.0$ to cluster-sized halos $\lgM = 15.2$ at the redshifts $z=0$ and $z=1$. The relevant cosmological parameters for the HR2 are  $$\Omega_m = 0.26\,, \ \Omega_b = 0.044\,, \ h = 0.72\,, \ \sigma_8 = 1/1.26 \simeq 0.794\,.$$
The HR2 catalogue will be our reference in Secs.\,\ref{sec:GSM} and \ref{sec:results}.
 Note that these parameters are slightly different from the DEUS simulation $$\hspace{-1.4cm}\Omega_m = 0.256\,, \Omega_b = 0.044\,,  \ h = 0.72\,, \ \sigma_8 = 0.793\,,$$
 which was the basis of our investigations in Sec.\,\ref{sec:displace}.

 \subsection{Estimating correlation functions}
\label{sec:corrfuncandpairw}

  We estimate the correlation functions using the Mo\&White estimator \cite{MW96}, which we will describe below. By re-sampling the whole box with 27 jackknife samples \cite{EG83}, we estimate the covariance matrix of the various correlation functions for model parameter fits and the error bars for the plots. Each of those 27 subsamples consists of the whole box, with 1 out of 27 sub-boxes of size 2400\,Mpc/$h$ removed. 
\subsubsection{Real space correlation function}
\label{sec:realspacecorr}
\begin{figure}[t!]
\includegraphics[width=0.46\textwidth]{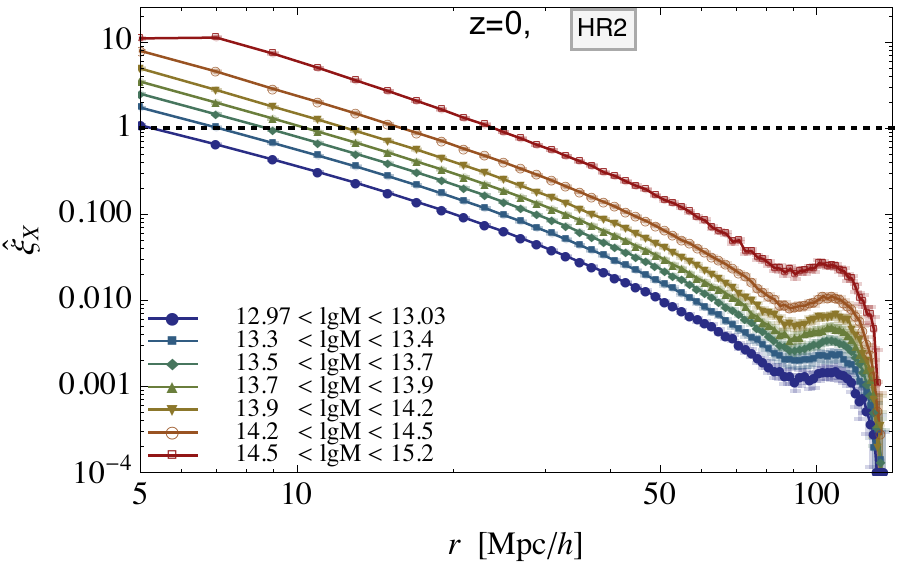}
\caption{The measured halo auto-correlation at $z=0$ function for masses specified in the legend. We connect the data points with lines for better visibility.}
\label{fig:allrealdirect}
\end{figure}

In real space the simplest estimator for the full box is given by \cite{MW96}
\begin{equation}
\label{MoWhest}
1+\hat \xi(r) = \frac{\Delta P(r)}{n_{\rm tot} 4 \pi r^2 \Delta r}\,,
\end{equation}
where $\Delta P(r)$ is the mean number of neighbour halos in a shell at distance $r$ with width $\Delta r$ around a halo at $r=0$, and $n_{\rm tot}$ is the mean number density of halos in the simulation at a given time, such that $n_{\rm tot} 4 \pi r^2 \Delta r$ gives the mean number of neighbour halos if the halos were evenly distributed. Therefore $\hat\xi(r)$ estimates the excess probability  to find a halo within an interval $I_r := [r - \Delta r/2,r + \Delta r/2]$ away from another halo. We determine $\Delta P(r)$ as $DD(r)/N_{ \rm tot}$, where $N_{\rm tot}$ is the total number of halos in the box and $DD(r)$ is the number of all halo pairings with distance $r \pm \Delta r/2$. In practice we calculate
\begin{equation}
DD(r) =  \sum_{k, r_{ki} \in  I_r }^{N_{\rm tot}} \sum_{i, r_{ik} \in  I_r }^{N_{\rm tot}}\,,
\end{equation}
where $r_{ik} = |\v{r}_i - \v{r}_k|$ is the distance between halo $i$ and halo $k$.
In order to estimate the error bars and the covariance matrix of $\hat\xi(r)$, we instead calculate
\begin{equation}
1+\hat\xi^{(j)}(r) = \frac{\Delta P^{(j)}(r)}{n_{\rm tot} 4 \pi r^2 \Delta r}\,,
\end{equation}
where $\Delta P^{(j)}(r)$ is the mean number of neighbour halos in a shell at distance $r \pm \Delta r/2$ around a halo in the subsample $j$, which is obtained be removing the subbox $j$. Therefore $\hat\xi^{(j)}(r)$ is the excess probability  to find a halo at distance $r \pm \Delta r/2$ away from another halo within the subsample. We determine $\Delta P^{(j)}(r)$ as $DD^{(j)}(r)/N^{(j)}_{ \rm tot}$, where $N^{(j)}_{\rm tot}$ is the total number of halos in the subsample $j$ and $DD^{(j)}(r)$ is the number of all halo pairings with distance $r \pm \Delta r/2$, with at least one partner lying in $j$
\begin{equation}
DD^{(j)}(r) =  \sum_{k, r_{ki} \in  I_r }^{N^{(j)}_{\rm tot}} \sum_{i, r_{ik} \in  I_r }^{N_{\rm tot}}\,.
\end{equation}
The correlation function is then given by
\begin{equation} \label{correst}
\hat{\bar{\xi}}(r) = \frac{1}{27} \sum_{j=1}^{27} \hat\xi^{(j)}(r)\,,
\end{equation}
with covariance matrix
\begin{equation}\label{covmatrix}
C_{\hat \xi}(r,r') = \frac{26}{27}\sum_{i=1}^{27} \left[\hat \xi^{(j)}(r) -\hat{\bar{\xi}}(r)\right]  \left[\hat\xi^{(j)}(r') -\hat{\bar{\xi}}(r')\right]\,.
\end{equation}
and error bar
\begin{equation}\label{correrror}
\sigma_{\hat\xi}(r) = \sqrt{\frac{26}{27}\sum_{i=1}^{27} \left[\hat \xi^{(j)}(r) -\hat{\bar{\xi}}(r)\right]^2} \,.
\end{equation}
We choose $\Delta r = 2\Mpc$ and cover 100 $r$-bins. 
Fig.\,\ref{fig:allrealdirect} shows the halo correlation for 7 mass bins at $z=0$. We indicate with the dotted line $\xi=1$ the nonlinear regime. For the largest mass bin corresponding to large galaxy clusters, a drop in the correlation function for  $r$ smaller than twice the virial radius is due to halo exclusion arising in a friend-of-friend halo finder that was used in HR2.

\subsubsection{Pairwise velocity statistics}
\begin{figure}[t!]
\includegraphics[width=0.48\textwidth]{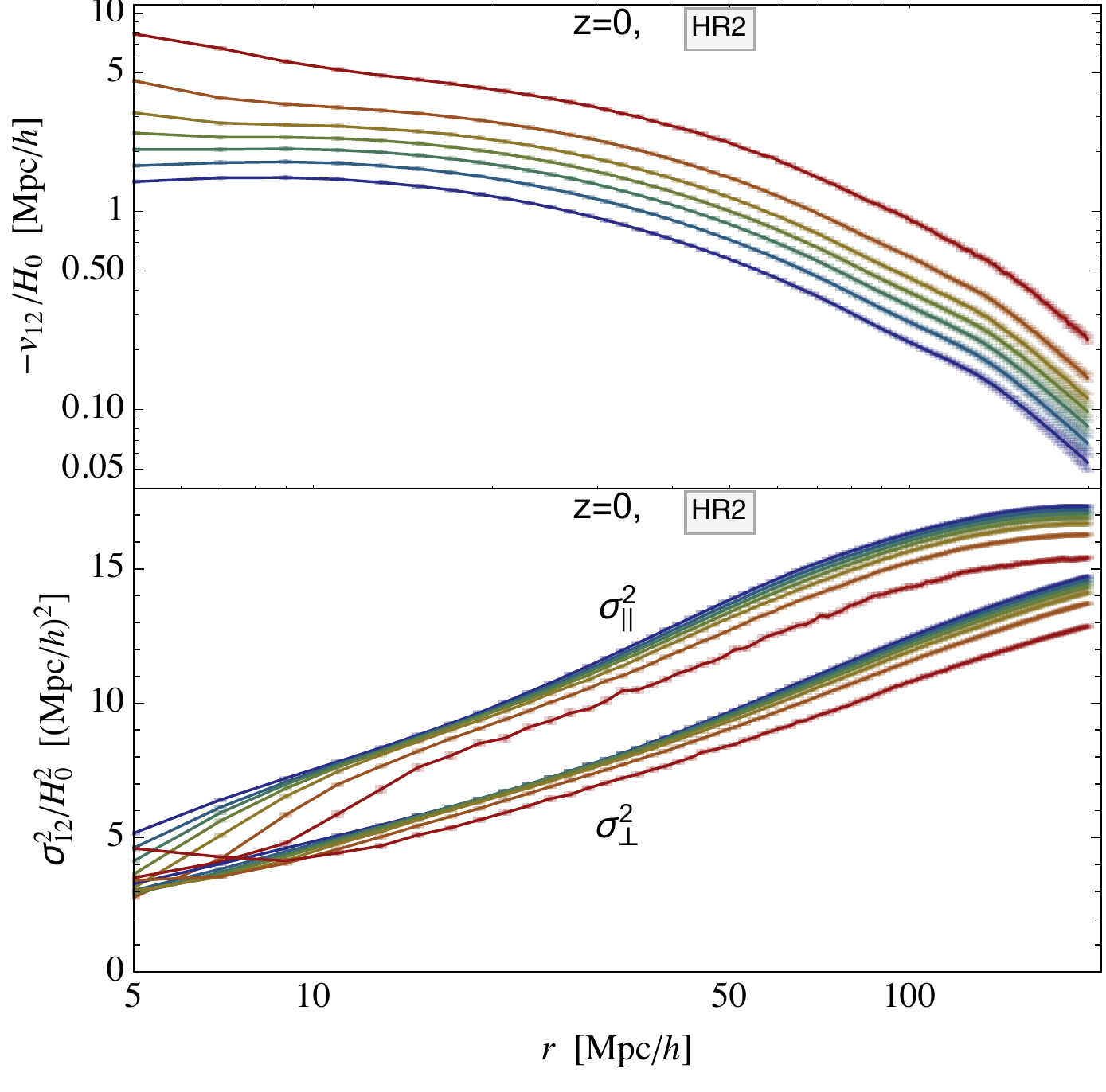}
\caption{The upper panel shows the measured $\hat v_{12}$. The lower panel shows the measured components of the pairwise velocity dispersion tensor $\v{\sigma}_{12}$, $\hat \sigma_{||}^2$ (upper lines) and $\hat \sigma_{\perp}^2$ (lower lines). The color coding is the same as in Fig.\,\ref{fig:allrealdirect}, but for better visibility we do not show the data symbols corresponding to the specific mass bins. The largest mass is on top for $\hat v_{12}$ and on the bottom for $\hat \sigma_{||}^2$ and  $\hat \sigma_{\perp}^2$ respectively.}
\label{fig:allv12sig12realdirect}
\end{figure}

To obtain the scale-dependent mean pairwise velocity $v_{12}(r)$  from the HR2 halo catalog, for fixed $r \in I_r$ and mass bin, we simply average over all velocity differences projected onto the pair separation vector. This mass-weighted average gives the mean pairwise velocity $\hat v_{12}(r)$
\begin{equation} \label{HR2v12}
\hat{v}_{12}(r) =  \frac{1}{DD(r)}\sum_{k, i, k\neq i }^{N_{\rm tot}} \delta_{r_{ik}, r} \delta_{r_{ki}, r}\,\Delta \v{v}_{ik}\cdot \hat{\v{r}}_{ik}\,,
\end{equation}
where $\hat{\v{r}}_{ik} = \v{r}_{ik}/ r_{ik}$ is the unit pair separation vector and $\Delta \v{v}_{ik}= \v{v}_i - \v{v}_k$ the relative halo velocity. 
$\hat{v}_{12}(r)$ is shown for the 7 different mass bins in the upper panel of Fig.\,\ref{fig:allv12sig12realdirect}.
 We proceed similarly for the two components of the pairwise velocity dispersion $\hat \sigma_{\perp}^2$ and $\hat \sigma_{||}^2$. In more detail we first construct $\hat{\tilde \sigma}_{12}$ by replacing in above expression $\Delta \v{v}_{ik}\cdot \hat{\v{r}}_{ik}$ by $(\Delta \v{v}_{ik}\cdot \hat{\v{r}}_{ik})^2$ and then obtain $\hat \sigma_{||}^2 = \hat{\tilde \sigma}_{||}^2 - \hat{v}_{12}^2 $. 
 We get $\hat \sigma_{||}^2$ by replacing in \eqref{HR2v12} the expression $\Delta \v{v}_{ik}\cdot \v{r}_{ik}$ by  $[|\Delta \v{v}_{ik}|^2 - (\Delta \v{v}_{ik}\cdot \hat{\v{r}}_{ik})^2 ]/2$.
The error estimates for $\hat v_{12},\hat \sigma_{\perp}^2$ and $\hat \sigma_{||}^2$ are handled as in the case of $\hat \xi_X$ using 27 Jack-knived samples.
\subsubsection{Redshift space correlation function} \label{zspacecomp}

\begin{figure}[t!]
\includegraphics[width=0.48\textwidth]{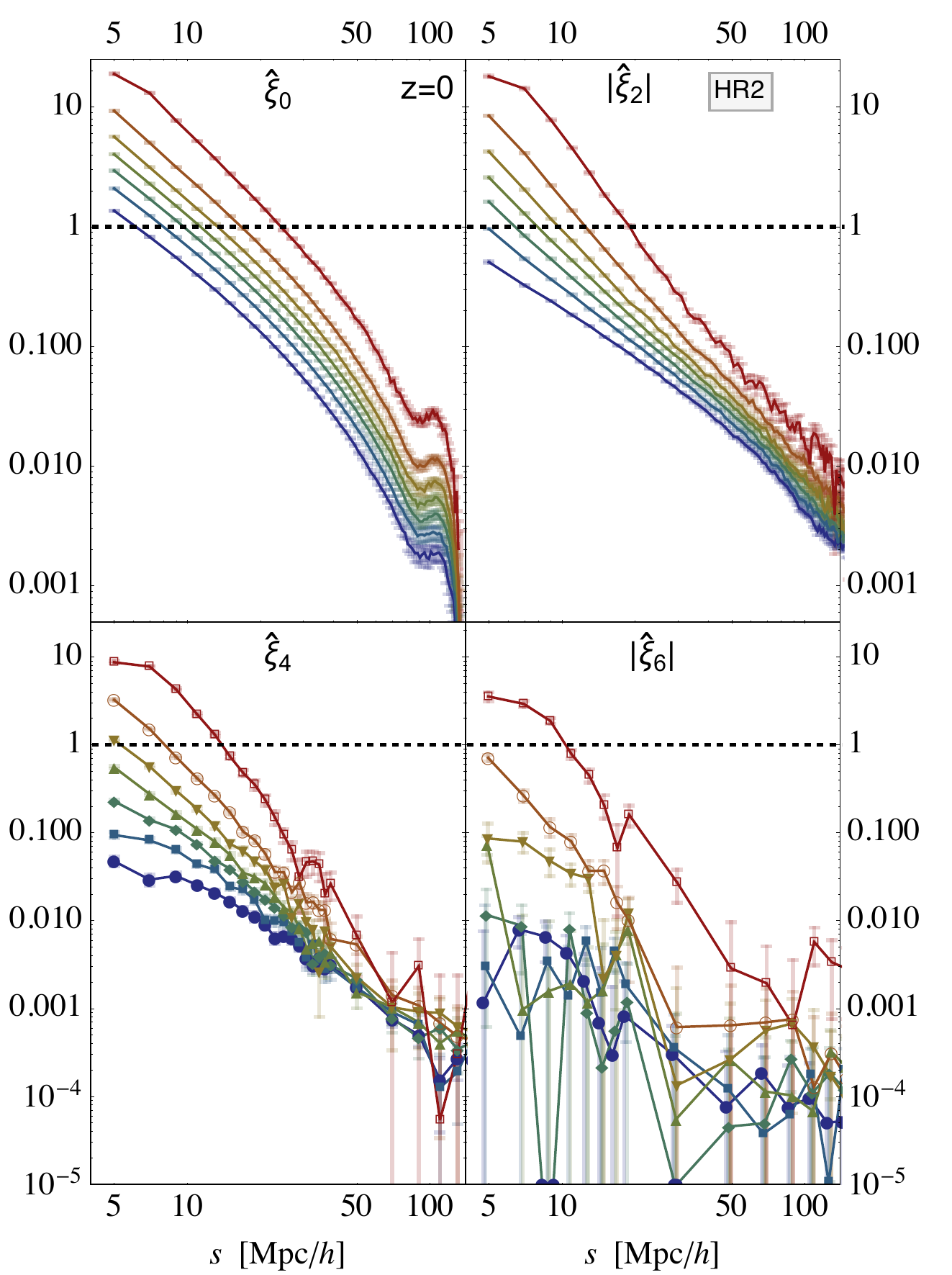}
\caption{The first 4 multipoles of the redshift space correlation function $\xi(s,\mu)$ measured in the HR2 halo catalog. We indicate with the dotted line at $\xi=1$ the nonlinear regime. For better visibility we connect the data points with lines and do not show the data symbols in the upper panels. For $\hat \xi_0$ and $|\hat \xi_2|$ the lines from  top to bottom have  decreasing mass.}
\label{fig:allzspacedirect}
\end{figure}

From the HR2 halo catalog we generate the redshift space by simply changing the $z$-coordinate $r_z$ of each halo to $s_z= r_z+ v_z/(aH)$, where $v_z$ is the $z$-component of the halo peculiar velocity $\v{v}$. This, of course, does not correspond to the real observed light cone, but it directly corresponds to the distant observer approximation used in the theoretical calculation used in Sec.\,\ref{sec:GSM} and is therefore best suited for comparison with analytical calculations. We choose again bins $I_s := [s - \Delta s/2,s + \Delta s/2]$ with width $\Delta s = 2\Mpc$, and $s$ the pair separation in redshift space.
In addition to $s$ we need to consider bins with respect to $\mu = \hat{\v{s}}\cdot \hat{\v{z}}$, the cosine of the angle between the line of sight and and the halo separation in redshift space.
 The analogue of the simple estimator \eqref{MoWhest} in redshift space is
\begin{equation}
1+\hat\xi(s,\mu) = \frac{\Delta P(s\,,\mu)}{n_{\rm tot} 2 \pi s^2 \Delta s \Delta \mu}\,.
\end{equation}
For convenience we display here only expressions for the full box while the Jackknife versions and covariance matrices can be obtained analogously to the real space $\hat\xi_X$ in Sec.\,\ref{sec:realspacecorr}. 

We expand the redshift-space halo correlation function $\xi_X(s_{||},s_\perp)=\xi_X(s,\mu)$, where $s^2 = s_{||}^2+s_\perp^2$ and $\mu = s_{||}/s$, into Legendre polynomials $L_n(\mu)$ using  
\begin{subequations}
\label{ximultipoles}
\begin{align} \label{legendremom}
\xi_X(s,\mu,t)&= \sum_{n=0}^\infty L_n(\mu) \xi_{X,n}(s,t)\,,\\
\label{xin}
\xi_{X,n} (s,t) &=\frac{1+2 n}{2}\, \int_{-1}^1 \xi_X(s,\mu,t) L_{n}(\mu) d \mu \,.
\end{align}
\end{subequations}
This is useful because the ampltiude of the redshift-space multipoles $\xi_{X,n}(s)$ quickly decays as a function of $n$. 
Due to the symmetry $\xi_X(s,\mu) = \xi_X(s,-\mu)$, the $\xi_{X,n}$ vanish for all odd $n$. 
In linear theory \cite{K87} the moments with $n>4$ even vanish identically. 
So, instead of sampling in $s$ and $\mu$, we integrate directly over $\tfrac{1}{2}(1+2 n) L_{n}(\mu) d \mu$ to obtain the multipoles
\begin{subequations}
\begin{align}
1+\hat \xi_0(s) &= \sum_{k, s_{ki} \in  I_s}^{N_{\rm tot}} \sum_{i, s_{ik} \in  I_s}^{N_{\rm tot}} \frac{1}{2}L_{0}(\mu_{ik}) \frac{N_{\rm tot}^{-1}}{n_{\rm tot} 2 \pi s^2 \Delta s}\\
&=  \frac{\Delta P(s)}{n_{\rm tot}\, 4 \pi s^2 \Delta s}\\
\hat \xi_n(s)&= \sum_{k, s_{ki} \in I_s}^{N_{\rm tot}} \sum_{i, s_{ik} \in  I_s}^{N_{\rm tot}} \frac{1+2 n}{2}L_{n}(\mu_{ik}) \frac{N_{\rm tot}^{-1}}{n_{\rm tot} 2 \pi s^2 \Delta s}\,,
\end{align}

where  $s_{ik} = |\v{s}_i - \v{s}_k|$ is the halo separation in redshift space. The third line holds only for $n>0$ and vanishes for all odd $n$, because $L_n$ is then an odd function and $\mu_{ik}$, the cosine of the angle between the halo separation vector $\v{s}_i - \v{s}_k$ and the line of sight $\hat{\v{z}}$, appears twice for each pair but with opposite sign. We calculated the first 6 moments using

\begin{align}
\notag L_0 =& 1\ , \
L_2= \frac{1}{2}(3\mu^2 -1)\ , \
L_4 = \frac{1}{8} (35 \mu^4 - 30 \mu^2 +3)\,,\\
L_6 = & \frac{1}{16} (231 \mu^6 - 315 \mu^4 + 105 \mu^2 -5) \label{Legendre} \,.
\end{align}
\end{subequations}
The hexacontatetrapole $\xi_6$, or for short the 64-pole, is a purely nonlinear effect since it vanishes in the linear, or Kaiser limit \cite{K87} and therefore contains purely nonlinear information, similar to the three-point correlation in absence of primordial non-Gaussianity. 
In Fig.\,\ref{fig:allzspacedirect} we display the measured $\hat{\xi}_0,\hat{\xi}_2,\hat{\xi}_4,\hat{\xi}_6$. 
It is interesting to observe that for the smallest mass bin with $\lgM \simeq 13$ (the bottom curves), the clustering on small scales quickly decreases about two orders of magnitude with increasing $n$, while for largest halos with  $\lgM \simeq 14.7$ (the top curves) it decreases quite slowly, less than one order of magnitude. 
Indeed, for the cluster-sized halos at the smallest scales the clustering is unchanged going from $n=0$ to $n=2$ and remains fully nonlinear even for the 64-pole. 
The core-like behavior of $\hat\xi$ for the largest halo mass (top curves in Figs.\,\ref{fig:allrealdirect} and \ref{fig:allzspacedirect})  is due to the halo separations $r$  becoming comparable to the virial radius $R_{\rm vir}$ such that the correlation function must drop due to halo exclusion which is an artifact of the Friend of Friend identification of halos.\footnote{Halos can overlap in reality and also in $N$-body simulations. Therefore when halos are very close or overlap a phase space method for halo identification would lead to a different shape of the correlation function once $r \lesssim R_{\rm vir}$.}

Note that in contrast to the matter correlation function, the halo correlation for fixed mass increases with redshift for the masses considered here. 
This is because massive halos, corresponding to large initial $\delta/\sigma$ have an inital proto-halo clustering that grows stronger with formation redshift than the gravitationally induced clustering of halos reduces with redshift.
This means that the so-called halo bias which we discuss in the next section is a very important ingredient in our understanding of halo correlation functions.

\subsection{Halo bias model}
\label{app:halobias}
Dark matter halos result from the non-linear collapse of initial density perturbations. 
The abundance and clustering of these virialized structures depends on both the properties of the initial matter density field and the subsequent dynamics. 
\subsubsection{Halo mass function}
The halo mass function is of interest because it determines the bias parameters through the conditional mass function.  

\paragraph*{Excursion set approach}
Following the seminal work of Press and Schechter \cite{PS74}, the excursion set approach \cite{BCEK91} computes the abundance of dark matter halos as a function of their mass. The method involves the at scale $R$ smoothed initial density field $\delta_R(\v{q})$ \eqref{deltaR} and the idea that once $\delta_R$ is above a threshold $\delta_c(z)$ for largest possible smoothing scale $R$, that the region will collapse to a halo of mass $M =4 \pi/3 \rho_0 R^3$, forming at redshift $z$.
The excursion set mass function $n(M)$ is obtained by equating the comoving density of halos $n(M)$ per mass range $[M, M+dM]$ to the fraction of collapsed comoving volume,
\begin{equation}
n(M)=f(\sigma) \frac{\rho_0}{M} \frac{d \ln \sigma^{-1}}{d M}\,,\label{nandf}
\end{equation}
where $\rho_0$ is the comoving background matter density, $\sigma(M)$, Eq.\,\eqref{sigma} with an extra factor $k^2$ in the integrand, the root-mean-square fluctuation of the initial density field and $f(\sigma)$ the so-called multiplicity function describing the fraction of collapsed volume per mass and $\sigma^2$ is the linearly extrapolated variance of the initial linearly density field smoothed at the Lagrangian scale
\begin{align}
\sigma^2(M) &=\frac{1}{2\pi^2}\int dk\, k^2\,\tilde{W}^2(k,R_{\rm L}(M)) P_{\rm lin}(k,z) \ , \label{sigma0}
\end{align}
The name excursion set stems from the fact that when varying $R$ at a fixed point in space, $\delta_R(\v{x})$ performs a random walk whose properties depend on the statistics of the underlying density field, which is usually assumed to be Gaussian and the choice of the filter function. Halos correspond to those random walks which first hit the absorbing barrier determined by the critical density threshold of collapse $\delta_c$. The Press-Schechter (PS) mass function 
\begin{equation}\label{fESP}
f_{\rm PS}(\sigma) = \sqrt{\frac{2}{\pi}}\frac{\delta_c}{\sigma} \exp\left[ -\frac{ \delta_c^2}{2\sigma^2}\right]
\end{equation}
can be derived by assuming that the window function is given by a sharp-k filter, such that $\delta_R$ performs a simple Markov random walk and that $\delta_c$ is the critical density threshold of spherical collapse.

\paragraph*{Extended excursion set approach}
This simple model can be generalized to non-spherical collapse for which the density threshold should be regarded itself as a random variable and stochastic influences have to be included. This has been done in \cite{MR2} in form of a diffusive barrier where the Gaussian diffusion $\langle (B-\langle B\rangle)^2 \rangle = D_B S$ is parametrized by a diffusion constant $D_B$ and the variance $S=\sigma^2(M)$. This idea has been taken up in \cite{AC1,AC2,CA2,CA1} and applied to ellipsoidal collapse by considering a diffusive stochastic barrier with linearly drifting average $\langle B\rangle = \delta_c\rightarrow \delta_c + \beta S$. The dependence on the variance $S(M)$ encodes that smaller halos, with larger variance $S$, are more sensitive to shear effects which counteract gravitational infall and thereby increase the collapse threshold \cite{ST02}.  In \cite{AWWR} it has been shown that the parameters of the barrier can be predicted from the initial condition of an $N$-body simulations. 
 
Taking all this into account the multiplicity function becomes
\begin{align}
f_0(\sigma)&=\frac{\delta_c}{\sigma}\sqrt{\frac{2a_B}{\pi}}\,\exp\left[-\frac{a_B}{2\sigma^2}(\delta_c+\beta\sigma^2)^2\right] \,,\label{fsigma0}
\end{align}
where $a_B=1/(1+D_B)$. To incorporate non-Markovian corrections that arise when the sharp $k$-filter is replaced by the more physical sharp $x$-filter, a path integral approach to compute the multiplicity function $f(\sigma)$ has been developed in \cite{MR1,CA2}. The magnitude of the non-Markovian corrections is given by a parameter $\kappa$, which depends on the linear matter power spectrum. For a standard $\Lambda$CDM Universe,  $\kappa\sim 0.65$. 
In \cite{ABPW} the solution was extended to a diffusive barrier with general shape. Using a simple drifting term, the multiplicity function to first order in $\kappa$ is given by 
\begin{equation}
f_{\rm ACMR}(\sigma)=f_0(\sigma)+f_{1,\beta=0}^{m-m}(\sigma)+
f_{1,\beta^{(1)}}^{m-m}(\sigma)+f_{1,\beta^{(2)}}^{m-m}(\sigma)\,,\label{ftot}
\end{equation}
where $f_0(\sigma)$ is given by \eqref{fsigma0} and
\begin{align*}
f_{1,\beta=0}^{m-m}(\sigma)&=-\kappa a_B\dfrac{\delta_c}{\sigma}\sqrt{\frac{2a_B}{\pi}}\left[\exp\left[-\frac{a_B \delta_c^2}{2\sigma^2}\right]-\frac{1}{2} \Gamma\left(0,\frac{a\delta_c^2}{2\sigma^2}\right)\right]\,,\\
f_{1,\beta^{(1)}}^{m-m}(\sigma)&=- a_B\,\delta_c\,\beta\left[\kappa a_B\,\text{Erfc}\left( \delta_c\sqrt{\frac{a_B}{2\sigma^2}}\right)+ f_{1,\beta=0}^{m-m}(\sigma)\right]\,,\\
f_{1,\beta^{(2)}}^{m-m}(\sigma)&=-a_B\,\beta\left[\frac{\beta}{2} \sigma^2 f_{1,\beta=0}^{m-m}(\sigma)+\delta_c \,f_{1,\beta^{(1)}}^{m-m}(\sigma)\right]\,.
\end{align*}
In \cite{achitouvetal} it was shown that the first order approximation in $\kappa$ is sufficient to reproduce the exact solution to $\sim 5\%$ accuracy, using parameter values $\beta=0.12$, $D_B=0.4$. A positive $\beta$ corresponds to the physically well understood picture that small masses require a larger density amplitude to collapse due to the more likely ellipticity of low variance peaks \cite{SMT}. 

\begin{figure}
\includegraphics[width=0.47\textwidth]{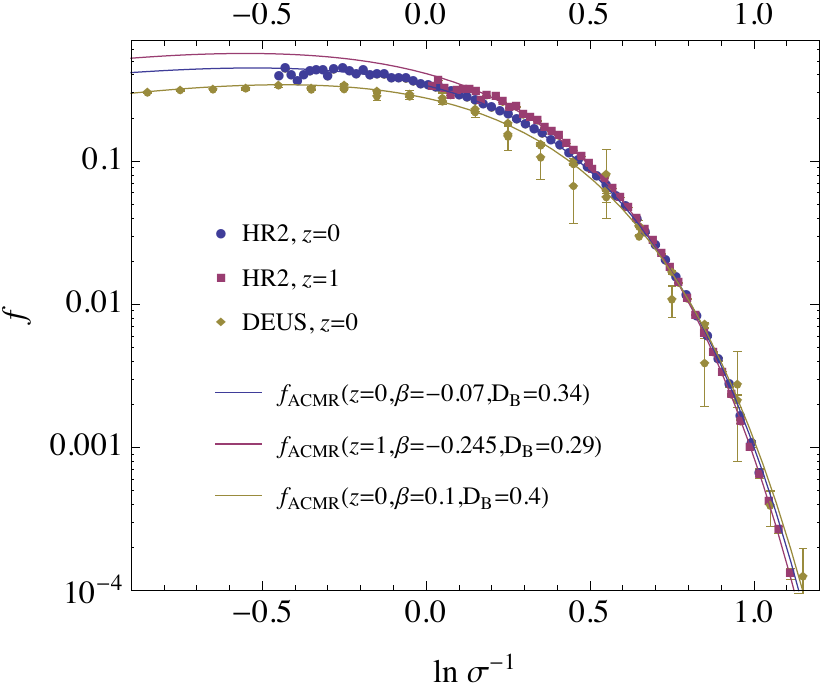}
\caption{Multiplicity function measured in HR2 for redshifts $z=1$ and $z=0$ and DEUS for $z=0$ compared to the fits for the multiplicity function based on \eqref{ftot}. Error bars show only Poisson noise. }
\label{multfuncs}
\end{figure}

\begin{figure}
\includegraphics[width=0.47\textwidth]{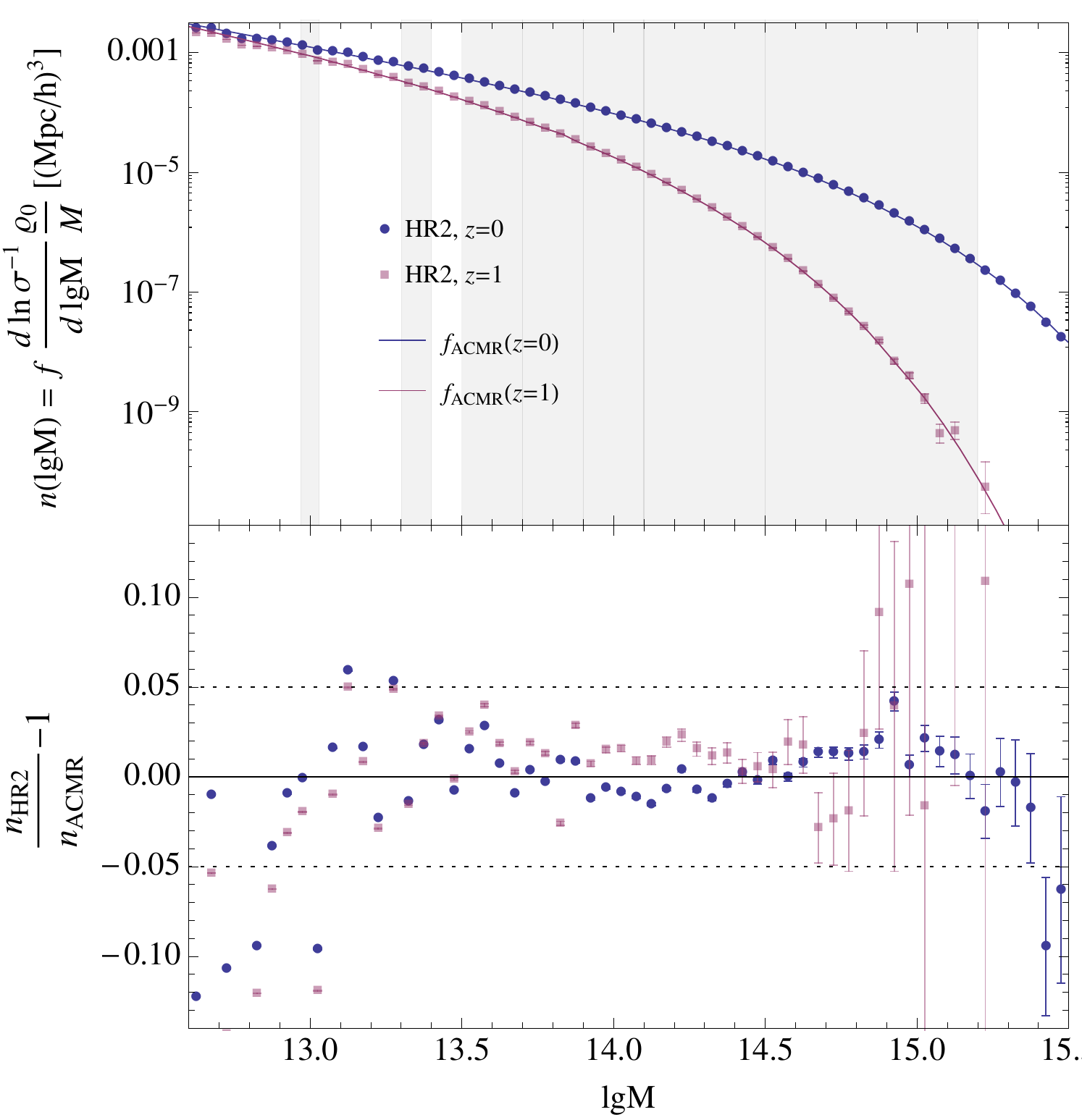}
\caption{Mass function measured from HR2 for redshifts $z=1$ and $z=0$ compared to fits for the multiplicity function based on \eqref{ftot}. }
\label{massfuncs}
\end{figure}

\paragraph*{Fitting the halo mass function}
We now use this result and fit the model parameters $\beta$ and $D_B$ to our simulations. For comparison we show in Fig.\,\ref{multfuncs} the multiplicity function from both the HR2 introduced here and the DEUS simulations described in Section \ref{sec:DEUSS}. Both simulations have similar cosmology and identified halos with a Friends-of-friends algorithm with linking length $b=0.2$. For the DEUS analysis we used three different box sizes, $2592\Mpc$, $648\Mpc$ and $162\Mpc$, which all contain $1024^3$ particles. The smaller box size also probes low mass halos compared to the HR2 requiring a positive $\beta$. For the Horizon Run 2 simulation we find $\beta=-0.07$, $D_B=0.34$ for $z=0$, and $\beta=-0.245$, $D_B=0.29$ for $z=1$ while for the DEUS simulation $\beta=0.1$, $D_B=0.4$ for $z=0$. The fact that $\beta<0$ and the slight redshift dependence of $f(\sigma)$ for HR2 observed in Fig.\,\ref{multfuncs} might signal a problem in our analysis or with the halo catalogue HR2. In Fig.\,\ref{massfuncs} we show the measured halo mass function from HR2 with the mass bins used in grey in comparison to the fit obtained for the multiplicity function.

\paragraph*{Peak background split} The conditional mass function
\begin{equation}
\label{condf}
f_0(S| \delta_{R_0}, S_0) = \sqrt{\frac{2}{\pi}}\frac{ (\delta_c - \delta_{R_0}) S/a_B}{\left(S/a_B- S_0 \right)^{3/2}} \exp\left[ -\frac{ (\delta_c+\beta S -\delta_{R_0})^2}{2\left(S/a_B-S_0\right)}\right]
\end{equation}
 is then obtained from the peak-background split \cite{BCEK91,Laceyetal} which considers density fluctuations to be embedded in a background region of size $R_0$ with a background over- or underdensity $\delta_{R_0}$ which locally reduces or increases the collapse threshold $\delta_c$.
This equation is only valid as long as $S>S_0$. 
The full non-Markovian $f$, Eq.\,\eqref{ftot}, in contrast to $f_0$, Eq.\,\eqref{fsigma0}, cannot simply be extended as a conditional mass function upon replacing  $S \rightarrow S - a_B S_0,  \delta_c \rightarrow \delta_c - \delta_{R_0}$.
Hence, for convenience we derive the bias parameters $b_n$, Eq.\,\eqref{IxBias}, from the Markovian $f_0$, Eq.\,\eqref{fsigma0} from which the corresponding conditional mass function \eqref{condf} can be easily obtained for moving barriers, see \cite{Laceyetal, ABPW}.
Another way to understand halo bias, besides the peak-background split, is in terms of the peak model \cite{BBKS86}, which takes into account that proto-halos are high (low) density peaks, which are naturally (anti-)clustered in a Gaussian random field. Both ideas have been combined into the excursion sets peaks model \cite{PS12}. However the parameters of the barrier that enter into this model are not consistent with the initial condition of the $N$-body simulations \cite{ParanjapeShethDesjacques2013}.

\subsubsection{Local Lagrangian bias model}
\label{LocLagrBiasModel}

We will use the formalism developed in \cite{M08, Mat12} to connect the halo fluctuation field 
\begin{align}
F_0[\delta_{R_0}(\v{q})]&= \frac{f_0(S| \delta_{R_0}(\v{q}), S_0)}{f_0(S)}= 1+\sum_{n=1}^\infty \frac{b_n(S,S_0)}{n!} \delta_{R_0}^n(\v{q}) \label{peakbackbias}
\end{align}
to the conditional mass function \eqref{nandf} and \eqref{condf} from which the Lagrangian bias parameters $b_n$ can be determined. 
That $F[\delta_{R_0}(\v{q})]$ really has the interpretation of a halo fluctuation field can be read off from the first equality, because the normalized local halo mass function, varies from place to place through its dependence on $\delta_{R_0}(\v{q})$.
 The second equality assuming that a Taylor expansion in $\delta_{R_0}$ is possible shows that the so-called Lagrangian bias parameters $b_n$ are all local in this model; they do not involve derivatives acting on $\delta_{R_0}$.

It is clear that the arbitrary scale $R_0$ should not enter the final result, and there are ways to absorb the background smoothing $R_0$ into a renormalized albeit nonlocal bias coefficients \cite{SJD13} such that the halo fluctuation field $F[\delta_{R_{\rm L}(M)}(\v{q})]$ only depends on the physical peak scale $R_{\rm L}(M)$ and not on the arbitrary background scale $R_0$ \cite{Mat12}.
It is important to realize that the peak smoothing scale is not an intermediate tool for regularization, that is set to zero at the end of the calculation, but rather a physical scale on which the final observables are expected to depend \cite{BBKS86, BM96, DS10}.  
  That is different for the background scale $R_0$ on which observables should not depend. 
  After renormalization the $R_0$ dependence disappears, while the expression for $F$ receives a nonlocal term proportional to the Laplacian of the density field and other correction terms which we neglect. Those terms appear in the effective theory of halo bias and also in the bias model based on peaks theory \cite{D08, SJD13, ABGZ14}. 
  Our main point here is that although one can get rid of the artificial smoothing scale $R_0$  appearing in Eq.\,\eqref{peakbackbias} through the process of renormalisation, the physical peak scale $R_{\rm L}(M)$ will still enter smoothing kernels of the initial density field such that
  \begin{align}
\label{renorm}
F[\delta_R(\v{q})]&= 1+\sum_{n=1}^\infty \frac{1}{n!}[b_n(S) \delta_R^n(\v{q}) + ...]\,,
\end{align}
where ellipses indicate the neglected non-local terms and $R=R_{\rm L}(M)$ and $S=\sigma^2(R_{\rm L}(M))$. 
While in all discussions of halo bias the mass dependence of the bias coefficients is kept  ($b_n=b_n(R_{\rm L}(M)) $), the mass dependent smoothing scale $R_{\rm L}(M)$ of the linear density field is usually dropped.
The significant effect of this smoothing scale will be discussed in the next section \ref{SmoothingscaleAsPeakscale}.
In the large scale limit $S \gg S_0$ the bias coefficients from \eqref{condf} also become independent of $S_0$ and can be calculated using $\del_{\delta_{R_0}}f_0(S|\delta_{R_0},S_0)|_{S_0\rightarrow 0} = - \del_{\delta_c} f_0(S)$
\begin{align} \label{IxBias}
b_n(M,z)& =  (-1)^n (\partial_{ \delta_c}^n f_0)/f_0\,,\\
\notag b_1(M,z) &= a_B \beta - \delta_c^{-1} + \sigma^{2}a_B \delta_c\,,\\
\notag b_2(M,z) &= (a_B \beta)^2 - 2 a_B \beta \delta_c^{-1}+\left(2a_B^2 \beta \delta_c  - 3 a_B\right) \sigma^{2}+ a_B^2 \delta_c^2 \sigma^{4} \,,
\end{align}
where all the mass dependence arises through $\sigma^2=\sigma^2(M)$, and therefore the window function.
In principle, the numerical values of the averaged bias parameters can be predicted using the mass function \eqref{nandf}
 \begin{equation} \label{averbias}
 \bar b_n =\frac{\int_{\lgM_{\rm min}}^{\lgM_{\rm max} }d \lgM\, n(\lgM) b_n(\lgM)}{\int_{\lgM_{\rm min}}^{\lgM_{\rm max} }d \lgM\, n(\lgM)}\,,
 \end{equation}
where $\lgM_{\rm min}$ and $\lgM_{\rm max}$ are the lower and upper value for the given mass bin of the measured halo correlation function $\bar \xi_X$, respectively. 
Not however that there is no a priori reason that the bias coefficients in the expansion \eqref{renorm}, used in this work, are similar to those appearing in \eqref{peakbackbias} in the limit $S_0 \rightarrow 0$. 
Therefore we won't use \eqref{averbias} and 
instead will assist the bias model following the procedure of \cite{CRW13, WRW14} by treating the mass $M$ appearing in \eqref{IxBias} as a free parameter and find the optimal mass $M_{\rm opt}$ by fitting the theoretical model to the real space correlation function.  
Following \cite{M08, CRW13, WRW14}, we will identify those fitted bias parameters with the statistical average $\langle F^{(n)} \rangle = b_n(M_{\rm opt})$ as they arise in integrated Lagrangian perturbation theory \cite{M08} and convolution Lagrangian perturbation theory \cite{CRW13}.

\begin{figure}
\includegraphics[width=0.45\textwidth]{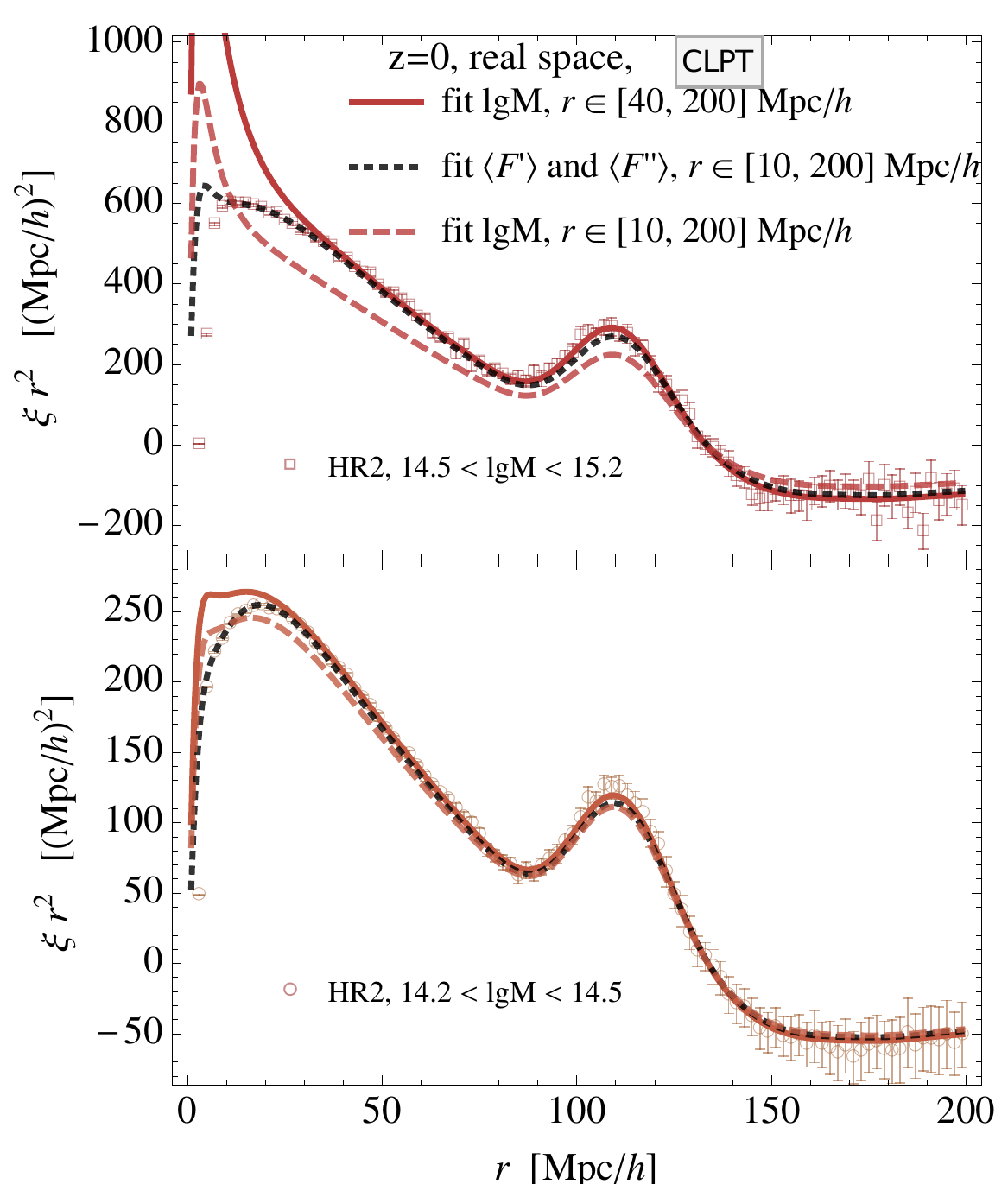}
\caption{Real space correlation function times $r^2$ for the largest mass bin; the mass function fit for $\lgM_{\rm opt}$ does not work anymore. Finding $\lgM_{\rm opt}$ while including correlation function data down to $10\Mpc$ the fit does not recover the linear scales. Fitting for both $\langle F'\rangle$ and $\langle F''\rangle$ leads to a good $\chi^2$, however it is unclear wether this is physically meaningful.}
\label{fig:fitfail}
\end{figure}

\subsubsection{Fitting the bias model}
\label{app:fitbias}
To determine the bias parameters that enter the evaluation of the streaming model ingredients we fix the TCLPT model for the real space correlation $\xi(r)$ to $R=1\,\Mpc$ and
\begin{equation}
\langle F'\rangle \equiv b_1(M_{\rm opt})\quad\mathrm{and}\quad
\langle F''\rangle \equiv b_2(M_{\rm opt})
\end{equation}
and fit for $M$. We report here on some details regarding the fitting procedure that has been used to obtain the best-fitting masses as well as the corresponding bias factors that are summarized in Table\,\ref{biasfittab}.

\paragraph*{Range for the fit} For the $\chi^2$-fit we did not use all 100 $r$-values per mass bin of the measured correlation function $\hat{\bar\xi}$, because we only want to force our model to match scales larger than $40\Mpc$. From the 80 remaining data points per mass bin we only used 26 linear combinations. The reason is that we measured the covariance matrix $C_{\hat \xi}(r_i,r_j)$, where $i,j$ label $r$-bins, only trough 27 samples and therefore the 80 eigenvalues Eval$_m$(C), decreasing with increasing $m$, are sharply dropping to zero after $m=26$. We therefore minimize
\begin{equation}\label{chisq}
\chi^2 = \sum_{m=1}^{26} \left[\sum_{j=21}^{100}\frac{\mathrm{ Evec}_{m j}(C)}{\mathrm{ Eval}_m(C)}\left(\xi(r_j,M) - \hat{\bar{\xi}}(r_j)\right)\right]^2\,.
\end{equation}
 Evec$_{m j}(C)$ is the eigenvector matrix of $C_{\hat \xi}(r_i,r_j)$, that projects the data into de-correlated linear combinations. Those linear combinations with the largest eigenvalues (smallest $m$) are called principal components and have the strongest impact on the best-fitting mass $M_{\rm opt}$.

\paragraph*{Issues for large cluster-sized halos}
As evident from Fig.\,\ref{fig:allreal}, the nonlinear bias model has problems on small scales $r<30\Mpc$ for the largest mass bins corresponding to galaxy clusters. This problem persists regardless of the chosen filter scale $R$ in TCLPT (including CLPT with $R=0$) and regardless of the chosen mass function; we tested also the Sheth-Tormen \cite{ST99} mass function that gives similar results for $b_1$ and $b_2$ but is problematic because $\lgM_{\rm opt}$ can be far away from 
\begin{align} \label{meanmass}
\overline \lgM= \frac{\int_{ \lgM_{\rm min}}^{ \lgM_{\rm max}} n(\lgM)\, \lgM\, d \lgM}{\int_{ \lgM_{\rm min}}^{ \lgM_{\rm max}} n(\lgM)\, d \lgM}\,.
\end{align}
This shows that if the mass function is calibrated, as we have done for $f_0$ by fitting for $D_B$ and $\beta$, then also the bias parameters become more consistent.

We also fitted for $\langle F'\rangle$ and $\langle F''\rangle$ independently to see whether they can be adjusted such that the problem for cluster-sized halos disappears. If only $r$-values larger than $40\Mpc$ are used, the independent fit for $\langle F'\rangle$ and $\langle F''\rangle$ gives values very close to $b_1(M_{\rm opt})$ and $b_2(M_{\rm opt})$ and thus the same problem occurs. For the largest halos $14.5 < \lgM <15.2$ we find when taking into account a larger $r$-range, $r \in [10,200]\Mpc$ for the fit, one obtains $\lgM_{\rm opt} = 14.66$ leading to $b_1(M_{\rm opt}) = 2.66$  destroying the agreement on large scales, see Fig.\,\ref{fig:fitfail}. We exclude scales smaller  than $10\Mpc$ because we expect there halo exclusion: two objects identified in HR2 through a friend-of-friend algorithm cannot come arbitrary close to each other. The two-parameter fit ($\langle F'\rangle$, $\langle F''\rangle$) also shows some artefacts below $r=10\Mpc$ for the largest mass bin but gives a good fit on larger scales for ($\langle F'\rangle_{\rm opt} = 3.01$, $\langle F''\rangle_{\rm opt} = 1.97$) on larger scales, compare last column in Table \ref{biasfittab}.

Note that for all theoretical curves in Fig.\,\ref{fig:fitfail} we used a smoothing scale of $1\Mpc$. If we had used instead the Lagrangian radius, the behavior of the correlation function at very small scales would be better under control, compare Figs.\,4 and 5 of \cite{UKH15}, but of course would then fail on large scales as we discussed in Sec.\,\ref{sec:realcorrfct}.
 We therefore expect that both the small and large scales of the real space correlation of cluster-sized halos can be better modelled with the inclusion of peak bias. Including the peak physics will retain the sharpness of the BAO peak despite the usage of the Lagrangian smoothing scale, while the very small scales will be better under control exactly because of the smoothing.

\bibliography{HusimiVlasovbib}

\end{document}